\documentclass[fleqn,usenatbib]{mnras}

\usepackage{newtxtext,newtxmath}

\usepackage[T1]{fontenc}
\usepackage{ae,aecompl}

\usepackage{graphicx}	
\usepackage{amsmath}	
\usepackage{amssymb}	
\usepackage[utf8]{inputenc}
\usepackage[italian, english]{babel}
\usepackage{xcolor}
\usepackage{bigints}
\usepackage{booktabs}
\usepackage{multirow}
\usepackage{lipsum}
\usepackage{siunitx}
\usepackage{xspace}
\usepackage[normalem]{ulem}
\usepackage{soul}

\newcommand{\kstar}{k_*}

\newcommand{\Mdyn}{M_\mathrm{dyn}}
\newcommand{\msun}{\mathrm{M_\odot}}

\newcommand{\kms}{\mathrm{km\,s^{-1}}}
\newcommand{\angstrom}{\text{\normalfont\AA}}
\newcommand{\sdss}{\mathrm{SDSS}}
\newcommand{\pdf}{\mathrm{P}}
\newcommand{\diff}{\mathrm{d}}
\newcommand{\obs}{\mathrm{obs}}

\newcommand{\logm}{\log M_*}

\newcommand{\logmobsi}{\log M_{*,i}^\mathrm{obs}}
\newcommand{\logmi}{\log M_{*,i}}

\newcommand{\logsigmae}{\log\sigma_{\mathrm e}}

\newcommand{\logsigmaeobsi}{\log\sigma_{\mathrm{e},i}^\mathrm{obs}}
\newcommand{\logsigmaei}{\log\sigma_{\mathrm{e},i}}
\newcommand{\errsigma}{\sigma^2_{\sigma_{\mathrm{e}},i}}
\newcommand{\errmass}{\sigma^2_{M_{*},i}}
\newcommand{\errmu}{\sigma^2_{\sigma,i}}
\newcommand{\pp}{\mathrm{P}}

\newcommand{\data}{\boldsymbol d}
\newcommand{\datai}{\boldsymbol d_i}

\newcommand{\pr}{{\rm P}}
\newcommand{\hyperp}{\boldsymbol\Phi}
\newcommand{\individ}{\boldsymbol\Theta}
\newcommand{\individi}{\boldsymbol\Theta_i}

\newcommand{\zetahom}{\zeta_{\mathrm{hom}}}
\renewcommand{\Re}{R_{\mathrm e}}
\newcommand{\aR}{a_{\mathrm R}}
\newcommand{\ak}{a_{\mathrm k}}

\newcommand{\sigmae}{\sigma_{\mathrm e}}
\newcommand{\Mconst}{$\mathcal{M}_\mathrm{const}$\xspace}
\newcommand{\Mevo}{$\mathcal{M}_\mathrm{evo}$\xspace}
\newcommand{\Msdss}{$\mathcal{M}^\mathrm{SDSS}$\xspace}
\newcommand{\Mconstfid}{$\mathcal{M}_\mathrm{const}^\mathit{fid}$\xspace}
\newcommand{\Mevofid}{$\mathcal{M}_\mathrm{evo}^\mathit{fid}$\xspace}
\newcommand{\Mconstext}{$\mathcal{M}_\mathrm{const}^\mathit{ext}$\xspace}
\newcommand{\Mevoext}{$\mathcal{M}_\mathrm{evo}^\mathit{ext}$\xspace}
\newcommand{\Mconstfidnes}{$\mathcal{M}_\mathrm{const,NES}^\mathit{fid}$\xspace}
\newcommand{\Mevofidnes}{$\mathcal{M}_\mathrm{evo,NES}^\mathit{fid}$\xspace}
\newcommand{\Mconstextnes}{$\mathcal{M}_\mathrm{const,NES}^\mathit{ext}$\xspace}
\newcommand{\Mevoextnes}{$\mathcal{M}_\mathrm{evo,NES}^\mathit{ext}$\xspace}
\newcommand{\Mconstnes}{$\mathcal{M}_\mathrm{const,NES}$\xspace}
\newcommand{\Mevones}{$\mathcal{M}_\mathrm{evo,NES}$\xspace}

\usepackage{etoolbox}
\makeatletter
\patchcmd\@combinedblfloats{\box\@outputbox}{\unvbox\@outputbox}{}{\errmessage{\noexpand patch failed}}
\makeatother
\newcommand{\minitab}[2][l]{\begin{tabular}{#1}#2\end{tabular}}

\def\equationautorefname#1#2\null{%
  equation#1(#2\null)%
}

\title[The stellar mass--velocity dispersion relation of early-type galaxies]{The cosmic evolution of the stellar mass--velocity dispersion relation of early-type galaxies}

\author[C.\ Cannarozzo et al.]{\noindent
Carlo Cannarozzo$^{1,2}$\thanks{E-mail: carlo.cannarozzo3@unibo.it},
Alessandro Sonnenfeld$^{3}$
and Carlo Nipoti$^{1}$
\\
$^{1}$Dipartimento di Fisica e Astronomia, Alma Mater Studiorum Università di Bologna, Via Piero Gobetti 93/2, I-40129 Bologna, Italy\\
$^{2}$INAF - Osservatorio di Astrofisica e Scienza dello Spazio di Bologna, Via Piero Gobetti 93/3, I-40129 Bologna, Italy\\
$^{3}$Leiden Observatory, Leiden University, Niels Bohrweg 2, 2333 CA Leiden, The Netherlands\\
}

\date{Accepted 2020 July 16. Received 2020 July 06; in original form 2019 October 12}
\pubyear{2020}

\begin{document}
\label{firstpage}
\pagerange{\pageref{firstpage}--\pageref{lastpage}}
   \maketitle
\begin{abstract}
We study the evolution of the observed correlation between  central stellar velocity dispersion $\sigmae$ and  stellar mass $M_*$ of massive ($M_*\gtrsim 3\times 10^{10}\,\mathrm{M_\odot}$) early-type galaxies (ETGs) out to redshift $z\approx 2.5$, taking advantage of a Bayesian hierarchical inference formalism. Collecting ETGs from state-of-the-art literature samples, we build a {\em fiducial} sample ($0\lesssim z\lesssim 1$), which is obtained with homogeneous selection criteria, but also a less homogeneous {\em extended} sample ($0\lesssim z\lesssim 2.5$). Based on the fiducial sample, we find that at $z\lesssim 1$ the $M_*$-$\sigmae$ relation is well represented by  $\sigmae\propto M_*^{\beta}(1+z)^{\zeta}$, with $\beta\simeq 0.18$ independent of redshift and $\zeta\simeq 0.4$ (at given $M_*$, $\sigmae$ decreases for decreasing $z$, for instance by a factor of $\approx1.3$ from $z=1$ to $z=0$).
When the slope $\beta$ is allowed to evolve, we find it increasing with redshift: $\beta(z)\simeq 0.16+0.26\log(1+z)$ describes the data as well as constant $\beta\simeq 0.18$. The intrinsic scatter of the $M_*$-$\sigmae$ relation is $\simeq0.08$ dex in $\sigmae$ at given $M_*$, independent of redshift. Our results suggest that, on average, the velocity dispersion of {\em individual} massive ($M_*\gtrsim 3\times 10^{11}\msun$) ETGs decreases with time while they evolve from $z\approx 1$ to $z\approx 0$.
The analysis of the extended sample, over the wider redshift range $0\lesssim z\lesssim 2.5$, leads to results similar to that of the fiducial sample, with slightly stronger redshift dependence of the normalisation ($\zeta\simeq 0.5$) and weaker redshift dependence of the slope
(${\rm d} \beta/{\rm d} \log (1+z)\simeq 0.18$)  when $\beta$ varies with time. At $z=2$ ETGs with $M_*\approx 10^{11}\,\msun$ have, on average, $\approx1.7$ higher $\sigmae$ than ETGs of similar stellar mass at $z=0$.
\end{abstract}




\section{Introduction}
\label{sec:intro}

Since the late 1970s it was found empirically that present-day early-type galaxies (ETGs) follow scaling relations, i.e.\ correlations among global observed quantities, such as the Faber-Jackson relation \citep{FaberJackson1976ApJ} between luminosity $L$ and central stellar velocity dispersion $\sigma_0$,  the Kormendy relation \citep{Kormendy1977ApJ} between effective radius $R_{\rm e}$ and surface brightness (or luminosity), and the fundamental plane \citep{DjorgovskiDavis1987ApJ,Dressler1987ApJ} relating $\sigma_0$, $L$ and $R_\mathrm{e}$.
When estimates of the stellar masses are available, analogous scaling relations are found, replacing $L$ with $M_*$: the $M_*$-$R_\mathrm{e}$ (stellar mass--size) relation,
the $M_*$-$\sigma_0$ (stellar mass--velocity dispersion) relation and the stellar-mass fundamental plane \citep[e.g.,][]{HydeBernardi2009MNRAS,Hyd09b,Auger2010ApJ,Zahid2016ApJ}.
These scaling laws are believed to contain valuable information on the process of formation and evolution of ETGs. Any successful theoretical model of galaxy formation should reproduce these empirical correlations of the present-day population of ETGs \citep{Som15,Naa17}. 

The observations strongly indicate that ETGs are not evolving passively. 
For instance, measurements of sizes and stellar masses of samples of quiescent galaxies at higher redshift imply that the $M_*$-$R_\mathrm{e}$ relation evolves with time: on average, for given stellar mass, galaxies were significantly more compact in the past \citep[e.g.][]{Fer04,Dam19}.
There are also indications that ETGs at higher redshift have, on average, higher stellar velocity dispersion than  present-day ETGs of similar $M_*$ \citep[e.g.][]{vandeSande2013ApJ,Belli2014ApJ,Gargiulo2016AAP,Belli2017ApJ,Tanaka2019arXiv}.
Interestingly, the stellar-mass fundamental plane, relating $M_*$, $\sigma_0$ and $R_\mathrm{e}$ appears to change little with redshift \citep{Bezanson2013bApJl,Bezanson2015ApJ,Zahid2016ApJ821}.
The observed behaviour of these scaling relations as a function of redshift represents a further challenge to models of galaxy formation and evolution. 

In the standard cosmological framework, structure formation in the Universe occurs as a consequence of the collapse and virialisation of the dark matter halos, in which baryons infall and collapse, thus forming galaxies.
In this framework, massive ETGs are believed to be the end products of various merging and accretion events. Given the old ages of the stellar populations of present-day ETGs, any relatively recent merger that these galaxies experienced must have had negligible associated star formation. Based on these arguments, a popular scenario for the late ($z\lesssim 2$) evolution of ETGs is the idea that these galaxies grow via dissipationless (or "dry") mergers. 
Interestingly, dry mergers make galaxies less compact: for instance, galaxies growing via parabolic dry merging increase their size as $R_{\rm e}\propto M_*^a$, with $a\gtrsim 1$, while their velocity dispersion evolves as $\sigma_0\propto M_*^b$, with $b\lesssim 0$ \citep*{Nipoti2003MNRAS,Naab2009ApJl,Hil13}. 
Thus, the transformation of individual ETGs via dry mergers is a possible explanation of the observed evolution of 
the $M_*$-$R_\mathrm{e}$, $M_*$-$\sigma_0$ and stellar-mass fundamental plane relations \citep{Nipoti2009bApJl,Nipoti2012MNRAS,Posti2014MNRAS,Oog13,Fri17}. Though this explanation is qualitatively feasible, it is not clear whether and to what extent dry mergers can explain quantitatively the observed evolution of these scaling laws. In this context, the stellar velocity dispersion $\sigma_0$ is a very interesting quantity to consider. Even for purely dry mergers of spheroids, $\sigma_0$ can increase, decrease of stay constant following a merger, depending on the merger mass ratio and orbital parameters \citep{Boy06,Naab2009ApJl,Nipoti2009aApJ,Nipoti2012MNRAS,Posti2014MNRAS}. Moreover, 
even slight amounts of dissipation and star formation during the merger can produce a non-negligible increase of the central stellar velocity dispersion with respect to the purely dissipationless case \citep{Rob06,Ciotti2007ApJ}.

In a cosmological context, the next frontier in the theoretical study of the scaling relations of ETGs is the comparison with observations of the evolution measured in hydrodynamic cosmological simulations. 
A quantitative characterisation of the evolution of the observed scaling relations of the ETGs is thus crucial to use them as test beds for theoretical models.  
On the one hand, the evolution of the observed stellar mass--size relation is now well established, being based on relatively large samples of ETGs out to $z\approx 3$ \citep*{Cimatti2012MNRAS,vanderWel14} . 
On the other hand, given that measuring the stellar velocity dispersion requires spectroscopic observations with relatively high resolution and signal-to-noise ratio, the study of the redshift evolution of correlations involving $\sigma_0$, such as the $M_*$-$\sigma_0$ relation and the stellar-mass fundamental plane,
is based on much smaller galaxy samples than those used to study the stellar mass--size
relation. This makes it more difficult to characterise quantitatively the evolution of these scaling laws out to significantly high redshift.

In this paper, we focus on the  stellar mass--velocity dispersion relation of ETGs with the aim of improving the quantitative characterisation of the observed evolution of this scaling law. 
We build an up-to-date sample of massive ETGs with measured stellar  mass and stellar velocity dispersion by collecting and homogenising as much as possible available state-of-the-art literature data.
In particular, we consider galaxies with stellar masses higher than $10^{10.5}\,\msun$ and we correct the observed stellar velocity dispersion to $\sigma_{\rm e}$, the central line-of-sight stellar velocity dispersion within an aperture of radius $R_{\rm e}$, so in our case $\sigma_0=\sigma_{\rm e}$.
We analyse statistically the evolution of the $M_*$-$\sigma_{\rm e}$ relation without resorting to binning in redshift and using a Bayesian hierarchical approach. 
As a result of this analysis we provide the posterior distributions of the hyper-parameters describing the $M_*$-$\sigma_{\rm e}$ relation in the redshift range $0\lesssim z \lesssim2.5$, under the assumption that, at given redshift, $\sigma_{\rm e}\propto M_*^\beta$. We explore both the case of redshift independent $\beta$ and the case in which $\beta$ is free to vary with redshift. 

The paper is organised as follows. \hyperref[sec:total_sample]{Section~\ref*{sec:total_sample}}
describes the galaxy sample and the criteria adopted to select ETGs. We present the statistical method in \autoref{sec:method} and our results in \autoref{sec:results}. Our results are discussed in \autoref{sec:discussion}.
\hyperref[sec:conclusions]{Section~\ref*{sec:conclusions}} concludes. Throughout this work, we adopt a standard $\Lambda$ cold dark matter cosmology with $\Omega_\mathrm{m}=0.3$, $\Omega_\mathrm{\Lambda}=0.7$ and $H_0=70\, \mathrm{km} \, \mathrm{s}^{-1} \mathrm{Mpc}^{-1}$. All stellar masses are calculated assuming a \citet{Chabrier2003PASP} initial mass function (IMF).

\section{Galaxy sample}\label{sec:total_sample}

To study the evolution of the stellar mass--velocity dispersion relation of ETGs we build a sample of galaxies consisting in a collection of various subsamples of ETGs in the literature. 
Our definition of what constitutes an ETG is based mainly on morphology, with the addition of cuts on emission line equivalent width of [OII] aimed at removing star-forming galaxies (as explained in the rest of this section).
Our goal is to build a sample spanning a redshift range as large as possible.
At the same time,
in order to make an accurate inference, it is important to 1) select galaxies and measure their stellar mass and velocity dispersion in a homogeneous way and 2) ensure that, at any given redshift and stellar mass, our selection criteria do not depend, either directly or indirectly, on velocity dispersion.
With our main focus on accuracy, we first define a {\em fiducial sample} of galaxies, for which conditions 1) and 2) above are satisfied.
We drew our fiducial sample from the Sloan Digital Sky Survey \citep[SDSS;][]{Eisenstein2011AJ} and the Large Early Galaxy Astrophysics Census \citep[LEGA-C;][]{LEGA-C}.
For the galaxies in this sample we strictly apply consistent selection criteria and measure their stellar masses using photometric data from the first data release of the Hyper Suprime-Cam \citep[HSC;][]{Miy++18} Subaru Strategic Program \citep[][DR1]{Aihara2018PASJ}.
The two surveys cover the redshift range $0 \lesssim z \lesssim 1$ and, most importantly, have well defined selection functions, which is critical to meet condition 2). 



We then define a second {\em high-redshift} sample, consisting of stellar mass and velocity dispersion measurements of galaxies at $0.8 \lesssim z \lesssim 2.5$ from various independent studies.
For the galaxies in this high-redshift sample, we only require that the definitions of stellar mass and stellar velocity dispersion are the same as those of the fiducial sample.
We also define an {\em extended} sample, obtained by combining the fiducial and high-redshift samples. 
In building our samples, we include only galaxies with stellar mass higher than a minimum mass $M_{\rm *,min}$, which in general depends both on the survey and on $z$ (see \autoref{ssec:fiducial} and \autoref{ssec:high}): in all cases $M_{\rm *,min} \geq 10^{10.5}\msun$, which we adopt as absolute lower limit in stellar mass.

Our strategy is to carry out our inference on both the fiducial and the extended samples. Given the way the samples are built, we expect our results at $z<1$ to be more robust (i.e. less prone to observational biases), but it is nevertheless very interesting to examine trends out to $z\approx 2.5$, as probed by our extended sample.
In the following two subsections we describe in detail how measurements for these samples are obtained.

\subsection{The fiducial sample}\label{ssec:fiducial}

Our fiducial sample consists of two sets of galaxies.
The first set is drawn from the data release 12 \citep[DR12;][]{Alam2015ApJS} of the SDSS. In particular, we consider only objects belonging to the main spectroscopic sample \citep{Str++02}.
The second set is selected from the LEGA-C survey DR2 \citep[][]{Straatman2018ApJS}. 
The LEGA-C DR2 contains spectra of 1,922 objects obtained with the Visible Multi-Object Spectrograph \citep[VIMOS;][]{LeF++03} on the Very Large Telescope (VLT). LEGA-C targets were selected by applying a cut in $K_{\rm s}$-band magnitude to a parent sample of galaxies with photometric redshift in the range $0.6 < z < 1.0$ drawn from the Ultra Deep Survey with the VISTA telescope \citep[UltraVISTA;][]{Muzzin2013ApJ}.

\subsubsection{ETG selection}

\begin{figure}
    \centering
    \includegraphics[width=1.\columnwidth]{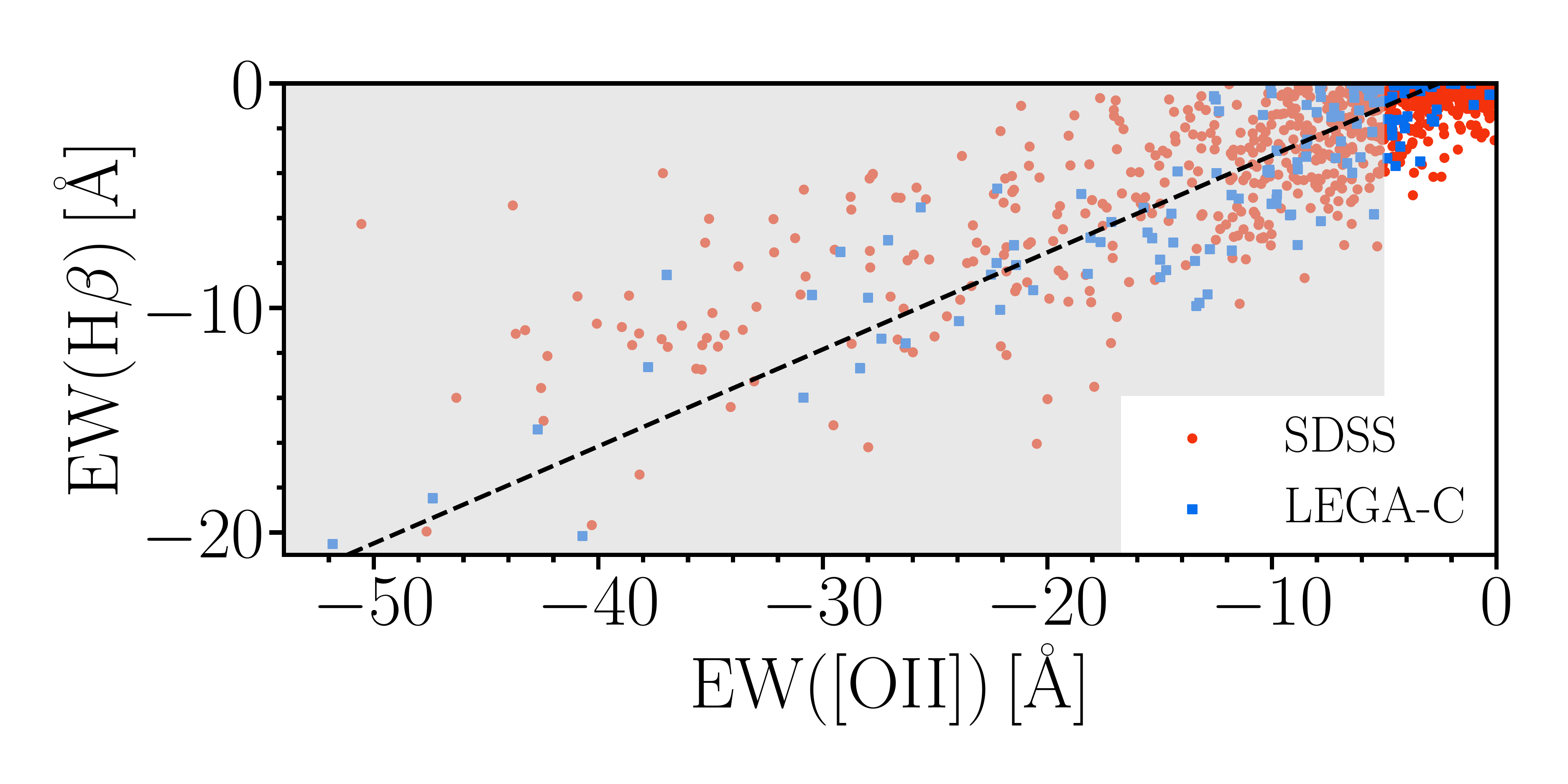}\vspace{-.8cm}
	\caption{Equivalent width of $\mathrm{H}\beta$, $\mathrm{EW(H\beta)}$, as a function of equivalent width of [OII], $\mathrm{EW([OII])}$, for galaxies drawn from the original catalogues of SDSS (circles) and LEGA-C (squares). For LEGA-C galaxies, we show only objects with signal-to-noise ratio $>10$. The black dashed line represents a linear fit to the data.  Galaxies in the shaded region  of the diagram ($\mathrm{EW([OII])}<-5\,\angstrom$) are excluded from our sample of ETGs.}
    \label{fig:oii_ew_vs_hb_ew}
\end{figure}

As anticipated, our definition of ETG is based mostly on morphology. For the morphological classification we opted for visual inspection because the number of galaxies of our sample is relatively small. Valid alternatives, which are necessarily preferable for larger data sets, are automated morphological classification algorithms (e.g.\ \citealt{Dom18}).
Before the visual inspection, we applied a pre-selection based on star formation activity: we removed star-forming galaxies from our sample, under the assumption that they are mostly associated with a late-type or irregular morphology.
We relied on the presence of emission lines in the spectra of our galaxies as an indicator of star formation activity.
In particular, we applied a selection based on the equivalent width of the forbidden emission line doublet of $\mathrm{[OII]}$,  $\mathrm{EW([OII])}\,\mathrm{\lambda\lambda}3726,3729$: we included only those galaxies that have $\mathrm{EW([OII])}\geq-5\,\angstrom$, where $\mathrm{EW([OII])}$ of SDSS and LEGA-C galaxies are obtained from the respective data release catalogues.
Although [OII] is not a perfect indicator of star formation activity, as it can suffer from contamination from emission by an active galactic nucleus, and other spectral lines could be used in its place (H$\beta$, for example), these lines are in general not accessible in the spectra of most LEGA-C galaxies, as they are redshifted outside the available spectral range.
For the sake of homogeneity in our selection criteria, and in order to keep the high end of the redshift distribution of the LEGA-C galaxies in our sample, we used [OII] as a first step towards obtaining a sample of ETGs.
Nevertheless, we found a good correlation between $\mathrm{EW([OII])}$ and $\mathrm{EW(H\beta)}$ for those galaxies drawn from the original catalogues of SDSS and LEGA-C for which both measurements are available (see \autoref{fig:oii_ew_vs_hb_ew}).
Although half of the LEGA-C galaxies do not have values of $\mathrm{EW([OII])}$ in the DR2 catalogue, these are for the most part objects at the low end of the redshift range, $z<0.8$.

The second step in our selection is to include only galaxies with an early-type morphology, according to visual inspection.
We used imaging data from the Wide layer of the HSC DR1, for this purpose.
The Wide layer of HSC covers approximately 108 square degrees. The number of SDSS main sample galaxies present in this dataset is $\approx2000$, which, while only a small fraction of the total number of SDSS galaxies, is still sufficiently large to carry out a statistical analysis of the stellar mass--velocity dispersion relation.
LEGA-C targets are located in a $\simeq1.3\,\mathrm{deg}^2$ region, for the most part overlapping with the Cosmic Evolution Survey \citep[COSMOS;][]{Sco++07} area.
HSC DR1 data from the Ultra Deep layer are available for most ($\approx1700$) of the objects in the LEGA-C DR2.

The motivation for using HSC data is in its high depth ($i$-band 26~mag detection limit for a point source in the Wide layer) and good image quality (typical $i$-band seeing is $0.6''$). This is particularly important for the LEGA-C galaxies, which are much fainter and have smaller angular sizes compared to the SDSS ones, due to their higher redshift.
For each galaxy with available HSC DR1 data, we obtained cutouts in the $g$, $r$, $i$, $z$ and $y$ filters, then visually inspected colour-composite RGB images made using the $g-, r-$ and $i-$band data.
We removed objects showing any presence of discs, spiral arms, as well as galaxies for which a single S\'{e}rsic model \citep{Sersic68} does not provide a qualitatively good description of the surface-brightness distribution (e.g., irregular galaxies). Such objects account for roughly 50\% of the inspected galaxies.
Additionally, a few percent of the objects were removed because of contamination from stars, and an even smaller fraction was eliminated because of the presence of close neighbours that make it difficult to carry out accurate photometric measurements.
Although this last step could in principle introduce a bias in the inferred $M_*$-$\sigmae$ relation in case this varies as a function of environment, given the small fraction of objects with close neighbours removed, any such bias will in any case be very small.

\begin{figure*}
    \centering
    \includegraphics[width=.99\textwidth]{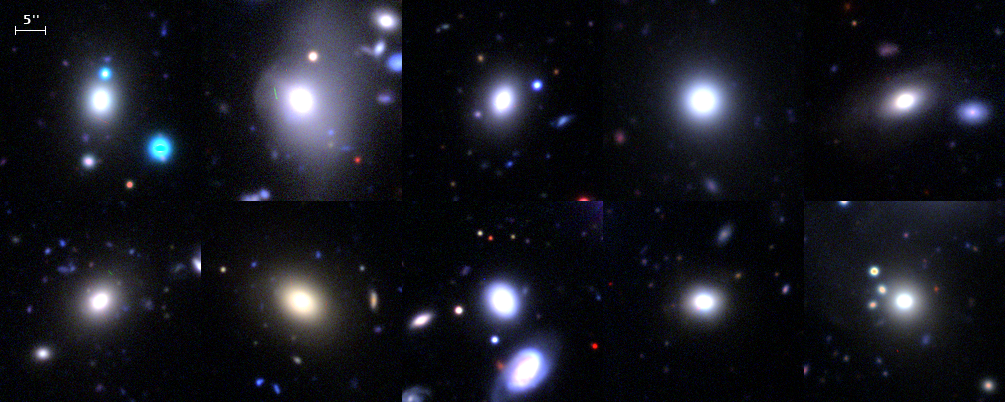}
    \caption{Colour-composite HSC images of a set of SDSS main sample ETGs that passed our selection in EW([OII]) and our visual inspection.}
    \label{fig:good}
\end{figure*}

\begin{figure*}
    \centering
    \includegraphics[width=.99\textwidth]{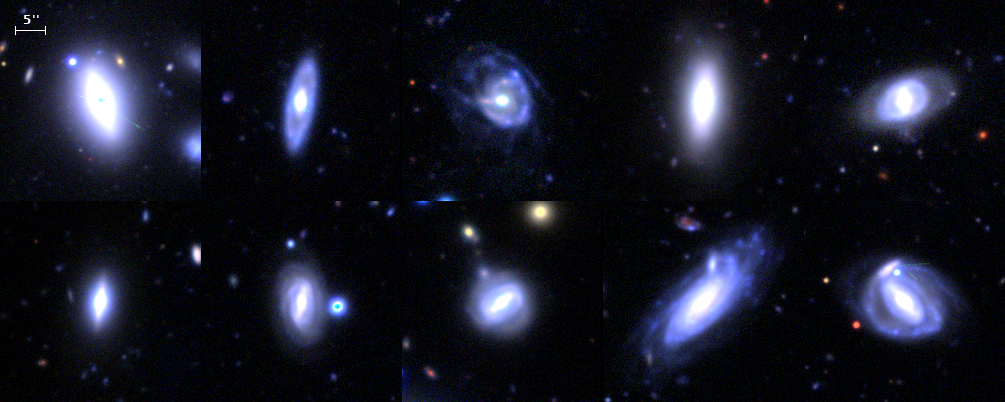}
    \caption{Colour-composite HSC images of a set of SDSS main sample galaxies that passed our selection in EW([OII]), but were rejected in our visual inspection step, due to the presence of disks and/or spiral arms.}
    \label{fig:bad}
\end{figure*}

In \autoref{fig:good} and \autoref{fig:bad} we show colour-composite images of example sets of SDSS galaxies included and excluded from our sample on the basis of our morphological classification.


\subsubsection{Photometric measurements}

Our procedure for measuring stellar masses of the galaxies in the fiducial sample consisted in fitting stellar population synthesis models to broadband photometric data.
Although photometric measurements for these galaxies are available from the literature, we chose to carry out new measurements using photometric data from the HSC survey.
The data from the HSC survey are much deeper and have a much higher image quality compared to the SDSS data.
This is important, because it allows for a cleaner detection and masking of foreground contaminants, and allows for a better characterisation of the faint extended envelope of massive galaxies \citep[see e.g.][]{Huang2018}.
Moreover, by using the same data and procedure to estimate the stellar masses of the galaxies in the SDSS and LEGA-C samples, our inference on the evolution of the $M_*$-$\sigmae$ relation is less prone to possible systematic effects related to the photometric measurements.


We estimated the $g$, $r$, $i$, $z$ and $y$ magnitudes of each galaxy by fitting a S\'{e}rsic surface brightness distribution to the data in these five bands simultaneously.
In particular, we obtained 201$\times$201 pixel ($\approx 34''\times34''$) sky-subtracted cutouts of each galaxy in each band, we fitted the five-band data simultaneously with a seeing-convolved S\'{e}rsic surface brightness profile with elliptical isophotes and spatially uniform colours, while iteratively masking out foreground or background objects using the software {\sc SExtractor} \citep{BertinArnouts1996A&AS}.

Saturated pixels were also masked, using the masks provided by HSC DR1.
We added in quadrature a $0.05$ magnitude systematic uncertainty to the observed flux in each band, to account for zero-point calibration errors in the HSC DR1 photometry, which have been shown to be on this order of magnitude or smaller \citep[see][]{Aihara2018PASJ}.

An important data reduction step on which our measurements rely is the sky subtraction.
We checked the robustness of the sky subtraction by repeating the analysis on a subset of galaxies, using the more recent data from the HSC data release 2\footnote{The HSC DR2 was released when the bulk of our analysis was complete.} \citep[DR2][]{Aihara2019}. The HSC DR2 used a substantially different sky subtraction method, compared to the DR1 \citep[see subsection 4.1 in][]{Aihara2019}.
The corresponding difference in flux leads to an average difference of $0.03$~dex on the stellar masses, with a $0.07$~dex scatter. While the scatter is well within the observational uncertainty on the stellar mass, this bias is a potential systematic effect that is difficult to correct for and should in principle be taken into account in our global error budget. However, it does not affect the conclusions of our study: our main goal is to measure the slope and evolution of the $M_*$-$\sigmae$ relation, which are robust to overall shifts in the stellar mass measurements of the sample.

\subsubsection{Stellar mass measurements}
\label{sssec: stellar_masses}
To infer stellar masses, we fitted the observed $g$, $r$, $i$, $z$ and $y$ fluxes with composite stellar population models. These were obtained using the BC03 stellar population synthesis (SPS) code \citep{BruzualCharlot2003MNRAS}, with semi-empirical stellar spectra from the BaSeL 3.1 library \citep{Wes++02}, Padova 1994 stellar evolution tracks \citep{Fagotto1994aA&AS,Fagotto1994bA&AS,Fagotto1994cA&AS} and a Chabrier IMF. We considered star formation histories with an exponentially declining star formation rate and we applied a prior on metallicity based on the mass--metallicity relation measured by \citet{Gal++05}. We sampled the posterior probability distribution of stellar mass, age (time since the initial burst of star formation), star formation rate decline timescale, metallicity and dust attenuation with a Markov Chain Monte Carlo (MCMC), following the method introduced by \citet{Auger2009ApJ}.
We then considered the posterior probability distribution in log-stellar mass, marginalised over the other parameters, and approximated it as a Gaussian with mean equal to 
\begin{equation}
\log{M_*^\obs} = \frac{\log{M_*^{(84)}} + \log{M_*^{(16)}}}{2}
\end{equation}
and standard deviation
\begin{equation}
\sigma_{M_*} = \frac{\log{M_*^{(84)}} - \log{M_*^{(16)}}}{2},
\end{equation}
where $\log{M_*^{(84)}}$ and $\log{M_*^{(16)}}$ are the 84 and 16 percentile of the distribution, respectively.
We refer to \citet{Sonnenfeld2019A&A} for more details.
In Appendix~\ref{app:comparison}, we compare our estimates of stellar mass with those of \citet[][M14 hereafter]{Mendel2014ApJS} for the SDSS galaxies of our sample.

\subsubsection{A complete sample}\label{ssec:complete}

In order to accurately infer the $M_*$-$\sigmae$ relation, it is necessary that the selection criteria used to define our sample do not introduce spurious correlations between these two variables.
A sufficient condition to achieve this is working with a sample that, at any given redshift, is highly complete in stellar mass, or is randomly drawn from a complete sample.
For the SDSS sample, we achieved this condition by first estimating, at each redshift $z$, the minimum stellar mass above which our sample is $99\%$ complete, $M_{\rm *,min}(z)$, and then removing from the sample all galaxies with stellar mass below this value.
To estimate $M_{\rm *,min}(z)$ of the SDSS sample we proceeded as follows.
The SDSS main sample, from which our galaxies are drawn, is complete down to an $r-$band Petrosian magnitude $r_\mathrm{P}$ of $17.77$ \citep{Str++02}.
At any redshift, this value of $r_\mathrm{P}$ corresponds to a range of values of the stellar mass, with a spread that is due to scatter in the stellar mass-to-light ratio and to a mismatch between the definition of Petrosian and S\'{e}rsic magnitudes.
We can nevertheless define the ratio between the observed stellar mass and the observed-frame SDSS $r-$band Petrosian luminosity $L_r$ and consider its distribution ${\rm P}(M_*/L_r)$.
We then made narrow redshift bins and, approximating ${\rm P}(M_*/L_r)$ as a Gaussian, used the mean and standard deviation of the sample of $M_*/L_r$ values in each bin to find the 99-th percentile of this distribution, $M_*/L_r|_{99}$. Finally, we obtained $M_{\rm *,min}(z)$ by multiplying $M_*/L_r|_{99}$ by the Petrosian luminosity corresponding to the limiting value $r_\mathrm{P}=17.77$.

\begin{figure}
\centering
    \includegraphics[width=.99\columnwidth]{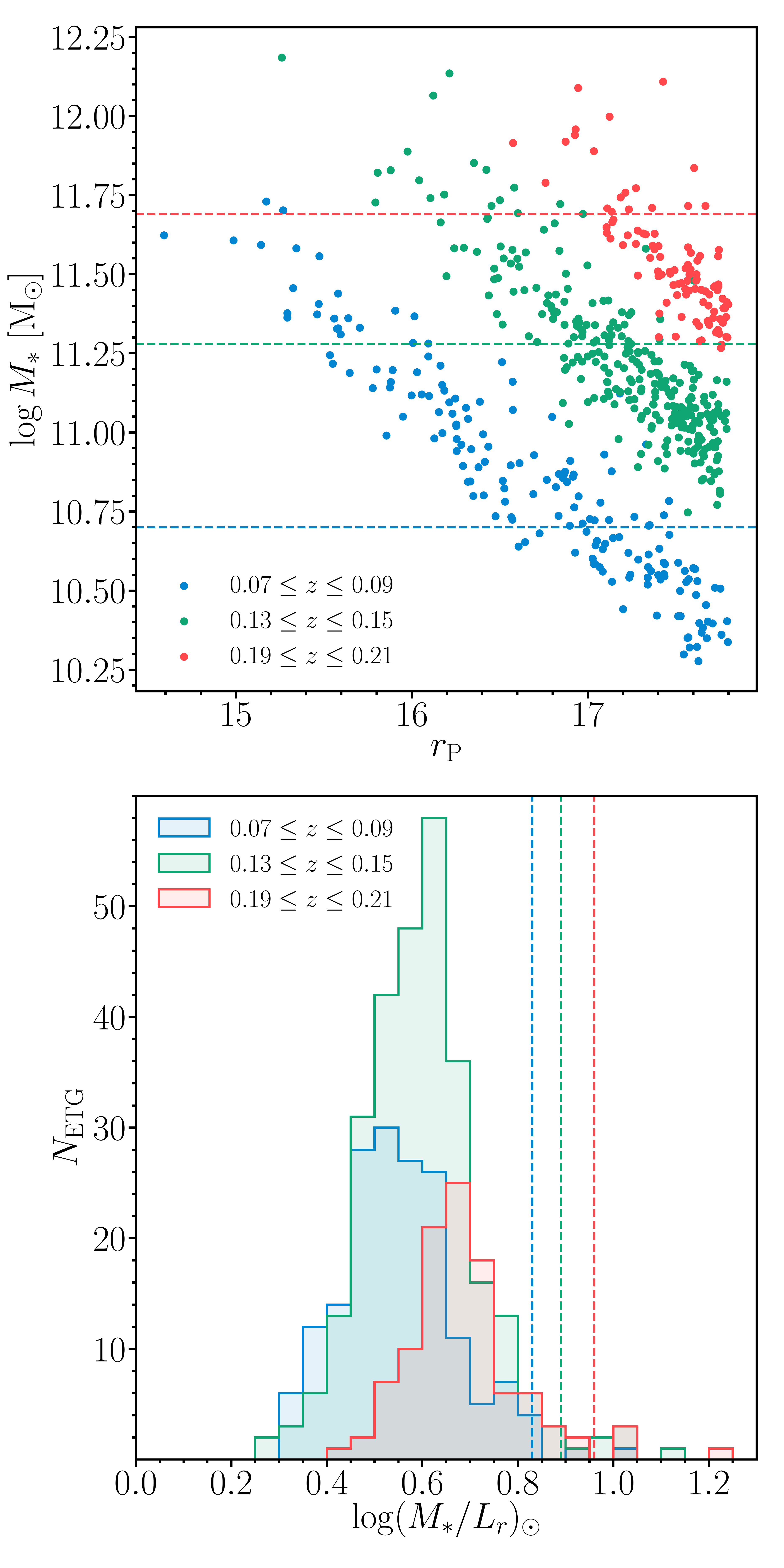}
    \caption{Stellar mass as a function of SDSS $r-$band Petrosian magnitude, for SDSS main sample ETGs in three narrow redshift bins (upper pannel). Horizontal dashed lines mark, in each redshift bin, the stellar mass above which an ETG drawn from the SDSS main sample has more than $99\%$ probability of entering our sample.
    Distribution in the ratio between stellar mass and observed-frame $r-$band Petrosian luminosity of the galaxies in the three redshift bins shown in the upper panel (lower panel). The 99-th percentile of each distribution is marked by a vertical dashed line. This value, multiplied by the Petrosian luminosity corresponding to the limiting $r-$band magnitude of the SDSS main sample, $r_\mathrm{P}=17.77$, gives the $99\%$ completeness limit shown in the upper panel.
    }
    \label{fig:mstarmin}
\end{figure}

In \autoref{fig:mstarmin}, we illustrate an application of this procedure on three redshift bins: in the upper panel, we show values of stellar mass as a function of $r_\mathrm{P}$, while in the lower panel we show the corresponding distributions in $M_*/L_r$. The 99-th percentile of the ${\rm P}(M_*/L_r)$ distribution and the corresponding value of $M_{\rm *,min}(z)$ are shown as dashed lines in the two panels.

\begin{figure}
        \centering
        \includegraphics[width=.99\columnwidth]{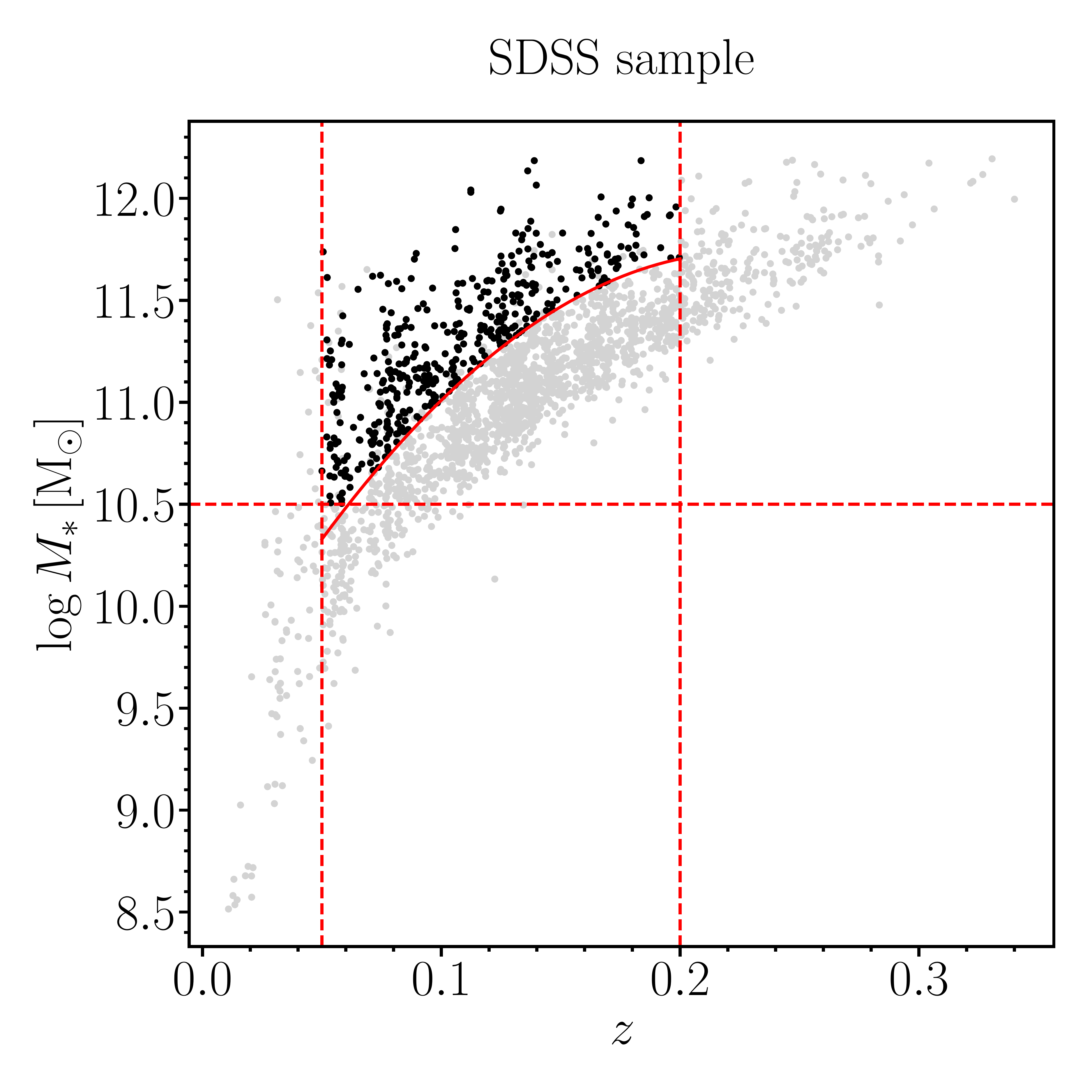}
        \centering
        \includegraphics[width=.99\columnwidth]{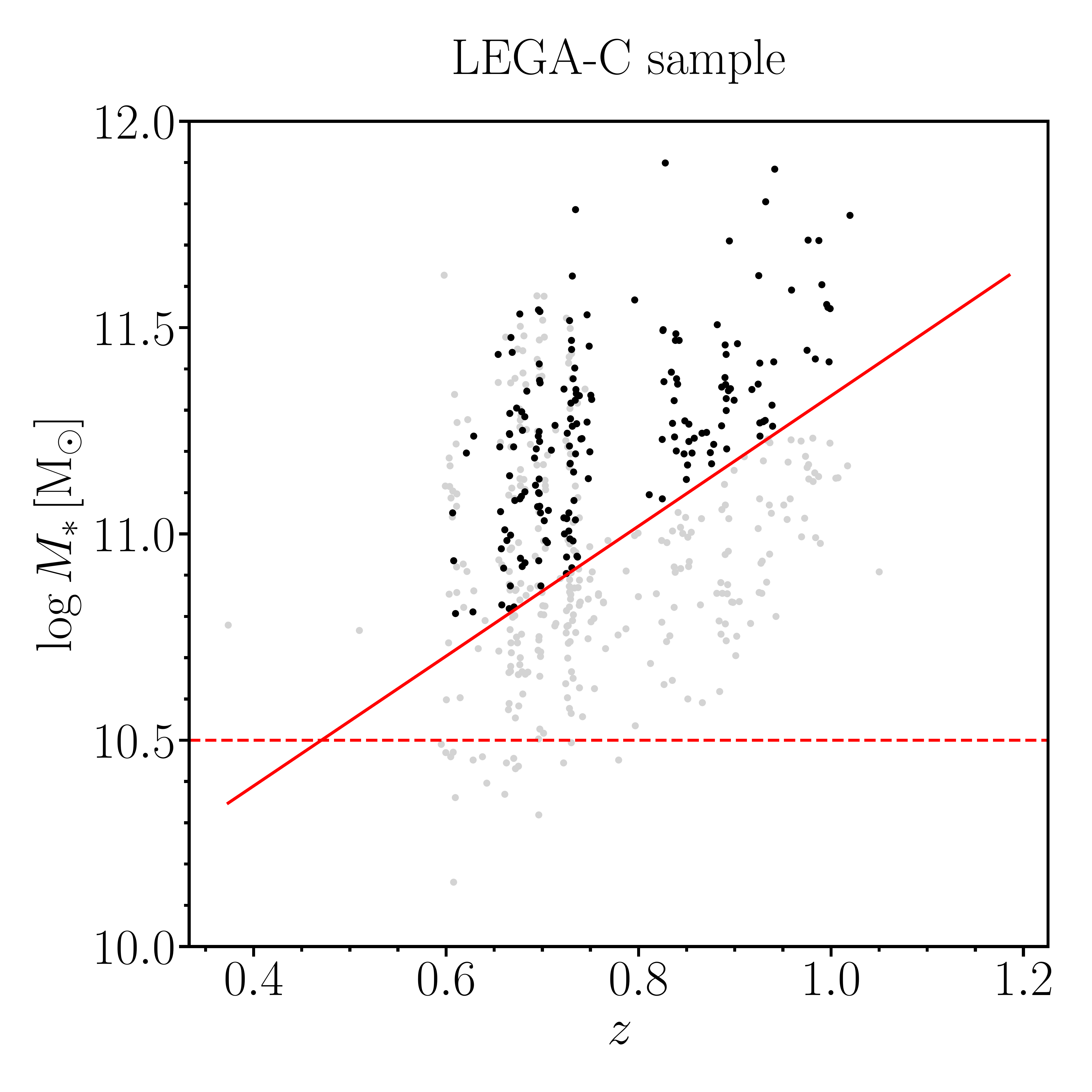}
  
		\caption{Stellar mass as a function of redshift for SDSS (upper panel) and LEGA-C (lower panel) galaxies. The solid curves represent the empirical 99\% mass-completeness limits. The horizontal dashed lines represent the absolute lower stellar mass limit $10^{10.5}\,\msun$, while the vertical dashed lines are the lowest ($z=0.05$) and highest ($z=0.20$) redshift limits imposed in the SDSS galaxy selection. In the upper panel, grey dots represent all the 2127 SDSS main sample galaxies morphologically selected and with $\mathrm{EW([OII])}>=-5\,\angstrom$, while black dots are the 413 objects above the mass-completeness limit, with $\log(M_*/\msun)>10.5$ in the redshift range $0.05<z<0.2$. In the lower panel, grey dots are the 492 LEGA-C galaxies selected in morphology, EW([OII]) and $K_s$-band magnitude, while the black dots represent the final LEGA-C sample of 178 ETGs above the mass-completeness limit.}
        \label{fig:completeness}
\end{figure}

We estimated $M_{\rm *,min}(z)$ in a series of bins in the redshift range $0.05 < z < 0.20$. Outside this interval, the number of galaxies per redshift bin becomes small, and it is more difficult to obtain an accurate estimate of $M_{\rm *,min}$.
We therefore only included SDSS galaxies in this redshift range, with a stellar mass larger than the value of $M_{\rm *,min}$ at the corresponding redshift. We approximated the function $M_{\rm *,min}(z)$ as a quadratic polynomial for this purpose.
In the upper panel of \autoref{fig:completeness}, we show the initial distribution in stellar mass as a function of redshift of our SDSS main sample ETGs composed by 2127 sources (grey dots), as well as the final sample (black dots), which consists of 413 objects, obtained after applying the cut in stellar mass. The solid curve shows $M_{\rm *,min}(z)$: our SDSS sample is more than 99\% complete above this stellar mass. 

LEGA-C primary targets have been selected on the basis of their photometric redshift and $K_s$-band magnitude, as obtained from the UltraVISTA survey photometric data \citep{Muzzin2013ApJ}.
Specifically, according to \citet{Straatman2018ApJS}, primary targets have been selected in the photometric redshift range $0.6<z_\mathrm{photo}<1$ and applying a redshift-dependent $K_s$-magnitude selection $K_s<K_{s,{\rm max}}(z_\mathrm{photo})$, with $K_{s,{\rm max}}(z_\mathrm{photo})=20.7-7.5\log\left[(1+z_\mathrm{photo})/1.8\right]$.
For the sake of robustness, in order to avoid contamination from objects with incorrect photo-$z$, we apply a more conservative selection adopting a constant $K_s$ limit, $K_s<20.36=K_{s,{\rm max}}(1)$.
We then obtained $M_{\rm *,min}(z)$ for the LEGA-C sample using the method described above for the SDSS sample, simply replacing $r_P$ with the UltraVISTA $K_s$-band magnitude. The resulting distribution in redshift and stellar mass is shown in the lower panel of \autoref{fig:completeness}.
From a sample of 492 galaxies selected in morphology, $K_s$-band magnitude and EW([OII]) (grey dots),
after selecting only galaxies with stellar mass above $M_{\rm *,min}(z)$, our LEGA-C sample of ETGs reduces to 178 objects (black dots).

The LEGA-C DR2 sample, however, does not include all galaxies brighter than the stated magnitude limit, as the survey was not finished at the time of that data release. Instead, the targets included in DR2 were selected according to a $K_s$-dependent probability, ${\rm P}(K_s)$. The resulting sample is therefore incomplete, but the incompleteness rate ${\rm P}(K_s)$ is a known quantity, provided in the LEGA-C DR2. In order to obtain an unbiased inference of the $M_*$-$\sigmae$ relation, it is then sufficient to re-weight each measurement by the inverse of ${\rm P}(K_s)$.
In \autoref{tab:sample_selection_steps}, we summarise the selection steps used to obtain the final SDSS and LEGA-C samples.

\begin{table}
\caption{Summary table of the selection steps adopted to build the final SDSS and LEGA-C samples. }
\begin{tabular}{cc}
\toprule
\toprule
\addlinespace
Selection step & $N_\mathrm{ETG}$ \\
\addlinespace
\midrule
\addlinespace
\multicolumn{2}{c}{SDSS sample} \\
\addlinespace
\midrule
\addlinespace
SDSS main sample galaxies selected on  &  \multirow{2}{*}{\minitab[c]{$2127$}} \\
morphology and $\mathrm{EW([OII])}$ &  \\
\addlinespace
SDSS galaxies at $0.05\leq z\leq0.2$  & \multirow{2}{*}{\minitab[c]{$413$}} \\ with $M_*>10^{10.5}\,\msun$ and $M_*>M_{\rm *,min}(z)$  & \\
\addlinespace
\midrule
\addlinespace
\multicolumn{2}{c}{LEGA-C sample} \\
\addlinespace
\midrule
\addlinespace
LEGA-C galaxies selected on & \multirow{2}{*}{\minitab[c]{$492$}} \\ 
morphology, $\mathrm{EW([OII])}$ and $K_s$-band magnitude  & \\
\addlinespace
LEGA-C galaxies & \multirow{2}{*}{\minitab[c]{$178$}} \\
with $M_*>10^{10.5}\,\msun$ and  $M_*>M_{\rm *,min}(z)$ \\
\addlinespace
\bottomrule
\bottomrule
\end{tabular}
\label{tab:sample_selection_steps}
\end{table}

\subsubsection{Velocity dispersion measurements}

For each SDSS galaxy, we obtain, from the DR12 catalogue, the value and relative uncertainty of the line-of-sight stellar velocity dispersion measured in the $1.5''$ radius fiber of the SDSS spectrograph, which we label $\sigma_\mathrm{ap}$.
We convert this measurement into an estimate of the central velocity dispersion integrated within an aperture equal to the half-light radius, $\sigma_{\mathrm{e}}$, by applying the following correction:
\begin{equation}\label{eq:aperture_correction}
\sigma_\mathrm{e} = \sigma_\mathrm{ap} \times \left(\frac{R_\mathrm{e}}{1.5''}\right)^{-\delta},
\end{equation}
where $R_\mathrm{e}$ is the half-light radius and $\delta=0.066$ \citep{Cappellari2006MNRAS}.

\begin{figure}
        \centering
        \includegraphics[width=.99\columnwidth]{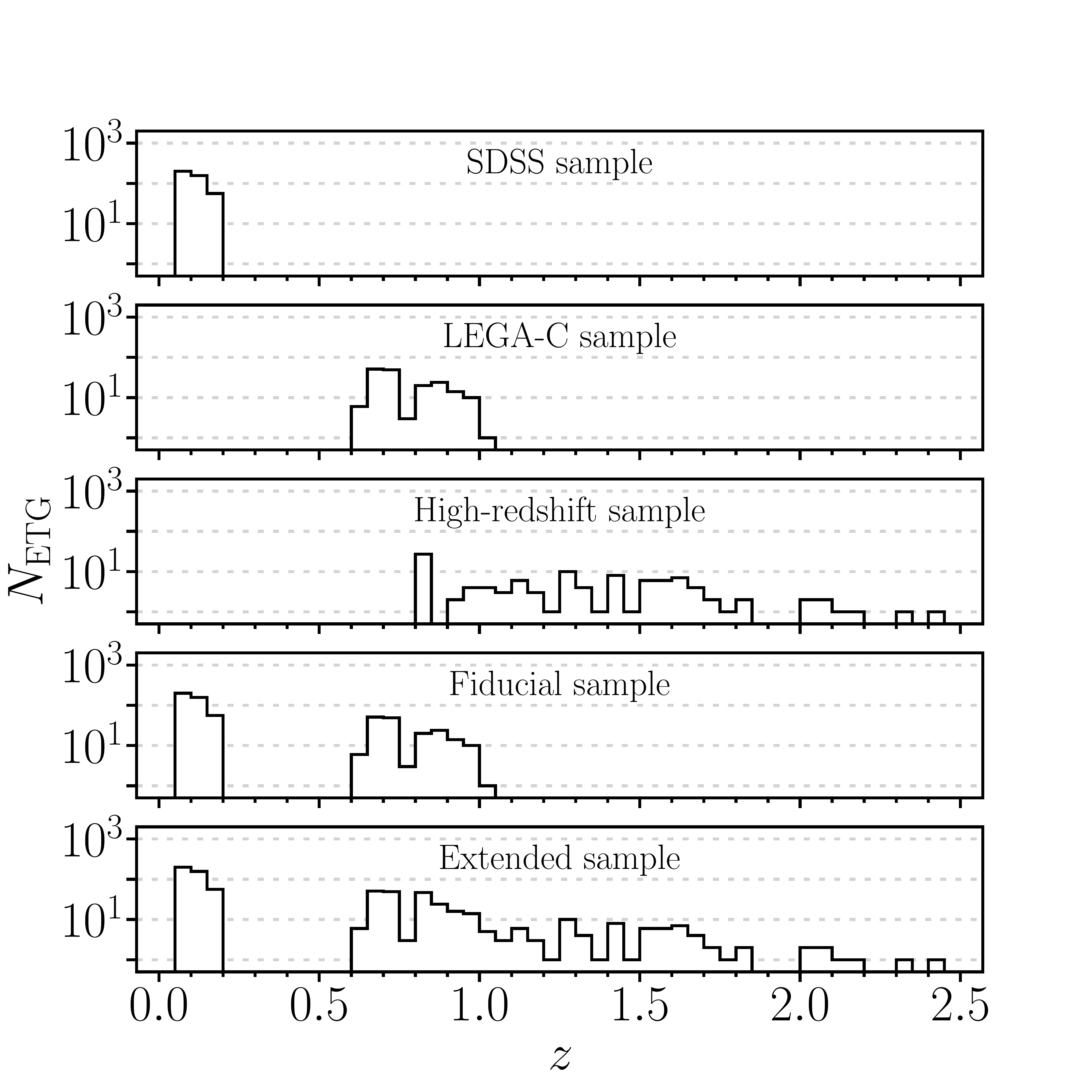}
        \centering
        \includegraphics[width=.99\columnwidth]{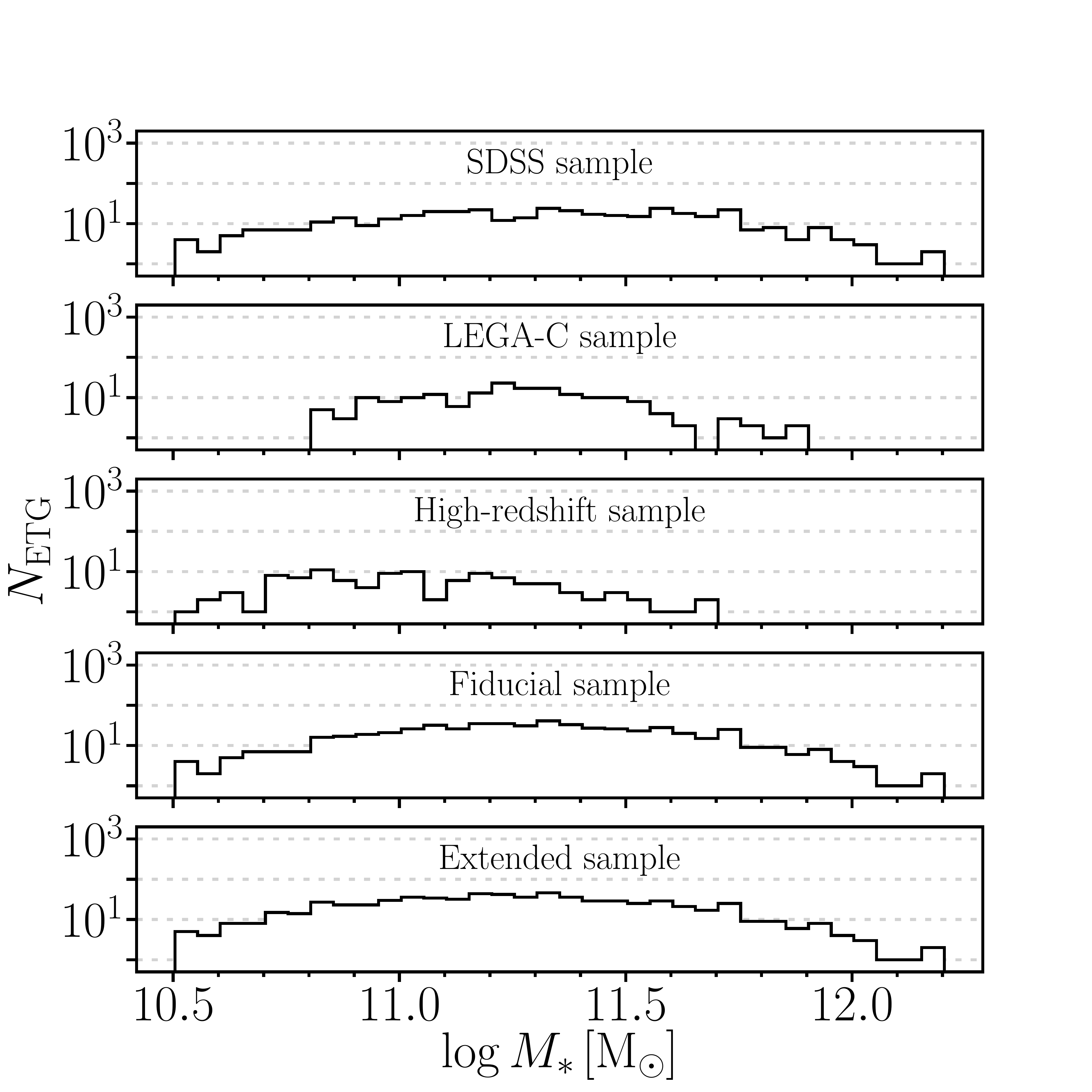}
		\caption{Distributions of the subsamples and samples of ETGs in redshift (upper panel) and stellar mass (lower panel). From the top to the bottom, the SDSS subsample, the LEGA-C subsample, the high-redshift sample (vdS13+B14+G15+B17 subsamples), the fiducial sample (SDSS+LEGA-C subsamples) and the extended sample (fiducial sample+high-resdhift sample) distributions are shown.}
        \label{fig:differential_distributions}
\end{figure}

Velocity dispersion measurements provided in the LEGA-C DR2 are converted  to values of the central velocity dispersion $\sigma_{\mathrm{e}}$ applying the aperture correction 
\begin{equation}
    \sigmae=1.05\,\sigma_\mathrm{ap},
\end{equation}
which is a good approximation for galaxies in the redshift range of the LEGA-C sample \citep{vandeSande2013ApJ,Belli2014ApJ}.
The distributions in redshift and in stellar mass of the SDSS and LEGA-C subsamples and of the fiducial sample are shown in \autoref{fig:differential_distributions} (see also \autoref{tab:samples}).

\begin{table}
\caption{Properties of the subsamples of ETGs used to build our fiducial (SDSS and LEGA-C) and high-redshift (vdS13, B14, G15 and B17) samples. Column 1: subsample name. Column 2: redshift range. Column 3: stellar mass range in logarithm. Column 4: number of galaxies. }
\begin{tabular}{cccc}
\toprule
\toprule
\addlinespace
Sample & $z$ & $\log (M_*/\msun)$ & $N_\mathrm{ETG}$ \\
\addlinespace
\midrule
\addlinespace
SDSS   & $(0.05;0.20)$ & $(10.50;12.19)$  & $413$ \\
\addlinespace
LEGA-C & $(0.60;1.02)$ & $(10.80;11.90)$  & $178$  \\
\addlinespace
vdS13  & $(0.81;2.19)$ & $(10.53;11.69)$  & $56$   \\
\addlinespace
B14    & $(1.02;1.60)$ & $(10.59;11.35)$  & $26$   \\
\addlinespace
G15    & $(1.26;1.41)$ & $(11.04;11.49)$  & $4$    \\
\addlinespace
B17    & $(1.52;2.44)$ & $(10.60;11.68)$  & $24$   \\
\addlinespace
\bottomrule
\bottomrule
\end{tabular}
\label{tab:samples}
\end{table}

\subsection{The high-redshift and extended samples}
\label{ssec:high}

Our high-redshift sample of ETGs is a sample of 110 galaxies with $\log (M_*/\msun)>10.5$ in the redshift range $0.8\lesssim z \lesssim 2.5$, built as follows. We obtain measurements of the stellar mass and stellar velocity dispersion of ETGs out to $z\approx 2.5$ from a variety of studies.
In order of increasing median redshift, we take 26 galaxies drawn from the LRIS sample presented in \citet[][hereafter B14]{Belli2014ApJ}, including only those galaxies for which $\mathrm{EW([OII])}\geq-5\,\angstrom$ (as done for the fiducial sample; \autoref{ssec:fiducial}), 56 galaxies from \citet[][hereafter vdS13]{vandeSande2013ApJ}, 4 galaxies from \citet[][hereafter G15]{Gargiulo2015A&A}, and 24 galaxies from \citet[][hereafter B17]{Belli2017ApJ}. The main properties of each of these subsamples are summarised in \autoref{tab:samples}.
Among the original sample of 73 galaxies of vdS13, only 5 galaxies are presented for the first time, while the remaining 68 sources are collected from different studies. We removed 17 of these 73 ETGs because they are already included as part of either B14's or B17's samples.
All the galaxies in the high-redshift samples are classified as ETGs, based on their $UVJ$ colours, morphology and/or spectra.
Of course, given the more heterogeneous selection, our extended sample is not as self-consistent as our fiducial sample, and, due to  the known correlations between $\sigmae$ and some structural or spectral properties of ETGs \citep{Zah17}, we cannot exclude that selection biases have non-negligible effects when the high-redshift sample is considered.
However, for the vast majority of these galaxies, stellar masses are measured by fitting SPS models to broadband imaging data and by scaling the total flux to match that measured by fitting a S\'{e}rsic surface brightness profile to high-resolution images from {\em Hubble Space Telescope} (HST).
The details of the SPS models are very similar to those we adopted in our measurement of the stellar masses of the fiducial sample. In all these subsamples stellar masses are computed assuming Chabrier IMF and central velocity dispersions are given within an aperture of radius $R_{\rm e}$.
Our extended sample, obtained by combining the fiducial and high-redshift samples, consists of 701 ETGs with $M_*\gtrsim10^{10.5}\,\msun$ in the redshift interval $0\lesssim z \lesssim 2.5$.
The distributions in redshift and in stellar mass of the high-redshift and extended samples are shown in \autoref{fig:differential_distributions}.

\section{Method}
\label{sec:method}

We use a Bayesian hierarchical method to infer the distribution of stellar velocity dispersion as a function of stellar mass and redshift for the ETGs in our samples.
This method allows us to properly propagate observational uncertainties, to disentangle intrinsic scatter from observational errors and to correct for Eddington bias \citep{Eddington13}, which is introduced when imposing a lower cutoff to the stellar mass distribution. Throughout this section stellar masses are expressed in units of $\msun$.

\subsection{Bayesian hierarchical formalism}

We describe each galaxy in our sample by its redshift, stellar mass and central stellar velocity dispersion. We refer to these parameters collectively as $\individ=\{\log M_*, \log \sigma_{\mathrm{e}},z\}$.
These represent the true values of the three quantities, which are in general different from the corresponding observed values.
We assume that the values of $\individ$ are drawn from a probability distribution, described in turn by a set of {\em hyper-parameters} $\boldsymbol\Phi$:
\begin{equation}
\pr(\individ) = \pr(\individ | \hyperp).
\end{equation}
Our goal is to infer plausible values of the hyper-parameters, which summarise the distribution of our galaxies in the $(\log M_*, \log \sigma_{\mathrm{e}}, z)$ space, given our data.
We will describe in detail the functional form of the distribution $\pr(\individ | \hyperp)$ in \autoref{subsec:pdf_and_hyperparameters}.

Using Bayes' theorem, the posterior probability distribution of the hyper-parameters given the data $\data$ is
\begin{equation}\label{eq:bayes}
\pr(\hyperp|\data) \propto \pr(\hyperp)\pr(\data | \hyperp),
\end{equation}
where $\pr(\hyperp)$ is the prior probability distribution of the model hyper-parameters and $\pr(\data | \hyperp)$ is the likelihood of observing the data given the model.

The data consist of observed stellar masses, stellar velocity dispersions and redshifts,
\begin{equation}
\data \equiv \{\log M_*^\obs, \log \sigma_{\mathrm{e}}^\obs, z^\obs\},
\end{equation}
and related uncertainties.
Since measurements on different galaxies are independent of each other, the likelihood term can be written as
\begin{equation}\label{eq:product}
\pr(\data | \hyperp) = \prod_i \pr(\datai| \hyperp),
\end{equation}
where $\datai$ is the data relative to the $i$-th galaxy.
For each galaxy in our sample, the likelihood of the data depends only on the true values of the redshift, stellar mass and velocity dispersion, $\boldsymbol\Theta$, and not on the hyper-parameters $\boldsymbol\Phi$.
In order to compute the $\pr(\datai|\hyperp)$ terms in  \autoref{eq:product}, then, we need to marginalise over all possible values of the individual object parameters $\individi$:
\begin{equation}\label{eq:integral}
\pr(\datai|\hyperp) = \int \diff\individi \pr(\datai,\individi|\hyperp) = \int \diff\individi \pr(\datai|\individi)\pr(\individi|\hyperp).
\end{equation}
This allows us to evaluate the posterior probability distribution, \autoref{eq:bayes}, provided that a model distribution $\pr(\individ|\hyperp)$ is specified, priors are defined and the shape of the likelihood is known.
The method is hierarchical in the sense that there exists a hierarchy of parameters: individual object parameters $\individi$ are drawn from a distribution that is, in turn, described by a set of hyper-parameters.

As explained in subsection \ref{ssec:fiducial}, the LEGA-C sample is not representative of a complete sample, but each galaxy was included with a $K_s$ magnitude-dependent probability $\pr(K_s)$, so that brighter galaxies are over-represented \citep[see figure 2 of][]{Straatman2018ApJS}.
To correct for this, we re-weight the contribution of each LEGA-C measurement to the likelihood by a factor proportional to $1/P(K_s)$: we transform \autoref{eq:product} to
\begin{equation}
\pr(\data | \hyperp) = \prod_i \pr(\datai| \hyperp)^{w_i},
\end{equation}
where $w_i$ is given by
\begin{equation}
w_i = \dfrac{1/P(K_{s,i})}{\left<1/P(K_s)\right>}
\end{equation}
for LEGA-C galaxies and $w_i=1$ otherwise.
The normalisation of the weights given in the equation above ensures that the effective number of LEGA-C data points equals the number of LEGA-C galaxies.
\subsection{The model}\label{subsec:pdf_and_hyperparameters}

The purpose of our model is to summarise the distribution in stellar mass and velocity dispersion of our samples of ETGs with a handful of parameters, $\hyperp$, that can provide an intuitive description of the $M_*$-$\sigma_{\mathrm{e}}$ relation.
In the absence of a well-established theoretically motivated model, we opt for an empirical one, that we describe in this subsection.

The dependent variable of our model is the central velocity dispersion, $\sigma_{\mathrm{e}}$, while stellar mass and redshift are independent variables.
As such, it is useful to write the probability distribution of individual galaxy parameters as
\begin{equation}\label{eq:model}
\pr(\individ|\hyperp) = \pr(\log M_*,z|\hyperp)\pr(\log \sigma_{\mathrm{e}}|\log M_*,z,\hyperp).
\end{equation}
Here, $\pr(\log M_*,z|\hyperp)$ describes the prior probability distribution for a galaxy in our sample to have logarithm of the true stellar mass $\log M_*$ and true redshift $z$. This probability depends on some hyper-parameters, which may vary between different subsamples.
Our galaxies have been selected by applying a lower cut to the observed stellar masses, ${M_*^\obs} > M_{\rm *,min}$. We then expect the probability distribution in the true stellar mass to go to zero for low values of $M_*$. We also expect $\pr(\log M_*,z|\hyperp)$ to vanish for very large values of $M_*$, as there are few known galaxies with $M_* > 10^{12}$.
For simplicity, we assume that $\pr(\log M_*,z|\hyperp)$ separates as follows:
\begin{equation}\label{eq:separate}
\pr(\log M_*,z|\hyperp) = \pr(\log M_*|\hyperp)\pr(z|\hyperp),
\end{equation}
where $\pr(\log M_*|\hyperp)$ is a skew Gaussian distribution in $\log{M_*}$,
\begin{equation}\label{eq:prior}
    \pr(\log M_{*}|\hyperp)\propto\frac{1}{\sqrt{2\pi\sigma^2_{\sigma_*}}}\,\, \mathrm{exp}\left\{-\frac{(\logm-\mu_*)^2}{2\sigma_*^2}\right\}\,\mathcal{E}(\logm|\Phi),
\end{equation}
with
\begin{equation}
    \mathcal{E}(\logm|\Phi)=1+\mathrm{erf}\left(\alpha_*\frac{\logm-\mu_*}{\sqrt{2\sigma_*}}\right),
\end{equation}
where the three hyper-parameters $\mu_*$, $\sigma_*$, $\alpha_*$ are modelled as
\begin{align}
    \label{eq:mu_star}
    &\mu_*=\mu_{*,0}+\mu_{*,s}\log\left(\frac{1+z}{1+z^\mathrm{piv}}\right),
    \\
    &\sigma_*=\sigma_{*,0}+\sigma_{*,s}\log\left(\frac{1+z}{1+z^\mathrm{piv}}\right)
    \label{eq:sigma_star}
    \\
    \mbox{and}\\
    &\alpha_*=const.
    \label{eq:sigma_star}
\end{align}

Since this is a prior on the stellar mass distribution, and since the typical uncertainty on the stellar mass measurements is much smaller than the width of this distribution (as shown in \autoref{sec:results}), the particular choice of the functional form of $\pr(\log M_*|\hyperp)$ does not matter in practice, because the likelihood term dominates over the prior.
The main role of the prior is downweighting extreme outliers and measurements with very large uncertainties.
The term $\pr(z|\hyperp)$ in \autoref{eq:separate} describes the redshift distribution of the galaxies in our sample. As we show below, this term does not enter the problem, because uncertainties on the observed redshifts can be neglected.

The second term on the right hand side of \autoref{eq:model} is the core of our model.
With it, we wish to capture the following features of the $M_*$-$\sigma_{\rm e}$ relation: its normalisation (i.e. the amplitude of the stellar velocity dispersion at a given value of the stellar mass) and its redshift evolution, the correlation between velocity dispersion and stellar mass, and the amplitude of the intrinsic scatter in $\sigma_{\rm e}$ at fixed $M_*$ and redshift.
With these requirements in mind, we assume that the logarithm of the stellar velocity dispersion is normally distributed, with a mean that can scale with redshift and stellar mass and with a variance that can evolve with redshift:
\begin{equation}
\pr(\log{\sigma_{\mathrm{e}}}|\log M_*,z,\hyperp) =\frac{1}{\sqrt{2\pi{\sigma_\sigma^2}}} \mathrm{exp}\left\{-\frac{(\logsigmae-\mu_{\sigma})^2}{2{\sigma_\sigma^2}}\right\}.
\label{eq:Mobs_z_sigma_Phi}
\end{equation}
We adopt the following functional form for the  mean of this distribution:
\begin{equation}\label{eq:mstar_z_sigma_relation}
\mu_\sigma=\mu_0^\sdss+\beta\log\left(\frac{M_*}{M_*^\mathrm{piv}}\right)+\zeta\log\left(\frac{1+z}{1+z^\mathrm{piv}}\right).
\end{equation}
In general, the slope $\beta$ is allowed to depend on $z$ as
\begin{equation}\label{eq:beta}
\beta=\beta_0^\sdss+\eta\log\left(\frac{1+z}{1+z^\mathrm{piv}}\right).
\end{equation}
We perform our analysis considering two different cases: the first is a \textit{constant-slope} case (model $\mathcal{M}_\mathrm{const}$), i.e.\ \autoref{eq:beta} with $\eta=0$; in the second, which we refer to as the \textit{evolving-slope} case (model $\mathcal{M}_\mathrm{evo}$), $\eta$ is a free hyper-parameter.
For the standard deviation $\sigma_{\sigma}$ in \autoref{eq:Mobs_z_sigma_Phi}, namely the intrinsic scatter of our relation, we adopt the form
\begin{equation}\label{eq:scatter}
    \sigma_\sigma=\psi_0^\sdss+\xi\log\left(\frac{1+z}{1+z^\mathrm{piv}}\right).
\end{equation}
In equations (\ref{eq:mstar_z_sigma_relation}-\ref{eq:scatter}) $M_*^\mathrm{piv}=10^{11.321}$ and $z^\mathrm{piv}=0.10436$, i.e.\ the median values of stellar mass and redshift of the SDSS ETGs, respectively, while the quantities $\mu_0^\sdss$, $\beta_0^\sdss$ and $\psi_0^\sdss$ are the median values of the hyper-parameters $\mu_0$, $\beta_0$ and $\psi_0$ obtained when fitting \autoref{eq:Mobs_z_sigma_Phi} to the ETGs of the SDSS subsample with 
\begin{equation}\label{eq:m_sdss}
\mu_\sigma=\mu_0+\beta_0\log\left(\frac{M_*}{M_*^\mathrm{piv}}\right) \qquad \textrm{and} \qquad \sigma_\sigma=\psi_0,
\end{equation}
i.e.\ neglecting any dependence on $z$. 
In order to prevent the redshift dependence of the relation from being influenced by any redshift dependence within the SDSS sample, which constitutes $\approx 60\%$ of the extended sample, we assume the model in \autoref{eq:m_sdss} as the \textit{zero point} at $z^\mathrm{piv}$ for our redshift-dependent models, because our main interest is to trace the evolution of the relation at higher redshift ($z\gtrsim 0.5$).
Hereafter, we will refer to the model  in \autoref{eq:m_sdss} applied to the SDSS subsample as model \Msdss.

Allowing for intrinsic scatter is an important feature of our model. Neglecting it leads typically to underestimating the slope of the $M_*$-$\sigmae$ relation \citep[see e.g.][]{Auger2010ApJ}. 
Our choice for the functional form of the distribution in velocity dispersion introduced above is somewhat arbitrary.
Although there could exist alternative distributions that fit the data equally well as our model or better, however, exploring such distributions is beyond the scope of this work.

\subsection{Sampling the posterior probability distribution functions of the model hyper-parameters}
\label{sec:sampling}

Our goal is to sample the posterior probability distribution function (PDF) of the model hyper-parameters $\hyperp$ given the data $\data$, $\pr(\hyperp|\data)$.
For this purpose, we use an MCMC approach, using a Python adaptation of the affine-invariant ensemble sampler of \citet{GW2010}, \textsf{emcee} \citep{Foreman-Mackey2013PASP}.
For each set of values of the hyper-parameters, we need to evaluate the likelihood of the data. This is given by the product over the galaxies in our sample of the integrals in \autoref{eq:integral}.
Using $\log{M_{*}}$, $\log{\sigma_{\mathrm{e}}}$ and $z$ as the integration variables and omitting the subscript $i$ in order to simplify the notation, \autoref{eq:integral} reads
\begin{equation}\label{eq:Delta_Phi}
\begin{aligned}
        \pdf(\log M_*^\obs,&\log \sigma_\mathrm{e}^\obs,z^\mathrm{obs}|\hyperp)=\\
        =\iiint&\diff\log M_{*}\, \diff\log\sigma_\mathrm{e}\,\diff z\,\times\\
        \times\,&\pdf(\log M_{*}^\obs,\log \sigma_{\mathrm e}^\obs,z^\obs|\log M_{*},\log \sigma_{\mathrm e},z)\,\times\\
        \times\,&\pdf(\log M_{*},\log \sigma_{\mathrm e},z|\hyperp)=\\
        =\,\,\,\iiint&\diff\log M_{*}\, \diff\log\sigma_\mathrm{e}\,\diff z\,\times\\
        \times\,&\pdf(\log M_*^\obs|\log M_*)\,\pdf(\log \sigma_\mathrm{e}^\obs|\log \sigma_\mathrm{e})\,\delta(z^\obs-z)\,\times\\
        \times\,&\pdf(\log M_{*}|\hyperp)\pdf(z|\hyperp)\pdf(\log \sigma_{\mathrm e}|\log M_*,z,\hyperp).
\end{aligned}
\end{equation}
In the last line, we have used equations~(\ref{eq:model}) and (\ref{eq:separate}), and we have approximated the likelihood of observing redshift $z^\obs$ as a delta function, in virtue of the very small uncertainties on the redshift (typical errors are $<10^{-4}$).
As a result, the redshift distribution term $\pr(z|\hyperp)$ becomes irrelevant, as it contributes to the integral only through a multiplicative constant that we can ignore.

Assuming a Gaussian likelihood in $\log{\sigma_{\mathrm{e}}^\obs}$ for the term $\pr(\log \sigma_{\mathrm{e}}^\obs|\log \sigma_{\mathrm{e}})$, the integral over $\diff\log{\sigma_{\mathrm{e}}}$ can be performed analytically, as we show in Appendix~\ref{app:likelihood}.
We also assume a Gaussian likelihood for the measurements of $\log{M_*^\obs}$,
\begin{equation}
    \pdf(\log M_{*}^\obs|\log M_*) = \frac{\mathcal{A}(\logm)}{\sqrt{2\pi\sigma_{M_*}^2}} \mathrm{exp}\left\{-\frac{(\log M_*-\log M_*^\obs)^2}{2\sigma_{M_*}^2}\right\},
\end{equation}
with one caveat: we are only selecting galaxies with $\log{M_*^\obs} > \log M_{\rm *,min}$, where $\log M_{\rm *,min}$ is derived from the mass-completeness limits at a given redshift for SDSS and LEGA-C galaxies (see \autoref{ssec:complete}) and it is assumed to be constant and equal to $10.5$ for all the ETGs of the high-redshift sample. The likelihood must be normalised accordingly:
\begin{equation}\label{eq:compute_normalization}
\bigintsss_{\log M_{\rm *,min}}^\infty \diff\log M_*^\mathrm{obs}\,\frac{\mathcal{A}(\logm)}{\sqrt{2\pi\sigma_{M_*}^2}} \mathrm{exp}\left\{-\frac{(\log M_*-\log M_*^\obs)^2}{2\sigma_{M_*}^2}\right\}=1.
\end{equation}
In other words, the probability of measuring any value of the stellar mass larger than ${M_{\rm *,min}}$, given that a galaxy is part of our sample, is one.
We perform the final integration over $\log{M_*}$ numerically with a Monte Carlo method (see Appendix~\ref{app:likelihood}). 
We assume flat priors on all model hyper-parameters.

\subsection{Bayesian evidence}
\label{sec:evidence}


In our analysis, we consider models with different numbers of free hyper-parameters.
To evaluate the performance of a given model in fitting the data, we rely on the \textit{Bayesian evidence} $\mathcal{Z}$ that is the average of the likelihood under priors for a given model $\mathcal{M}$:
\begin{equation}
    \mathcal{Z}=\pdf(\data|\mathcal{M})=
    \int\diff\individ\,\pdf(\data|\individ,\mathcal{M})\,
        \pdf(\individ|\mathcal{M}).
\end{equation}
We remark that, in our approach, the parameters $\individ$ are described by a set of global hyper-parameters $\hyperp$.
When comparing two models, say models $\mathcal{M}_1$ and $\mathcal{M}_2$, we are interested in computing the ratio of the posterior probabilities of the models
\begin{equation}
    \frac{\pdf(\mathcal{M}_1|\data)}
    {\pdf(\mathcal{M}_2|\data)} = 
    \mathcal{B} \, \frac{\pdf(\mathcal{M}_1)}
    {\pdf(\mathcal{M}_2)},
\end{equation}
where 
\begin{equation}
    \label{eq:bayes_factor}
    \mathcal{B}\equiv
    \frac{\pdf(\data|\mathcal{M}_1)}
    {\pdf(\data|\mathcal{M}_2)}=\frac{\mathcal{Z}_1}{\mathcal{Z}_2}
\end{equation}
is the \textit{Bayes factor}.
When $\mathcal{B}\gg1$, $\mathcal{M}_1$ provides a better description of the data than $\mathcal{M}_2$, and vice versa when $\mathcal{B}\ll 1$.
The value of the Bayes factor is usually compared with the reference values of the empirical \textit{Jeffreys' scale} \citep{Jeffreys1961ToP}, reported in \autoref{tab:jeffreys}.
\begin{table}
\caption{Jeffreys' scale \citep{Jeffreys1961ToP}, giving the strength of evidence in the comparison of two models having Bayes factor $\mathcal{B}$ (equation \ref{eq:bayes_factor}).}
\begin{tabular}{cc}
\toprule
\toprule
\addlinespace
$|\ln\mathcal{B}|$ & Strength of evidence \\
\addlinespace
\midrule
\addlinespace
$0-1$   & Inconclusive \\
$1-2.5$ & Weak evidence \\
$2.5-5$ & Strong evidence \\
$>5$    & Decisive evidence \\
\bottomrule
\bottomrule
\end{tabular}
   \label{tab:jeffreys}
\end{table}
Given two different models, the quantity $|\ln{\mathcal{B}}|$ is a measure of the strength of evidence that one of the two models is preferable. 
We compute the Bayesian evidence $\mathcal{Z}$ of a model exploiting the \textit{nested sampling} technique \citep{Skilling2004AIP}. Briefly, the nested sampling algorithm estimates the Bayesian evidence reducing the $n$-dimensional evidence integral (where $n$ is the number of the parameters of a given model) into a 1D integral that is less expensive to evaluate numerically. In practice, we evaluate $\mathcal{Z}$ for a model using the \textsc{MultiNest} algorithm \citep[see][]{FerozHobson2008MNRAS,FerozHobsonBridges2009MNRAS} included in the Python module \textsf{PyMultiNest} \citep{Buchner2014A&A}.
For details about the estimates of the Bayesian evidence and the algorithm exploited to compute them, we refer the interested readers to \citet{FerozHobson2008MNRAS} and \citet{Buchner2014A&A}.

\begin{table*}
\centering
\caption{Hyper-parameters used in the models. Column 1: name of the model. Column 2: name of the hyper-parameter. Column 3: description of the hyper-parameter. Column 4: uniform
priors used in the models ("low" and "up" indicate, respectively, the lower and upper bounds). For those hyper-parameters showing two ranges for prior assumptions, the first refers to the fiducial sample and the second to the extended sample. $M_*^\mathrm{piv}$ and $z^\mathrm{piv}$ are the median values of stellar mass and redshift of the SDSS ETGs (\autoref{subsec:pdf_and_hyperparameters}).}
\begin{tabular}{cccc}
\toprule\toprule
\addlinespace
Model & Hyper-parameter & Description & Prior (low; up) \\
\addlinespace
\midrule
\addlinespace
\multirow{8}{*}{\minitab[c]{\Msdss}}
                         & $\mu_0$ & Median value of $\logsigmae$ at $M_*^\mathrm{piv}$ &
                          ($1$; $3$) \\
                         & $\beta_0$ & Index of the $M_*$-$\sigma_\mathrm{e}$ relation: $\sigma_\mathrm{e}\propto M_*^{\beta_0}$ &
                          ($0$; $1$)\\
                         & $\psi_0$ & Intrinsic scatter in $\log\sigma_\mathrm{e}$ &
                          ($0$; $1$)\\
                         & $\mu_{*,0}$ & Normalisation of the mean of Gaussian prior of $\logm$ &  ($10$; $13$)\\
                         & $\mu_{*,s}$ & Slope of the mean of Gaussian prior of $\logm$ &  ($15$; $30$)\\
                         & $\sigma_{*,0}$ & Normalisation of the standard deviation in the Gaussian prior of $\logm$ &  ($0$; $2$)\\
                         & $\sigma_{*,s}$ & Slope of the standard deviation in the Gaussian prior of $\logm$ &  ($-10$; $10$)\\
                         & $\alpha_*$ & Skewness parameter in the Gaussian prior of $\logm$ &  ($-5$; $15$)\\
\addlinespace
\midrule
\addlinespace
\addlinespace
\multirow{11}{*}{\minitab[c]{$\mathcal{M}_\mathrm{evo}$}}
                           & $\mu_0^\sdss$ & Median value of $\log\sigma_\mathrm{e}$ at $M_*=M_*^\mathrm{piv}$ and $z=z^\mathrm{piv}$ & $\simeq2.287$\\
                           & $\beta_0^\mathrm{SDSS}$ & Index of the $M_*$-$\sigma_\mathrm{e}$ relation  at $z=z^\mathrm{piv}$:  $\sigma_\mathrm{e}\propto M_*^{\beta_0^\mathrm{SDSS}}$  & $\simeq0.176$ \\
                           & $\eta$ & Index of the $\beta-(1+z)$ relation: $\beta\propto(1+z)^\eta$ &  ($-2$; $2$)\\
                           & $\zeta$ & Index of the $\sigma_\mathrm{e}-(1+z)$ relation: $\sigma_\mathrm{e}\propto(1+z)^\zeta$ &  ($-2$; $2$) \\
                           & $\psi_0^\mathrm{SDSS} $ & Median value of $\psi_0$ of the intrinsic scatter at $z=z^\mathrm{piv}$ & $\simeq0.075$ \\
                           & $\xi$   & Index of the $\sigma_\sigma-(1+z)$ relation: $\sigma_\sigma\propto(1+z)^\xi$ &  ($-2$; $2$) \\
                           & $\mu_{*,0}$ & Normalisation of the mean of Gaussian prior of $\logm$ & ($7$; $13$)\\      & $\mu_{*,s}$ & Slope of the mean of Gaussian prior of $\logm$ & ($0$; $10$) || ($-5$; $5$)\\
                           & $\sigma_{*,0}$ & Normalisation of  the standard deviation in the Gaussian prior of $\logm$ &  ($-5$; $5$)\\
                           & $\sigma_{*,s}$ & Slope of the standard deviation in the Gaussian prior of $\logm$ &  ($-10$; $10$)\\
                           & $\alpha_*$ & Skewness parameter in the Gaussian prior of $\logm$ &  ($-15$; $15$)\\                           
\addlinespace
\midrule
\addlinespace
\multirow{2}{*}{\minitab[c]{\Mconst}} & & Same as \Mevo, but with $\eta=0$ & \\
& $\alpha_*$ & Skewness parameter in the Gaussian prior of $\logm$ &  ($-10$; $10$) || ($-15$; $15$)\\
\addlinespace
\midrule
\addlinespace
\Mevones & & Same as \Mevo, but with $\xi=0$ & \\
\addlinespace
\midrule
\addlinespace
\multirow{3}{*}{\minitab[c]{\Mconstnes}} & & Same as \Mevo, but with $\eta=\xi=0$ & \\
& $\sigma_{*,s}$ & Slope of the standard deviation in the Gaussian prior of s$\logm$ &  ($-10$; $10$) || ($-5$; $5$)\\
& $\alpha_*$ & Skewness parameter in the Gaussian prior of $\logm$ &  ($-10$; $10$) || ($-15$; $15$)\\
\addlinespace
\bottomrule\bottomrule
\end{tabular}
\label{tab:table_hyperparameters}
\end{table*}

\section{Results}
\label{sec:results}
In this section we present the results obtained applying the Bayesian method described in \autoref{sec:method} to our fiducial  and extended samples of ETGs (see \autoref{sec:total_sample}). 

In \autoref{subsec:pdf_and_hyperparameters} we have introduced three models: model \Msdss (representing the present-day $M_*$-$\sigmae$ relation), model \Mconst (representing the evolution of the $M_*$-$\sigmae$ relation with redshift-independent slope $\beta$) and model \Mevo (representing the evolution of the $M_*$-$\sigmae$ relation  with redshift-dependent slope $\beta$).
In models \Mconst and \Mevo the intrinsic scatter of the $M_*$-$\sigmae$ relation is allowed to vary with redshift. 
In addition to these models, we also explore simpler models in which the intrinsic scatter is assumed to be independent of redshift. These models are named \Mconstnes and \Mevones, where NES stands for \textit{non-evolving scatter}.
In summary, we take into account five models:
model \Msdss, represented by \autoref{eq:m_sdss}, models  \Mconst and \Mevo, described by \autoref{eq:mstar_z_sigma_relation} (the former obtained by assuming $\eta=0$ in \autoref{eq:beta}), and  the models \Mconstnes and \Mevones, which are the same as \Mconst and \Mevo, respectively, but with $\xi=0$ in \autoref{eq:scatter}. A description of the hyper-parameters used for each model is provided in \autoref{tab:table_hyperparameters}.
Model \Msdss is applied to the SDSS subsample. The other four models are applied twice, once to the fiducial sample and once to the extended sample (we use the superscripts \textit{fid} and \textit{ext} to indicate that a model is applied, respectively, to the fiducial and extended samples). 

\begin{figure*}
    \centering
    \includegraphics[width=1.8\columnwidth]{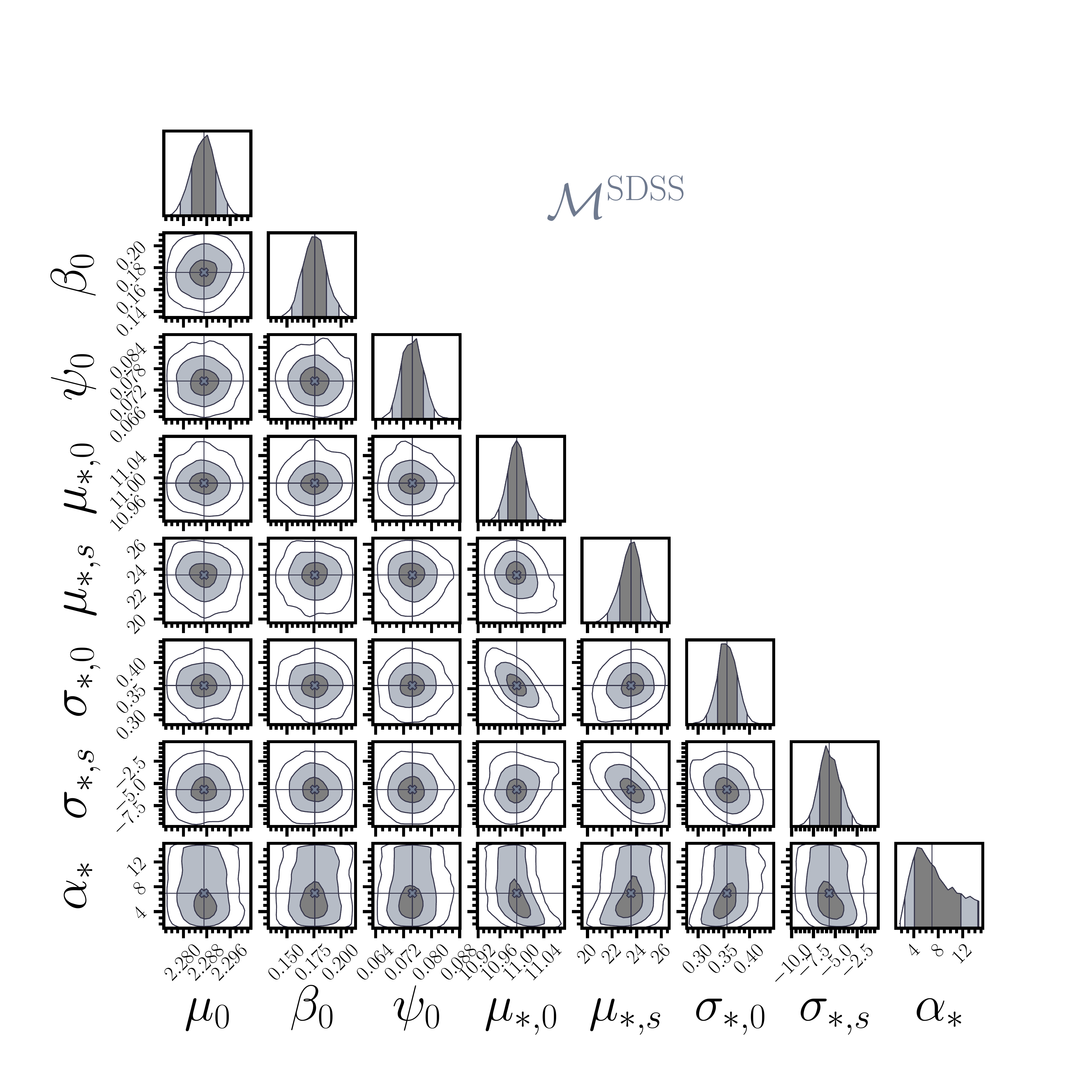}
    \caption{Posterior probability distributions of the hyper-parameters for model \Msdss (see Tables \ref{tab:table_hyperparameters} and \ref{tab:mcmc_median_errors}). In the 1D distributions (upper panel of each column) the vertical solid lines and colours delimit the 68, 95 and 99.7-th quantile based posterior credible interval. In the 2D distributions (all the other panels) the contours enclose the 68, 95 and 99.7 percent posterior credible regions. The dashed lines indicate the median values of the hyper-parameters.}
    \label{fig:mcmc_sdss}
\end{figure*}

The model-data comparison is performed as described in \autoref{sec:method}. We validated our method by applying it to a mock data set similar to the our SDSS data set (see Appendix \ref{sec:mock}). 
Each MCMC  run (see \autoref{sec:sampling}) uses $50$ random walkers running for 1000 steps to reach the convergence of the hyper-parameter distribution. 
The resulting inferences on the hyper-parameters used in model \Msdss are shown in \autoref{fig:mcmc_sdss}.  The SDSS galaxies are described by $\sigmae\propto M_*^{\beta_0}$ with $\beta_0\simeq0.176$. The normalisation $\mu_0\simeq2.287$ is such that galaxies with $M_*=10^{11}\msun$ have $\sigmae\simeq 170\,\kms$ and the intrinsic scatter is $\simeq0.075$ dex in $\sigmae$ at fixed $M_*$. The posterior distributions of the hyperparameters $\mu_0$, $\beta_0$ and $\psi_0$ are relatively narrow (\autoref{fig:mcmc_sdss}), with $1\sigma$ scatter of at most few percent  (\autoref{tab:mcmc_median_errors}), so our SDSS sample of ETG is sufficiently numerous for our purposes, even if it contains only a small fraction of the massive ETGs of the entire SDSS sample.
Our results on the present-day $M_*$-$\sigmae$ relation are broadly consistent with previous analyses
(see \autoref{sec:comparison} for details).

\begin{figure*}
    \centering
    \includegraphics[width=1.8\columnwidth]{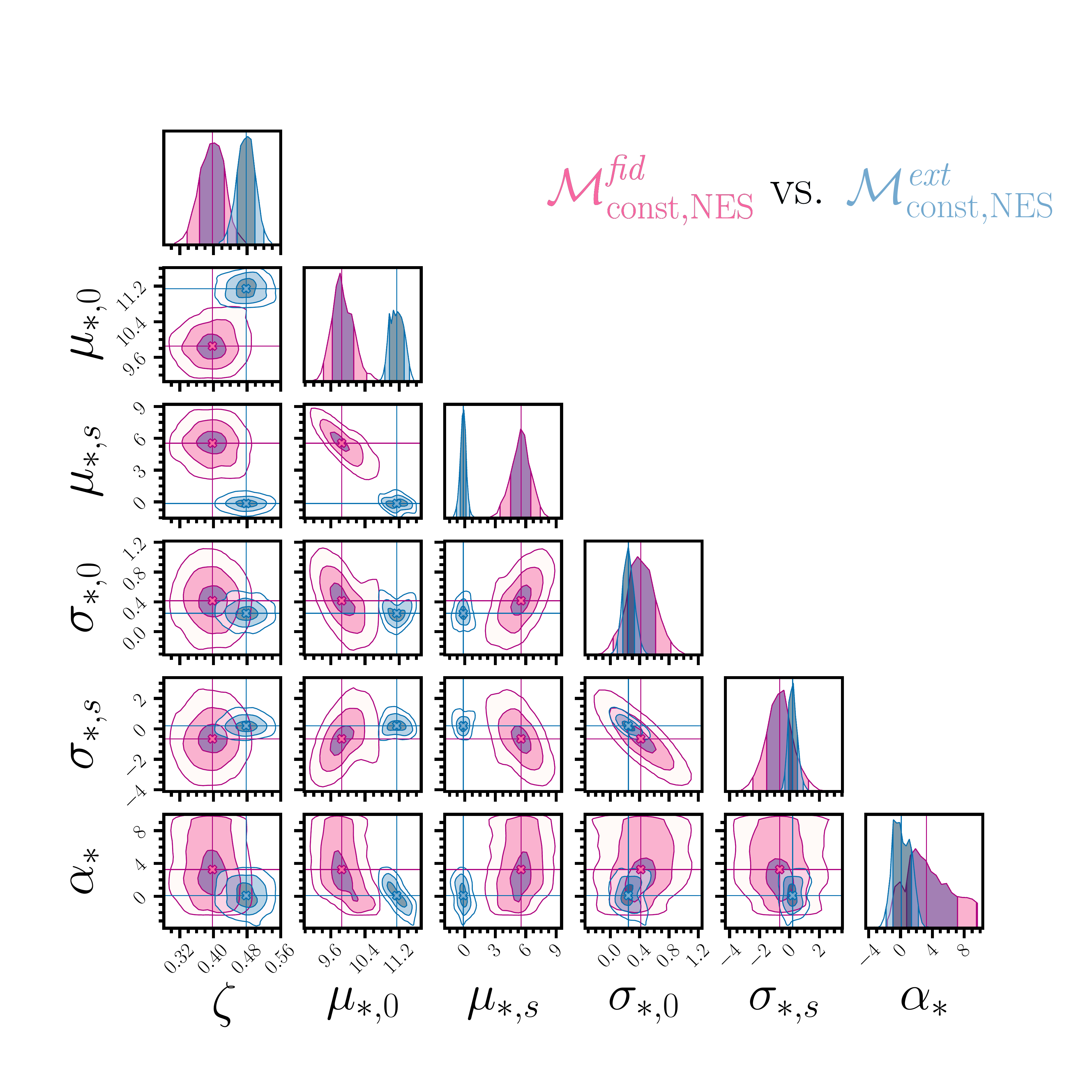}
    \caption{Same as \autoref{fig:mcmc_sdss}, but for models \Mconstfidnes (pink contours) and \Mconstextnes (azure contours; see Tables \ref{tab:table_hyperparameters} and \ref{tab:mcmc_median_errors}).}
    \label{fig:mcmc_constant_nes}
\end{figure*}

\begin{figure*}
    \centering
    \includegraphics[width=1.8\columnwidth]{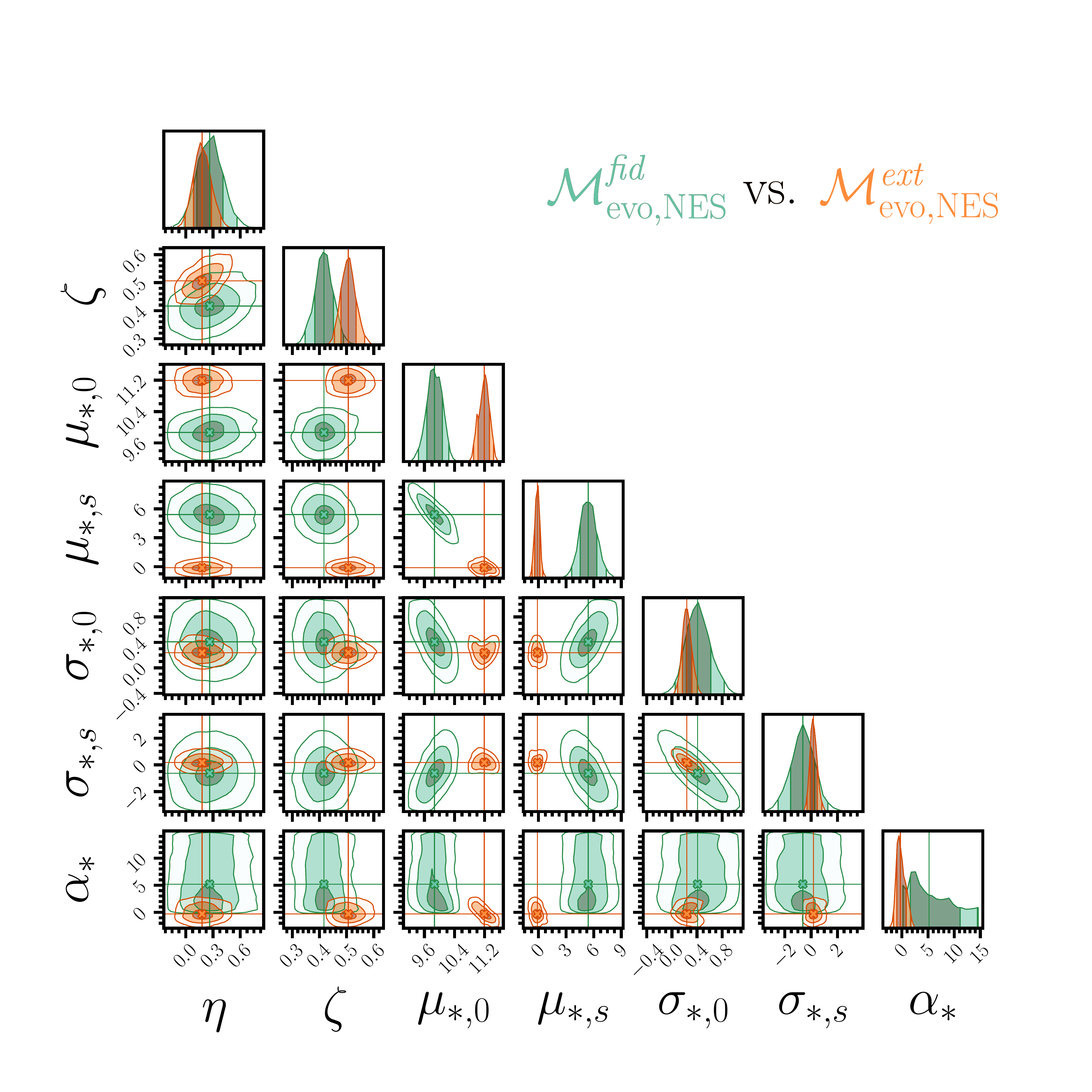}
    \caption{Same as \autoref{fig:mcmc_sdss}, but for models \Mevofidnes (green contours) and \Mevoextnes (orange contours; see Tables \ref{tab:table_hyperparameters} and \ref{tab:mcmc_median_errors}).}
    \label{fig:mcmc_evo_nes}
\end{figure*}

The median values of the hyper-parameters of all models, with the corresponding $1\sigma$ uncertainties, are listed in \autoref{tab:mcmc_median_errors}.
In order to compare the models we compute the Bayesian evidence $\mathcal{Z}$ of each model (\autoref{sec:evidence}), using a configuration of 400 live points in the nested sampling algorithm. The resulting $\mathcal{Z}$ and the Bayes factors  are listed in \autoref{tab:evidences}.
The performances of models \Mconst and \Mevo  are relatively poor when applied to both the fiducial and the extended samples, so in the following we focus on model \Mconstnes and \Mevones: in \autoref{fig:mcmc_constant_nes} and \autoref{fig:mcmc_evo_nes}, we show the inferences of these two models applied to both the fiducial and the extended samples.
\begin{table*}
\caption{Inferred median and $68\%$ posterior credible intervals of the hyper-parameters of the models.}
\begin{tabular}{ccccccccc}
\toprule
\toprule
\addlinespace
Model       &    
                 $\mu_0$    & 
                 $\beta_0$  & 
                 $\psi_0$   & 
                 $\mu_{*,0}$&
                 $\mu_{*,s}$& 
                 $\sigma_{*,0}$& 
                 $\sigma_{*,s}$&
                 $\alpha_*$ \\
                \addlinespace
                \midrule
                \addlinespace
$\mathcal{M}^\mathrm{SDSS}$ &
$2.287_{-0.004}^{+0.004}$ &
$0.176_{-0.011}^{+0.011}$ &
$0.075_{-0.003}^{+0.003}$ &
$10.990_{-0.016}^{+0.017}$ &
$23.533_{-0.910}^{+0.792}$ &
$0.356_{-0.018}^{+0.020}$ &
$-5.690_{-1.095}^{+1.358}$ &
$6.969_{-2.868}^{+4.725}$ \\
\addlinespace
\bottomrule
\bottomrule
\addlinespace
\addlinespace
\addlinespace
\addlinespace
\toprule
\toprule
\addlinespace
Model       &    
                 $\eta$    &
                 $\zeta$    & 
                 $\xi$  &
                 $\mu_{*,0}$&
                 $\mu_{*,s}$& 
                 $\sigma_{*,0}$& 
                 $\sigma_{*,s}$&
                 $\alpha_*$ \\
                 \addlinespace
                \midrule
                \addlinespace
\Mconstfid & 
$-$ & 
$0.390_{-0.031}^{+0.031}$ & 
$0.033_{-0.023}^{+0.026}$ & 
$9.827_{-0.213}^{+0.259}$ & 
$5.616_{-1.054}^{+0.955}$ & 
$0.457_{-0.193}^{+0.199}$ & 
$-0.841_{-0.908}^{+0.937}$ & 
$3.993_{-2.579}^{+4.089}$ \\
\addlinespace
\Mevofid & 
$0.248_{-0.156}^{+0.150}$ & 
$0.415_{-0.035}^{+0.034}$ & 
$0.028_{-0.024}^{+0.027}$ &
$9.855_{-0.220}^{+0.226}$ & 
$5.455_{-0.913}^{+0.995}$ & 
$0.428_{-0.191}^{+0.211}$ & 
$-0.684_{-0.950}^{+0.920}$ & 
$5.727_{-4.083}^{+5.763}$ \\
\addlinespace
\Mconstext & 
$-$ &
$0.474_{-0.023}^{+0.024}$ & 
$0.025_{-0.019}^{+0.019}$ & 
$11.147_{-0.158}^{+0.168}$ & 
$-0.172_{-0.348}^{+0.322}$ & 
$0.249_{-0.075}^{+0.088}$ & 
$0.210_{-0.271}^{+0.301}$ & 
$0.062_{-1.126}^{+1.421}$ \\
\addlinespace
\Mevoext & 
$0.179_{-0.106}^{+0.101}$ & 
$0.505_{-0.029}^{+0.028}$ & 
$0.019_{-0.022}^{+0.021}$ & 
$11.156_{-0.121}^{+0.140}$ & 
$-0.197_{-0.319}^{+0.300}$ & 
$0.240_{-0.079}^{+0.062}$ &
$0.209_{-0.254}^{+0.286}$ & 
$0.061_{-0.952}^{+1.031}$ \\
\addlinespace
\Mconstfidnes &
$-$ & 
$0.398_{-0.031}^{+0.028}$ &
$-$ & 
$9.853_{-0.220}^{+0.282}$ &
$5.538_{-1.029}^{+0.951}$ &
$0.415_{-0.181}^{+0.202}$ &
$-0.657_{-0.868}^{+0.846}$ &
$3.267_{-2.494}^{+3.894}$\\
\addlinespace
\Mevofidnes &
$0.264_{-0.142}^{+0.146}$ &
$0.417_{-0.034}^{+0.035}$ &
$-$ & 
$9.867_{-0.204}^{+0.211}$ &
$5.408_{-0.870}^{+0.934}$ &
$0.408_{-0.179}^{+0.211}$ &
$-0.625_{-0.921}^{+0.900}$ &
$5.195_{-3.571}^{+5.947}$ \\
\addlinespace
\Mconstextnes &
$-$ & 
$0.478_{-0.021}^{+0.021}$ &
$-$ & 
$11.142_{-0.171}^{+0.174}$ &
$-0.148_{-0.302}^{+0.272}$ &
$0.246_{-0.076}^{+0.084}$ &
$0.203_{-0.274}^{+0.325}$ &
$0.093_{-1.036}^{+1.289}$ \\
\addlinespace
\Mevoextnes &
$0.180_{-0.095}^{+0.103}$ & 
$0.506_{-0.027}^{+0.029}$ &
$-$ &
$11.192_{-0.166}^{+0.138}$ & 
$-0.120_{-0.285}^{+0.271}$ & 
$0.238_{-0.071}^{+0.081}$ & 
$0.177_{-0.246}^{+0.292}$ & 
$-0.281_{-0.704}^{+1.103}$ \\
\addlinespace
\bottomrule
\bottomrule
\end{tabular}
\label{tab:mcmc_median_errors}
\end{table*}

\begin{table}
\caption{Logarithms of the Bayesian evidences, $\ln\mathcal{Z}$, and logarithms of the Bayes factors, $\ln\mathcal{B}$, of the models. The values of $\mathcal{B}$ are relative to the Bayesian evidences of model \Mconstfidnes  for the fiducial sample and of model \Mconstextnes for the extended sample, i.e.\ the models with the highest evidences for given sample.}
\begin{tabular}{ccc}
\toprule
\toprule
\addlinespace
Model & $\ln{\mathcal{Z}}$ & $\ln{\mathcal{B}}$\\
\addlinespace
\midrule
\addlinespace
\Mconstfid & $241.791\pm0.204$ & $-2.269\pm0.391$ \\
\addlinespace
\Mevofid & $240.645\pm0.216$ & $-3.415\pm0.403$ \\
\addlinespace
\Mconstfidnes & $244.060\pm0.187$ & $-$ \\
\addlinespace
\Mevofidnes & $243.710\pm0.197$ & $-0.350\pm0.384$ \\
\addlinespace
\midrule
\addlinespace
\Mconstext & $284.705\pm0.225$ & $-2.653\pm0.431$    \\
\addlinespace
\Mevoext & $283.908\pm0.235$ & $-3.450\pm0.441$ \\
\addlinespace
\Mconstextnes & $287.358\pm0.206$ & $-$ \\
\addlinespace
\Mevoextnes & $286.584\pm0.222$ & $0.774\pm0.428$ \\
\addlinespace
\bottomrule
\bottomrule
\end{tabular}
\label{tab:evidences}
\end{table}


\subsection{Fiducial sample ($0\lesssim z\lesssim 1$)}
\label{sec:results_fiducial}
The model with the highest Bayesian evidence, among those applied to the fiducial sample, is \Mconstfidnes (see \autoref{tab:evidences}). Model \Mevofidnes, though with slightly lower evidence,   describes the data as well as model \Mconstfidnes, according to Jeffreys' scale (\autoref{tab:jeffreys}), while models \Mconstfid and \Mevofid are rejected with strong evidence. Thus, based on our analysis of the fiducial sample, we conclude that at $z\lesssim 1$ the normalisation of the $M_*$-$\sigmae$ relation changes with $z$, while the intrinsic scatter is independent of redshift; the slope $\beta$ is either constant or increasing with redshift (see \autoref{fig:evolution_normalisation}).

The median $M_*$-$\sigmae$ relations found for models \Mconstfidnes and  \Mevofidnes at three representative redshifts are shown in the left panels of \autoref{fig:mstar_sigma_relations}. Quantitatively, according to model \Mconstfidnes, in the redshift interval $0\lesssim z\lesssim1$, the $M_*$-$\sigmae$ relation is well described by a power law $\sigmae\propto M_*^\beta$ with redshift-independent slope $\beta\simeq0.18$ and intrinsic scatter $\simeq0.08\,\mathrm{dex}$ in $\sigmae$ at given $M_*$. At fixed $M_*$, $\sigmae\propto(1+z)^\zeta$, with $\zeta\simeq0.40$, so galaxies of given $M_*$ tend to have higher $\sigmae$ at higher redshift: the median velocity dispersion at fixed $M_*$ is a factor $\approx1.3$ higher at $z=1$ than at $z=0$. According to model \Mevofidnes, $\sigmae$ varies with $M_*$ and $z$ as $\sigmae\propto{M_*}^{\beta(z)}(1+z)^\zeta$, with $\zeta\simeq0.42$ and  $\beta(z)\simeq 0.16+0.26\log(1+z)$. For instance, at $z=1$, the slope of the $M_*$-$\sigmae$ relation is $\beta\simeq 0.24$.  The time variation of $\sigmae$ at given $M_*$ depends on $M_*$: at $M_*=10^{11}\,\msun$ it is similar to that inferred according model \Mconstfidnes (\autoref{fig:evolution_normalisation}, upper panel).

In summary, the evolution of the $M_*$-$\sigmae$ relation in the redshift range $0\lesssim z\lesssim 1$ can be roughly described by
\begin{equation}
\log\left(\frac{\sigmae}{\kms}\right)\simeq 2.21+0.18 \log\left(\frac{M_*}{10^{11}\msun}\right)+0.40\log(1+z),
\end{equation}
based on the median values of the hyper-parameters of model \Mconstfidnes, or
\begin{equation}
\begin{split}
\log\left(\frac{\sigmae}{\kms}\right)\simeq& \left[0.16+0.26\log(1+z)\right]\log\left(\frac{M_*}{10^{11}\msun}\right)+\\
&+0.42\log(1+z)+2.22,
\end{split}
\end{equation}
based on the median values of the hyper-parameters of model \Mevofidnes, 
with redshift-independent intrinsic scatter $\simeq 0.08 \, \mathrm{dex}$ in $\sigmae$ at a given $M_*$.

\subsection{Extended sample ($0\lesssim z\lesssim 2.5$)}
\label{sec:results_extended}

\begin{figure}
        \centering
        \includegraphics[width=1\columnwidth]{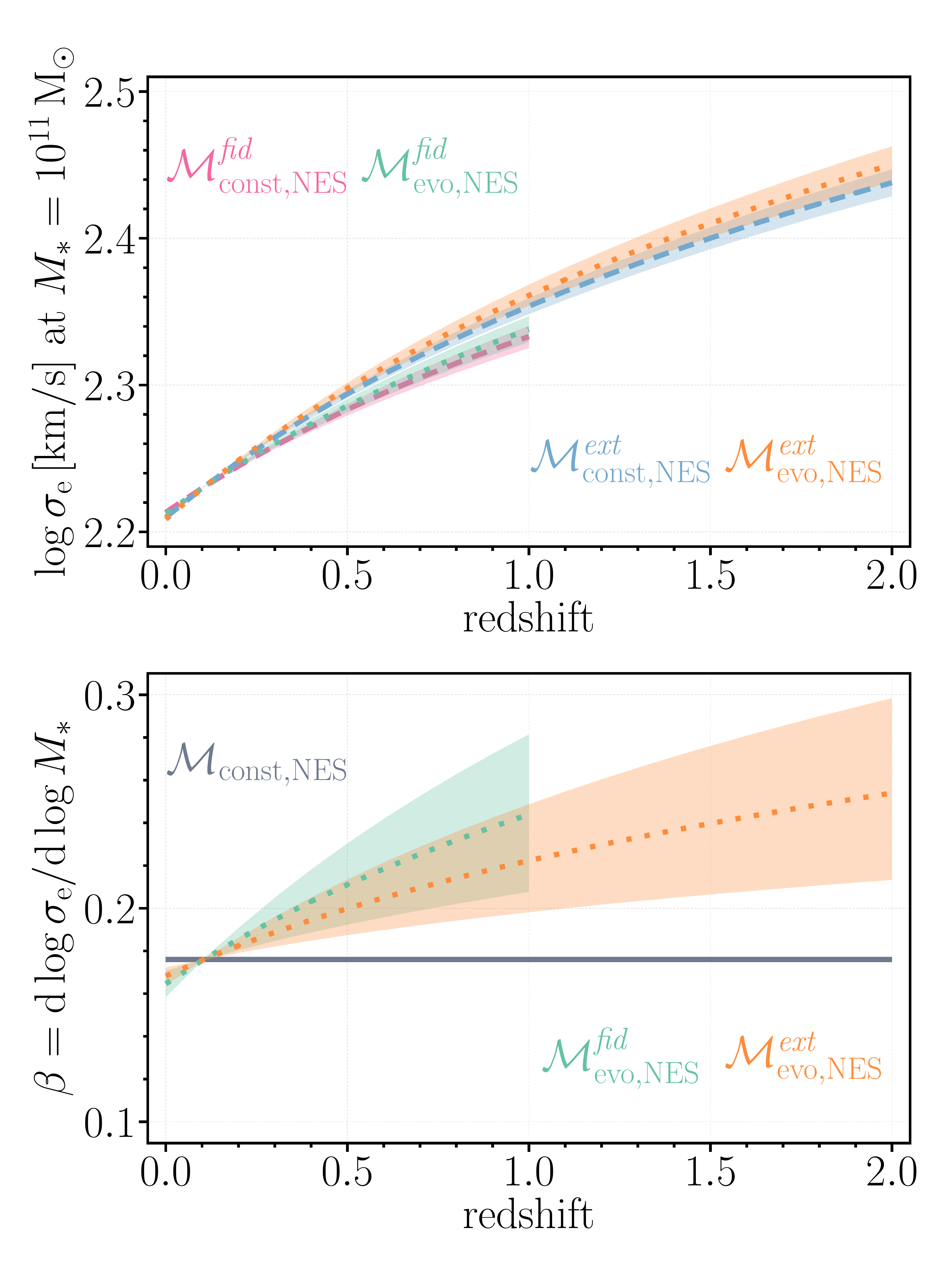}
    \caption{Median central stellar velocity dispersion $\sigmae$ at $M_*=10^{11}\,\msun$ (upper panel) and slope $\beta$ of the $M_*$-$\sigmae$ relation (lower panel) as functions of redshift for models \Mconstfidnes (pink dashed curve), \Mevofidnes (green dotted curves), \Mconstextnes (azure dashed curve) and \Mevoextnes (orange dotted curves). In the lower panel the curves for models \Mconstfidnes and \Mconstextnes are identical, and are represented by the grey solid line.  
    The curves are obtained by computing, at given $x$, the median value of $y$ (where $x$ and $y$ are the quantities in abscissa and ordinate, respectively) among all the values sampled by the posterior distribution obtained with the MCMC; similarly, the shaded bands, which we will refer to as $1\sigma$ uncertainty bands, are defined by computing the $16\%$ and the $84\%$ of the distribution of $y$, at given $x$, for the same sampling. }
	\label{fig:evolution_normalisation}
\end{figure}

The results obtained for the extended sample are very similar to those obtained for the fiducial sample. The model with the highest Bayesian evidence is model \Mconstextnes (\autoref{tab:evidences}), but the performance of model \Mevoextnes is comparable on the basis of Jeffreys' scale (see \autoref{tab:jeffreys}). 
Models \Mconstext and \Mevoext are rejected with strong evidence.
Thus, on the basis of our data, over the redshift range $0\lesssim z\lesssim 2.5$ the $M_*$-$\sigmae$ relation of ETGs evolves in time by changing its normalisation, with
redshift-independent intrinsic scatter, and  with slope either constant or increasing with redshift (see \autoref{fig:evolution_normalisation}).

\begin{figure*}
    \begin{minipage}{2\columnwidth}
        \includegraphics[width=0.5\columnwidth]{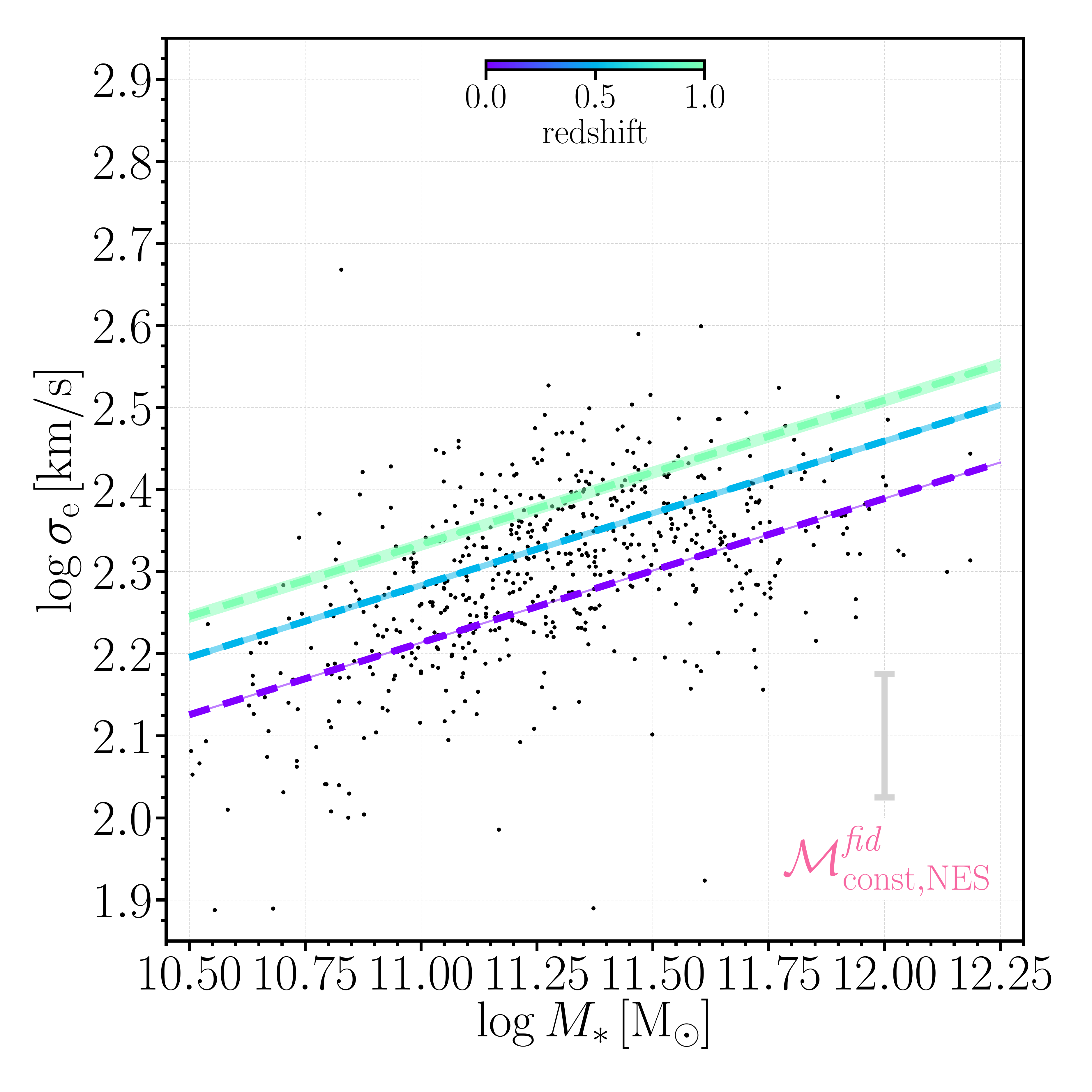}
        \includegraphics[width=0.5\columnwidth]{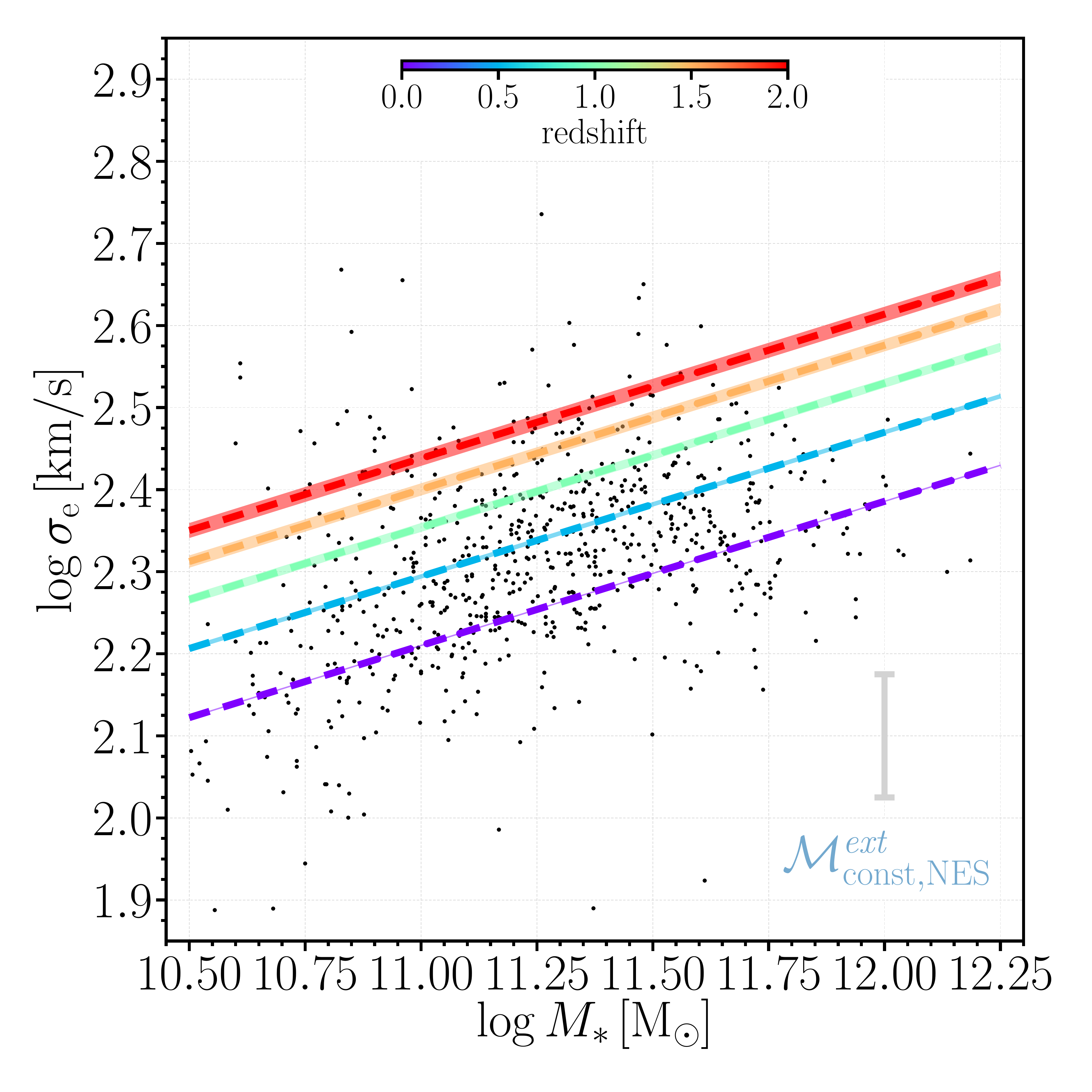} 
        \includegraphics[width=0.5\columnwidth]{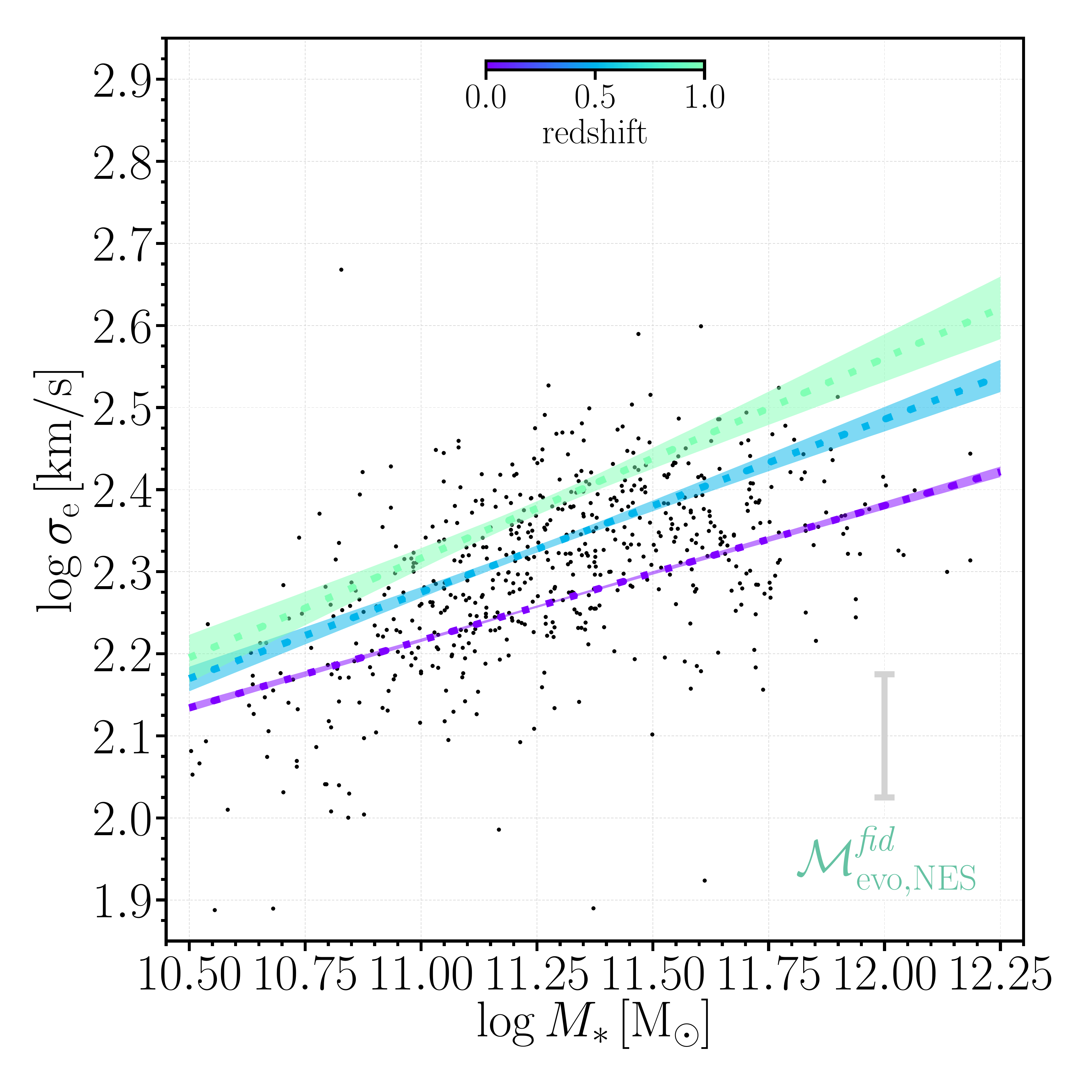}
        \includegraphics[width=0.5\columnwidth]{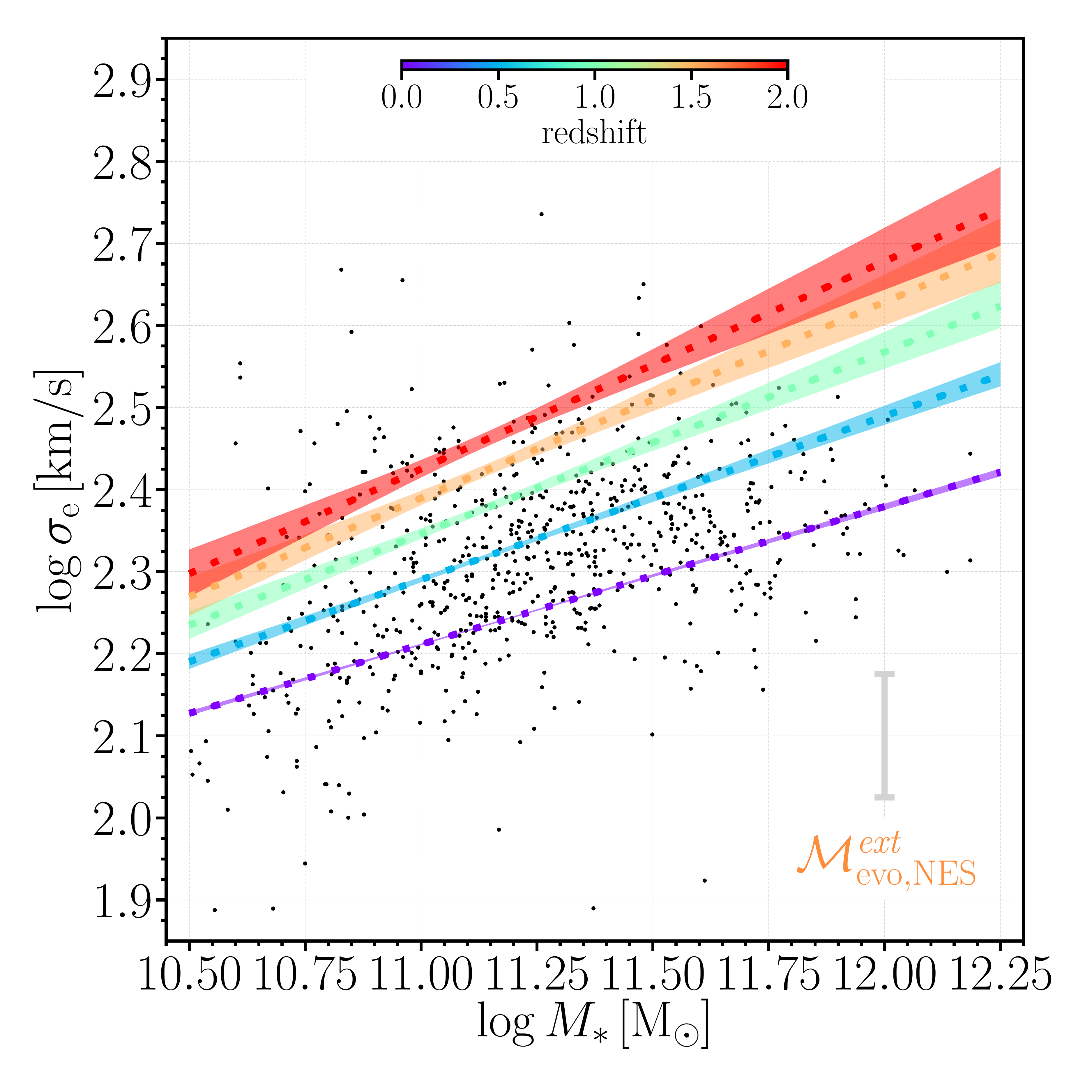} 
    \end{minipage} 
\caption{Central stellar velocity dispersion as a function of stellar mass. In the left panels the curves represent the median relations of the fiducial-sample models \Mconstfidnes (upper panel) and  \Mevofidnes (lower panel), 
at $z=0$, $z=0.5$ and $z=1$. In the right panels the curves represent the median relations of the extended-sample models \Mconstextnes (upper panel) and  \Mevoextnes (lower panel), 
at $z=0$, $z=0.5$, $z=1$, $z=1.5$ and $z=2$. The bands indicate the $1 \sigma $ uncertainty ranges. The median relations and the $1\sigma$ bands are computed as described in the caption of \autoref{fig:evolution_normalisation}.  The grey error bars represent the redshift-independent intrinsic scatter. The dots indicate the positions in these diagrams of the ETGs of the fiducial  (left panels) and  extended  (right panels) samples.}
	\label{fig:mstar_sigma_relations}
\end{figure*}

The median $M_*$-$\sigmae$ relations of models \Mconstextnes and \Mevoextnes at five representative redshifts are shown in the right panels of \autoref{fig:mstar_sigma_relations}. Quantitatively, according to model \Mconstextnes, in the redshift interval $0\lesssim z\lesssim 2.5$, the $M_*$-$\sigmae$ relation is well described by a power law $\sigmae\propto M_*^\beta$ with slope $\beta\simeq0.18$ and intrinsic scatter $\simeq0.08\,\mathrm{dex}$ in $\sigmae$ at given $M_*$: at fixed $M_*$,  $\sigmae\propto(1+z)^\zeta$, with $\zeta\simeq0.48$, so galaxies of given $M_*$ tend to have higher $\sigmae$ at higher redshift. For instance, the median velocity dispersion at fixed $M_*$ is a factor $\approx1.7$ higher at $z=2$ than at $z=0$. According to model \Mevoextnes,  $\sigmae$ varies with $M_*$ and $z$ as $\sigmae\propto M_*^{\beta(z)}(1+z)^\zeta$, with $\zeta\simeq0.51$ and $\beta(z)\simeq 0.17+0.18\log(1+z)$ (so $\beta\simeq 0.26$ at $z=2$; \autoref{fig:evolution_normalisation}, lower panel). The time variation of $\sigmae$ at given $M_*$ depends on $M_*$, but at $M_*=10^{11}\,\msun$ is similar to that obtained with model \Mconstextnes (\autoref{fig:evolution_normalisation}, upper panel)

In summary,  the evolution of the $M_*$-$\sigmae$ relation in the redshift range $0\lesssim z\lesssim 2.5$ can be roughly described by
\begin{equation}
\log\left(\frac{\sigmae}{\kms}\right)\simeq 2.21+0.18 \log\left(\frac{M_*}{10^{11}\msun}\right)+0.48\log(1+z),
\end{equation}
based on the median values of the hyper-parameters of model \Mconstextnes, or
\begin{equation}
\begin{split}
\log\left(\frac{\sigmae}{\kms}\right)\simeq\,& [0.17+0.18\log(1+z)]\log\left(\frac{M_*}{10^{11}\msun}\right)+\\&+0.51\log(1+z)+2.21,
\end{split}
\end{equation}
based on the median values of the hyper-parameters of model \Mevoextnes, with redshift-independent intrinsic scatter $\simeq0.08 \, \mathrm{dex}$ in $\sigmae$ at a given $M_*$.

\subsection{Comparing the results for the fiducial and extended samples}
\label{sec:results_comparison_fiducial_extended}

Among the hyper-parameters of model \Mconstnes, only $\zeta$, which quantifies the redshift dependence of $\sigmae$ at given $M_*$, contains physical information on the $M_*$-$\sigmae$ relation: the other five hyper-parameters describe properties of the galaxy sample. Thus, when comparing the inferences obtained for model \Mconstnes applied to the fiducial and extended samples (\autoref{fig:mcmc_constant_nes}), we must focus on the inference on $\zeta$.
For model \Mevones, instead, the physical information is contained in the hyper-parameters $\eta$ and $\zeta$, which must be considered  when comparing the inferences for models \Mevofidnes and \Mevoextnes (\autoref{fig:mcmc_evo_nes}). While the differences in the distributions of $\eta$ between \Mevofidnes and \Mevoextnes are well within 1$\sigma$, the differences in the distributions of $\zeta$ are between 1$\sigma$ and 2$\sigma$ for both pairs of models (\Mconstfidnes-\Mconstextnes and  \Mevofidnes-\Mevoextnes). 
Thus, while we find no significant differences in $\eta$, the extended-sample data prefer a somewhat higher value of $\zeta$ than the fiducial-sample data, suggesting that the evolution of $\sigmae$ at a given $M_*$ might be stronger at higher redshift.

However, we recall that the extended sample is not as homogeneous and complete as the fiducial sample, so the aforementioned difference in $\zeta$ could be produced by some observational bias. 
For instance, while for the fiducial sample we selected ETGs on the basis of morphology and strength of emission lines, in some of the subsamples of the high-redshift sample (\citealt{Belli2014bApJ} and \citealt{Belli2017ApJ}), ETGs were selected using also the so-called $UVJ$ colour-colour diagram, which is a useful tool to separate passive and star-forming galaxies 
\citep[e.g.][]{Mor++13}. 
To quantify the effect of these different selection criteria, we performed the following test. 
Using $UVJ$ colour measures from the UltraVISTA survey (\citealt{Muzzin2013ApJ}), we placed the
LEGA-C galaxies of our fiducial sample in the $UVJ$ diagram, finding that $90\%$ of them  lie in the locus of passive galaxies (see \citealt{Can20}). 
We then modified our fiducial sample by excluding the remaining $10\%$ of the LEGA-C galaxies and applied 
model \Mconstfidnes to this modified fiducial sample, finding inferences on the hyper-parameters (in particular  $\zeta=0.408_{-0.031}^{+0.032}$) in agreement within $1\sigma$ with those  of \Mconstfidnes shown in \autoref{fig:mcmc_constant_nes}. This test indicates that the results obtained for the extended sample 
should be independent of whether the $UVJ$-colour selection is used as additional criterion to define the sample of ETGs. Of course the selection of the extended sample is heterogeneous also in other respects, so we cannot exclude that there are other non-negligible biases.

As a more general comment, we note that, for both the fiducial and the extended samples,
the results on the evolution of the $M_*$-$\sigmae$ relation hold within the assumption that the slope and the normalisation vary as power laws of $1+z$. Our inferences on the redshift intervals in which we have no galaxies ($0.2<z<0.6$) or very few galaxies ($z>1.75$) in our samples (\autoref{fig:differential_distributions}) clearly rely on this assumption and are thus driven by the properties of galaxies in other redshift intervals.

\section{Discussion}
\label{sec:discussion}

\subsection{Potential systematics}

Our inference relies on measurements of the stellar mass and central velocity dispersion of galaxies. Both these quantities are subject to systematic effects that could in principle affect our results. The biggest systematics are those affecting the stellar mass measurements, which we discuss here.

Stellar mass measurements are the result of fits of S\'{e}rsic profiles to broad band photometric data, from which luminosities and colours are derived and subsequently fitted with stellar population synthesis models.
One possible source of systematics is a deviation of the true stellar density profile of a galaxy from a S\'{e}rsic profile.
For instance, as shown by \citet{Sonnenfeld2019A&A} in their study of a sample of massive ellipticals at $z\sim0.6$, it is difficult to distinguish between a pure S\'{e}rsic model and a model consisting of the sum of a S\'{e}rsic and an exponential component, even with relatively deep data from the HSC survey: differences between the two models only arise at large radii and can lead to variations in the estimated luminosity on the order of $0.1$~dex.
Secondly, our models assume implicitly that the stellar population parameters of a galaxy are spatially constant. However, if these vary as a function of radius, a bias on the inferred stellar masses could be introduced.
More generally, the stellar population synthesis models on which our $M_*$ measurements are based are known to be subject to systematics \citep[see e.g.][]{Con++13}. Most importantly, uncertainties on the stellar IMF can lead to a global shift of the stellar mass distribution, affecting the inference on the normalisation of the $M_*$-$\sigmae$ relation $\mu_0$, and/or the slope of the relation $\beta_0$, in case the IMF varies as a function of mass.
Additionally, gradients in the IMF can also introduce biases: along with $M_*/L$ gradients at fixed IMF, these are particularly relevant if our measurements are used to quantify the stellar component to the dynamical mass of a galaxy \citep[see e.g.][and related discussions]{Li++17,Ber++18,Son++18b,Dom++19}.

All of these systematic effects are common to virtually all estimates of the $M_*$-$\sigma_0$ relation in the literature and are difficult to address, given our current knowledge on the accuracy of our models of galaxy stellar profiles and stellar populations. Nevertheless, they should be taken into consideration when comparing our observations with theoretical models.

\subsection{Comparison with previous works}
\label{sec:comparison}

In this section we compare our results on the $M_*$-$\sigmae$ relation with previous works in the literature, which we briefly describe in the following.

\begin{itemize}
\item \citet{Auger2010ApJ} study a sample of 59 ETGs (morphologically classified as ellipticals or S0s) identified as strong gravitational lenses in the Sloan Lens ACS Survey (SLACS) \citep{Bolton2008ApJ,Auger2009ApJ} with a mean redshift $z\approx0.2$. The stellar masses of these ETGs span the range $11<\log(M_*/\msun)<12$. \citet{Auger2010ApJ} report  fits of the $M_*$-$\sigma_{\mathrm{e}/2}$  relation both allowing and not allowing for the presence of intrinsic scatter ($\sigma_{\mathrm{e}/2}$ is the velocity dispersion within an aperture $R_\mathrm{e}/2$).

\item \citet{HydeBernardi2009MNRAS} extract 46410 ETGs from the SDSS DR4 with parameters updated to the DR6 values \citep{Adelman-McCarthy2008ApJS}, selecting galaxies
with $60 < \sigma_{\mathrm{e}/8}/(\mathrm{km\,s^{-1}}) < 400$, where $\sigma_{\mathrm{e}/8}$ is the stellar velocity dispersion measured within an aperture $R_{\mathrm{e}}/8$, in the redshift range $0.07<z\leq0.35$. \citet{HydeBernardi2009MNRAS} fit the distribution of $\log \sigma_{\mathrm{e}/8}$ as a function of $\log M_*$ both with a linear function, over the stellar mass range $10.5<\log(M_*/\msun)<11.5$, and with a quadratic function, in the range $9.5<\log(M_*/\msun)<12$.

\item \citet{Damjanov2018ApJS} estimate the $M_*$-$\sigma_0$ relation of 565 quiescent galaxies of the hCOS20.6 sample, with $10.5<\log(M_*/\msun)<11.4$, in the redshift range $0.2<z<0.5$. The velocity dispersions, corrected to an aperture of $3\,\mathrm{kpc}$, can be  taken as good approximations (to within $3-4\%$; I.\ Damjanov, private communication) of measurements of $\sigmae$.

\item \citet{Zahid2016ApJ} analyse the $M_*$-$\sigma_0$ relation for massive quiescent galaxies out to $z\approx 0.7$. For our comparison, we use their power-law fit obtained for a subsample of 1316 galaxies drawn from the Smithsonian Hectospec Lensing Survey (SHELS; \citealt{Geller2005ApJ}) at $0.3<z<0.4$.
Also in this case the velocity dispersions, which are corrected to an aperture of  $3\,\mathrm{kpc}$, can be taken as measurements of $\sigmae$.

\item \citet{Belli2014ApJ} measure $\sigma_\mathrm{e}$ and $M_*$ for a sample of galaxies with median redshift $z\simeq 1.23$. 
We take from  \citet{Zahid2016ApJ} the best fit parameters of the $M_*$-$\sigma_\mathrm{e}$ relation for the sample of \citet{Belli2014ApJ}.

\item \citet{Mason2015ApJ} study the redshift evolution of the $M_*$-$\sigmae$ relation, assuming redshift-independent slope determined by the low-$z$ relation measured by \citet{Auger2010ApJ}, finding that $\sigmae$ at fixed $M_*$ increases with redshift as $(1+z)^{0.2}$
In particular, we consider here the fit of \citeauthor{Mason2015ApJ} evaluated at $z=0.35, 1.23$ and $2$, taking as reference the fit of \citet{Auger2010ApJ} with non-zero intrinsic scatter.
\end{itemize}

\begin{figure*}
    \begin{minipage}{2\columnwidth}
        \includegraphics[width=.5\columnwidth]{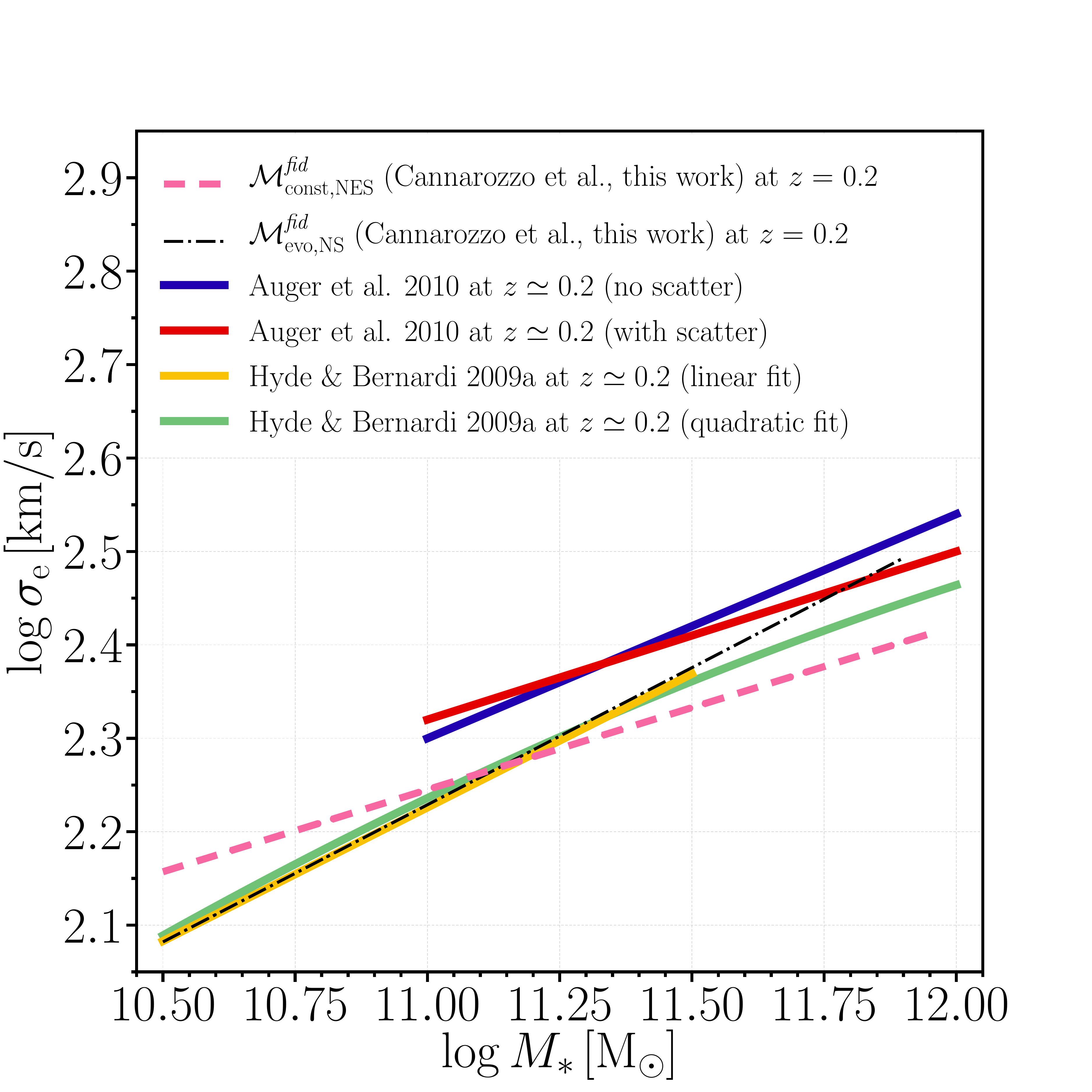}
        \includegraphics[width=.5\columnwidth]{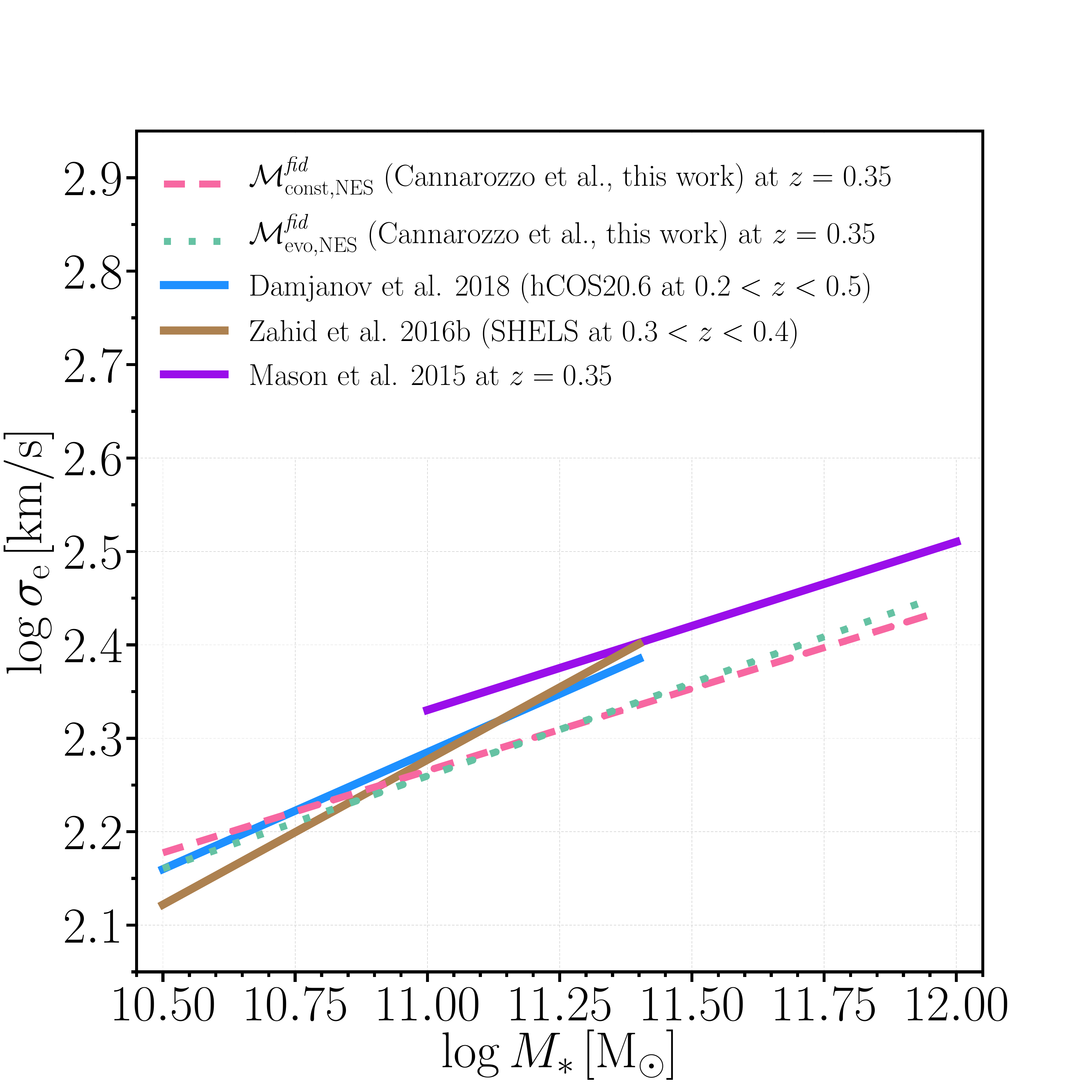}
        \includegraphics[width=.5\columnwidth]{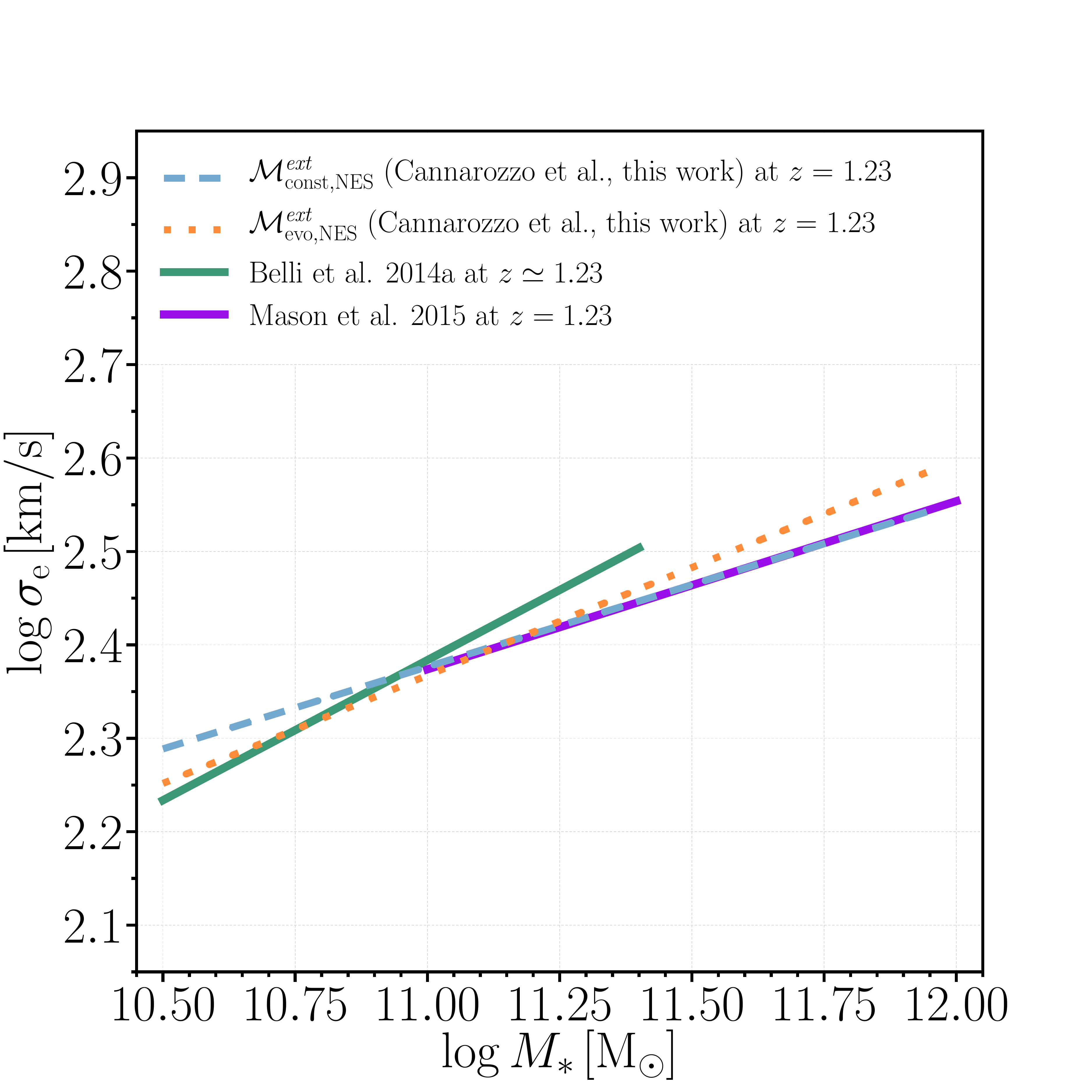}
        \includegraphics[width=.5\columnwidth]{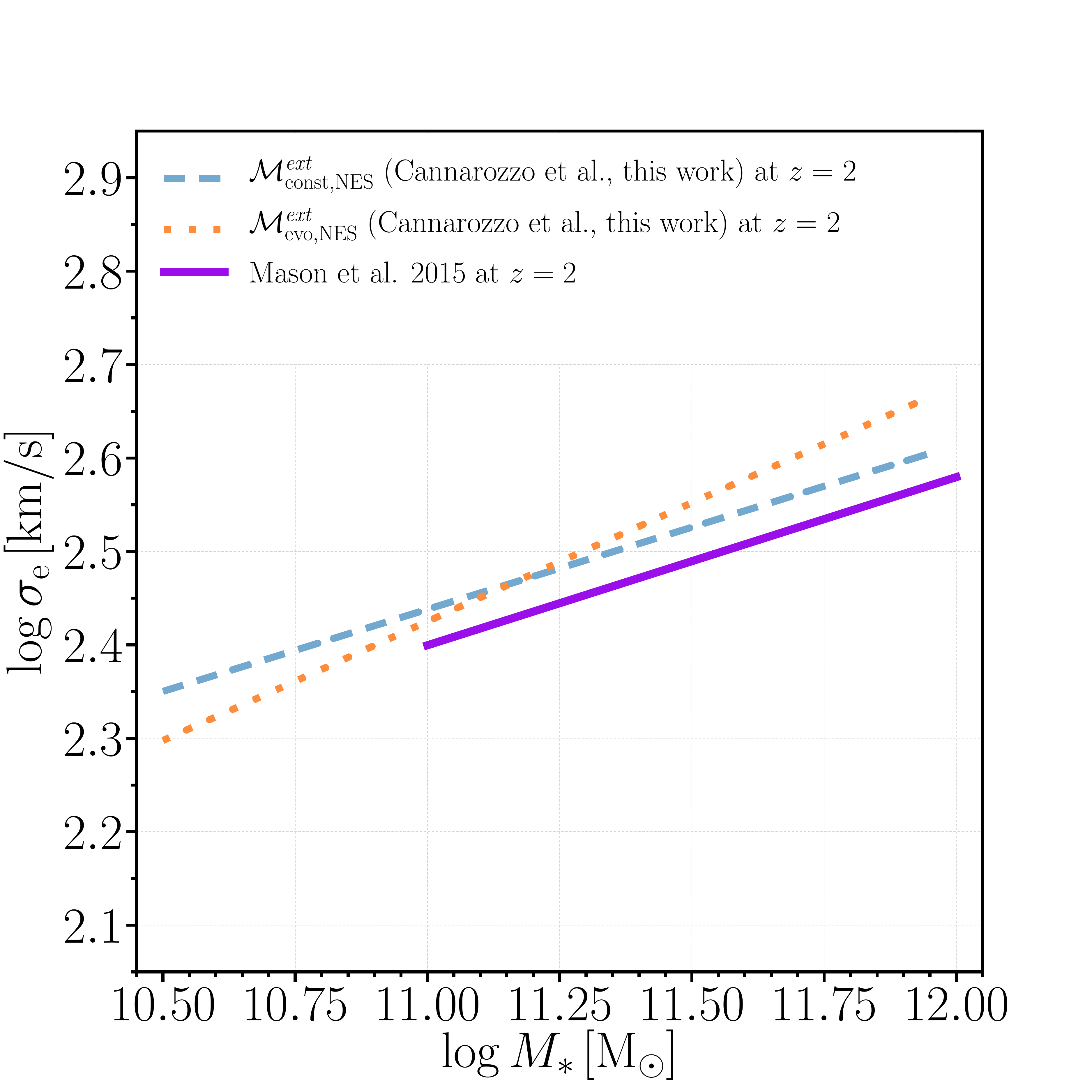}
    \end{minipage}
    \caption{Comparison between the median $M_*$-$\sigmae$ relations of our models \Mconstnes (dashed curves) and \Mevones (dotted curves),  and fits from the literature (solid curves) at $z=0.2$ (upper left panel), $z=0.35$ (upper right panel), $z=1.23$ (lower left panel) and $z=2$ (lower right panel). Upper left panel: the red and blue solid curves are the linear fits of \citet{Auger2010ApJ}, including and not including the intrinsic scatter, respectively, while the gold and light green solid curves are the linear and the quadratic fits of \citet{HydeBernardi2009MNRAS}, respectively.
    The thin dash-dotted curve is the median relation of model $\mathcal{M}_\mathrm{evo,NS}^\mathit{fid}$, in which we assume zero  intrinsic scatter (see text). Upper right panel: the azure curve is the linear fit for the hCOS20.6 galaxies ($0.2<z<0.5$) by \citet{Damjanov2018ApJS}, the brown curve is the linear fit of \citet{Zahid2016ApJ} for the SHELS sample ($0.3<z<0.4$) and the violet curve is the fit of \citet{Mason2015ApJ} evaluated at $z=0.35$.
    Lower left panel: the green curve is the linear fit of \citet{Belli2014ApJ} and the violet curve is the fit of \citet{Mason2015ApJ} evaluated at $z=1.23$ . Lower right panel: the violet curve is the fit of \citet{Mason2015ApJ} evaluated at $z=2$. Each curve is shown over the stellar mass range spanned by the considered data set.}
	\label{fig:comparison}
\end{figure*}

In order to compare the results of the different works, we express all the fits in the form
\begin{equation}\label{eq:comparison}
    \log\left(\frac{\sigma_{\rm e}}{\kms}\right)=\mu+\beta\log\left(\frac{M_*}{M_*^\mathrm{piv}}\right)+\gamma\left[\log\left(\frac{M_*}{M_*^\mathrm{piv}}\right)\right]^2,
\end{equation}
where $M_*^\mathrm{piv}=10^{11.321}\,\msun$ (Chabrier IMF). 
We correct for aperture the fits of \citet{Auger2010ApJ} and \citet{HydeBernardi2009MNRAS} using \autoref{eq:aperture_correction}, so $\logsigmae=\log\sigma_\mathrm{e/8}-0.06$ and $\logsigmae=\log\sigma_\mathrm{e/2}-0.02$. 
Except for the quadratic fit of \citet{HydeBernardi2009MNRAS}, all the other fits assume $\gamma=0$ in \autoref{eq:comparison}.  
The values of the parameters of \autoref{eq:comparison} for the considered literature works are reported in \autoref{tab:comparison}.
In \autoref{fig:comparison}, we show the comparison between our models and the previous works at $z=0.2$, 0.35, 1.23 and 2.

\begin{table}
\caption{Values of the parameters of \autoref{eq:comparison}, according to the fits of the literature works that we compared with our model. 
$\gamma=0$ in all cases, but in the case of the quadratic fit of \citet{HydeBernardi2009MNRAS}, for which $\gamma=-0.044$.}
\begin{tabular}{clcc}
\toprule
\toprule
\addlinespace
Redshift & Reference & $\mu$ & $\beta$ \\
\addlinespace
\midrule
\addlinespace
\multirow{9}{*}{\minitab[c]{$z\simeq0.2$}} &\citet{Auger2010ApJ}                      & $2.38$ & $0.24$\\
\addlinespace
&\citet{Auger2010ApJ}        & \multirow{2}{*}{\minitab[c]{$2.38$}} & \multirow{2}{*}{\minitab[c]{$0.18$}}\\
&with intrinsic scatter                               & &\\
\addlinespace
&\citet{HydeBernardi2009MNRAS}         & \multirow{2}{*}{\minitab[c]{$2.32$}} & \multirow{2}{*}{\minitab[c]{$0.29$}}\\
&Linear fit                                   & &\\
\addlinespace
&\citet{HydeBernardi2009MNRAS}         & \multirow{2}{*}{\minitab[c]{$2.32$}} & \multirow{2}{*}{\minitab[c]{$0.24$}}\\
&Quadratic fit                                   & &\\
\addlinespace
\midrule
\addlinespace
\multirow{5}{*}{\minitab[c]{$z\simeq0.35$}} 

&\citet{Damjanov2018ApJS}  & \multirow{2}{*}{\minitab[c]{$2.37$}} & \multirow{2}{*}{\minitab[c]{$0.25$}} \\
&hCOS20.6 ($0.2<z<0.5$)&&\\
\addlinespace
&\citet{Zahid2016ApJ}  & \multirow{2}{*}{\minitab[c]{$2.38$}} & \multirow{2}{*}{\minitab[c]{$0.31$}} \\
&SHELS ($0.3<z<0.4$)&&\\
\addlinespace
$z=0.35$&\citet{Mason2015ApJ} & $2.39$ & $0.18$\\
\addlinespace
\midrule
\addlinespace
$z\simeq1.23$&\citet{Belli2014ApJ} & $2.48$  & $0.30$ \\
\addlinespace
$z=1.23$&\citet{Mason2015ApJ} & $2.43$ & $0.18$\\
\addlinespace
\midrule
\addlinespace
$z=2$&\citet{Mason2015ApJ} & $2.46$ & $0.18$\\
\addlinespace
\bottomrule
\bottomrule
\end{tabular}
     \label{tab:comparison}
\end{table}

At $z=0.2$ (upper left panel of \autoref{fig:comparison}) we compare our results with \citet{Auger2010ApJ} and \citet{HydeBernardi2009MNRAS}.
The curve of model \Mconstfidnes at $z=0.2$ intersects the linear fit of \citet{HydeBernardi2009MNRAS}, but has shallower slope, similar to the fits of \citet{Auger2010ApJ}, which however have $\approx 20\%$ higher normalisation.  At $z=0.2$ the median relation of model \Mevofidnes, not shown in the plot, is very similar to that of model \Mconstfidnes.
The steeper slope $\beta$ of the fits of \citet{HydeBernardi2009MNRAS} can be ascribed to two main reasons: they exclude the highest-mass galaxies and they do not allow for intrinsic scatter in their model. The fact that the correlation is shallower at higher $M_*$ is apparent from the shape of the quadratic fit of \citet{HydeBernardi2009MNRAS}. The difference between the two fits of \citet{Auger2010ApJ} illustrates the effect on $\beta$ of allowing for intrinsic scatter.
As a further test of the importance of considering the intrinsic scatter in the model, we applied to the fiducial sample the analysis described in \autoref{sec:method}, but assuming zero intrinsic scatter ($\sigma_\sigma=0$ in equation~\ref{eq:scatter}) for models \Mconst and \Mevo. Based on the Bayesian evidence, in this case the best model is $\mathcal{M}_\mathrm{evo,NS}^\mathit{fid}$, i.e.\ an evolving-slope model with a null scatter (NS) that can be approximately described by
\begin{equation}
    \log\left(\frac{\sigmae}{\kms}\right)\simeq 2.21+\beta(z)\left(\frac{M_*}{10^{11}\msun}\right)+0.46\log(1+z),
\end{equation}
with $\beta(z)=0.22+0.89\log(1+z)$.
This model evaluated at $z=0.2$  (dash-dotted curve in upper left panel of \autoref{fig:comparison}) has slope $\beta\simeq 0.29$ and overlaps almost perfectly with the linear fit of \citet{HydeBernardi2009MNRAS}.

In the  upper right panel of \autoref{fig:comparison}  we compare our models \Mconstfidnes and \Mevofidnes at $z=0.35$ with the fits obtained  by \citet{Mason2015ApJ} at the same redshift,  by \citet{Damjanov2018ApJS} at $0.2<z<0.5$ and  by \citet{Zahid2016ApJ} for SHELS galaxies at $0.3<z<0.4$. Taking into account the differences in the stellar-mass range, and that \citet{Damjanov2018ApJS} and \citet{Zahid2016ApJ} do not allow for the presence of intrinsic scatter, there is reasonable agreement among the five curves.

In the lower left panel of \autoref{fig:comparison} we compare our models \Mconstextnes and \Mevoextnes  at $z=1.23$ (mean redshift of the sample of \citealt{Belli2014ApJ}) with the linear fit obtained by \citet{Belli2014ApJ} and that of \citet{Mason2015ApJ} at the same redshift. Considering that  \citet{Belli2014ApJ} do not allow for the presence of intrinsic scatter, the four curves are broadly consistent.

In the lower right panel of \autoref{fig:comparison} we compare the median relations of our models \Mconstextnes and \Mevoextnes with the estimate of \citet{Mason2015ApJ} at $z=2$, finding that our relations predict a higher velocity dispersion at the same stellar mass, which is a consequence of the fact that the \citet{Mason2015ApJ} find a weaker redshift dependence of the normalisation than our models.


Overall, we do find a satisfactory agreement among our results and previous works in the literature. Some of the differences pointed out above may be ascribed to different redshift distributions of the galaxy sample, stellar mass ranges, data and models used in the measurements of the stellar masses, selection criteria or fitting methods. 
For instance, it is apparent from  \autoref{fig:comparison} that different studies consider different stellar mass intervals. Studies focusing on lower stellar masses tend to find steeper slopes than those focusing on higher stellar masses.  Furthermore, allowing for the presence of intrinsic scatter when modelling the data leads to shallower slopes.  Models allowing for the presence of intrinsic scatter, such as those presented in this work, are expected to provide a more correct description of the $M_*$-$\sigmae$ correlation.

\subsection{Connection with the size evolution of ETGs}
\label{sec:size_evolution}

It is useful discuss the results here obtained for the evolution of the $M_*$-$\sigmae$ relation of ETGs in light of the well known evolution of the $M_*$-$\Re$ relation: the redshift dependence of the median effective radius at fixed stellar mass can be parameterised as $\Re\propto (1+z)^{\aR}$.
The value of $\aR$ for ETGs appears to depend somewhat on the considered sample, mass and redshift intervals, ranging from $\aR\approx -1.5$ (\citealt{vanderWel14}; $0\lesssim z\lesssim 3$) to $\aR\approx -0.6$ (\citealt{Cimatti2012MNRAS}; $0\lesssim z\lesssim 2$).
If all the ETGs in the considered redshift range were structurally and kinematically homologous (see, e.g., section 5.4.1 of \citealt*{CFN19}), we would have $\sigmae^2\propto{M_*}/{\Re}$ and thus, at fixed stellar mass $M_*$,
$\sigmae\propto{\Re^{-1/2}}\propto (1+z)^{\zetahom}$, with $\zetahom=-\aR/2$. For $-1.5\lesssim \aR\lesssim -0.6$, we get $0.3 \lesssim \zetahom \lesssim 0.75$. This toy model is consistent with our observational finding $\sigmae\propto (1+z)^{\zeta}$ with $0.4\lesssim \zeta \lesssim 0.5$. 

It must be stressed that the observed value of $\zeta$ must not be necessarily equal to $\zetahom$. From a theoretical point of view, an observed evolution in $\sigmae$ different than predicted by the above toy model can be expected if ETGs do not evolve maintaining homology. For instance, dry merging, which is one of the processes believed to be responsible for the size and velocity dispersion evolution of ETGs (see \autoref{sec:intro}), is known to produce non-homology, because it varies the
shape and the kinematics of the stellar distribution, and the mutual density distributions of luminous and dark matter  \citep{Nipoti2003MNRAS,Hil13,Fri17}.
We can quantify the effect of non-homology by defining the dimensionless parameter
\begin{equation}
    \kstar\equiv \frac{G M_*}{ \sigmae^2\Re},
    \label{eq:kvir}
\end{equation}
such that galaxies that are structurally and kinematically homologous have the same value of $\kstar$. If, at fixed $M_*$, $\Re\propto(1+z)^{\aR}$ and $\sigmae\propto(1+z)^\zeta$, the average value of $\kstar$ must vary with redshift as
$\kstar\propto(1+z)^{\ak}$ with $\ak=-(2\zeta+\aR)$. Thus, we have $\zeta\neq -\aR/2$ if $\ak\neq 0$, i.e.\ if, on average, galaxies at different redshift have different $\kstar$.
However, from an observational point of view, a significant evolution of $\kstar$ seems to be excluded.
Defining the dynamical mass $\Mdyn\equiv 5 \sigmae^2 \Re/G$, 
the average ratio $M_*/\Mdyn\propto \kstar$ is found to increase mildly  with redshift (or even remain constant; \citealt{vandeSande2013ApJ,Belli2014ApJ}),
and the zero point of the stellar-mass fundamental plane  (which also can be seen as a measure of the average $\kstar$) varies only little with redshift \citep{Bezanson2013ApJ,Zahid2016ApJ821}.

\section{Conclusions}
\label{sec:conclusions}

We have studied the evolution of the correlation between central stellar velocity dispersion $\sigmae$ (measured within $R_\mathrm{e}$) and stellar mass $M_*$ for massive  ($M_*\gtrsim 10^{10.5}\,\mathrm{M_\odot}$) ETGs observed in the redshift range $0\lesssim z\lesssim 2.5$. We have modelled the evolution of this scaling law using a Bayesian hierarchical method.
This allowed us to optimally exploit the available observational data, without resorting to binning in either redshift or stellar-mass space. The main conclusions of this work are the following.
\begin{itemize}
    \item On average, the central velocity dispersion  of massive ($M_*\gtrsim 10^{10.5}\,\mathrm{M_\odot}$) ETGs increases with stellar mass following a power-law relation $\sigmae\propto M_*^\beta$ with either $\beta\simeq0.18$, independent of redshift, or $\beta$ increasing with redshift as $\beta\simeq 0.16+0.26\log(1+z)$ in the redshift range $0\lesssim z\lesssim 1$ probed by our fiducial sample.
    
    \item The normalisation of the $M_*$-$\sigmae$ relation increases with redshift: for instance,  when $\beta\simeq 0.18$ independent of redshift,  at fixed stellar mass $\sigmae\propto (1+z)^\zeta$ with $\zeta\simeq 0.4$ out to $z\approx 1$. In other words, a typical ETG of $M_*\approx 10^{11}\,\msun$ at $z\approx 0$ has $\sigmae$ lower by a factor $\approx 1.3$ than ETGs of similar stellar mass at $z\approx 1$.
    
    \item The intrinsic scatter of the $M_*$-$\sigmae$ relation is $\simeq0.08$ dex in $\sigmae$ at given $M_*$, independent of redshift.
    
    \item Over the wider redshift range $0\lesssim z\lesssim 2.5$, probed by our extended sample, we find results similar to those found for the $z\lesssim 1$ fiducial sample, with slightly stronger redshift dependence of the normalisation ($\zeta\simeq 0.5$) and weaker redshift dependence of the slope (${\rm d} \beta/{\rm d} \log (1+z)\simeq 0.18$) when $\beta$ varies with time.  On average, the velocity dispersion of ETGs of $M_*\approx 10^{11}\,\msun$ at $z=2$ is a factor of $\approx1.7$ higher than that of $z=0$ ETGs of similar stellar mass.
    
 \end{itemize}
 
The results of this work confirm and strengthen previous indications that the $M_*$-$\sigmae$ relation of massive ETGs evolves with cosmic time. The theoretical interpretation of the observed evolution is not straightforward. Of course, the stellar mass of an individual galaxy can vary with time: it can increase as a consequence of mergers and star formation and decrease as a consequence of mass return by ageing stellar populations. In the standard paradigm, the first effect is dominant, so we expect that, as cosmic time  goes on, an individual galaxy moves in the $M_*$-$\sigmae$ plane in the direction of increasing $M_*$. As pointed out in \autoref{sec:intro}, the variation of $\sigmae$ for an individual galaxy is more uncertain: even pure dry mergers can make it increase or decrease depending on the merging orbital parameters and mass ratio. It is then clear that, at least qualitatively, the evolution shown in \autoref{fig:mstar_sigma_relations} could be reproduced by individual galaxies evolving at decreasing $\sigmae$, but, at least at the low-mass end, even an evolution of individual galaxies at constant or slightly increasing $\sigmae$ is not excluded. Remarkably, our results suggest that, on average, the stellar velocity dispersion of individual galaxies with $M_*\gtrsim 3\times 10^{11}\msun$ at $z\approx 1$ must decrease from $z\approx 1$ to $z\approx 0$ for them to end up on the median present-day $M_*$-$\sigmae$ relation.

An additional complication to the theoretical interpretation of the evolution of the scaling laws of ETGs is that it is not guaranteed that the high-$z$ (say $z\approx 2$) ETGs are representative of the progenitors of all present-day ETGs. If the progenitors of some of the present-day ETGs were star-forming at $z\approx 2$, they would not be included in our sample of $z\approx 2$ ETGs: this is the so-called \emph{progenitor bias}, which must be accounted for when interpreting the evolution of a population of objects.
However, the effect of progenitor bias should be small at least for the most massive ETGs in the redshift range  $0\lesssim z \lesssim 1$, in which the number density of  quiescent galaxies shows little evolution \citep{Lop12}.

The theoretical interpretation of the evolution of the scaling relations of ETGs can benefit from the comparison of the observational data with the results of cosmological simulations of galaxy formation. In this approach, the progenitor bias can be taken into account automatically if simulated and observed galaxies are selected with consistent criteria. Moreover, in the simulations we can trace the evolution of individual galaxies, which is a crucial piece of information that we do not have for individual observed galaxies. The method presented in this paper is suitable to be applied to samples of simulated as well as observed galaxies. In the near future we plan to apply this method to compare the observed evolution of the $M_*$-$\sigmae$ relation of ETGs with the results of state-of-the-art cosmological simulations of galaxy formation.

\section*{Acknowledgements}
We are grateful to S.\ Belli and M. Maseda for sharing data and for helpful suggestions, and to 
A.\ Cimatti, I.\ Damjanov, S.\ Faber, M.\ Moresco and B.\ Nipoti for useful discussions. We thank the anonymous referees for their advice and comments, which considerably contributed to improve the quality of the manuscript. AS acknowledges funding from the European Union's Horizon 2020 research and innovation programme under grand agreement No.\ 792916.

\section*{Data Availability}
The data underlying this article will be shared on reasonable request to the corresponding author.



\bibliographystyle{mnras}
\bibliography{bibliography} 

\begin{thebibliography}{}
\makeatletter
\relax
\def\mn@urlcharsother{\let\do\@makeother \do\$\do\&\do\#\do\^\do\_\do\%\do\~}
\def\mn@doi{\begingroup\mn@urlcharsother \@ifnextchar [ {\mn@doi@}
  {\mn@doi@[]}}
\def\mn@doi@[#1]#2{\def\@tempa{#1}\ifx\@tempa\@empty \href
  {http://dx.doi.org/#2} {doi:#2}\else \href {http://dx.doi.org/#2} {#1}\fi
  \endgroup}
\def\mn@eprint#1#2{\mn@eprint@#1:#2::\@nil}
\def\mn@eprint@arXiv#1{\href {http://arxiv.org/abs/#1} {{\tt arXiv:#1}}}
\def\mn@eprint@dblp#1{\href {http://dblp.uni-trier.de/rec/bibtex/#1.xml}
  {dblp:#1}}
\def\mn@eprint@#1:#2:#3:#4\@nil{\def\@tempa {#1}\def\@tempb {#2}\def\@tempc
  {#3}\ifx \@tempc \@empty \let \@tempc \@tempb \let \@tempb \@tempa \fi \ifx
  \@tempb \@empty \def\@tempb {arXiv}\fi \@ifundefined
  {mn@eprint@\@tempb}{\@tempb:\@tempc}{\expandafter \expandafter \csname
  mn@eprint@\@tempb\endcsname \expandafter{\@tempc}}}

\bibitem[\protect\citeauthoryear{{Adelman-McCarthy} et~al.,}{{Adelman-McCarthy}
  et~al.}{2008}]{Adelman-McCarthy2008ApJS}
{Adelman-McCarthy} J.~K.,  et~al., 2008, \mn@doi [\apjs] {10.1086/524984},
  \href {https://ui.adsabs.harvard.edu/abs/2008ApJS..175..297A} {175, 297}

\bibitem[\protect\citeauthoryear{{Aihara} et~al.,}{{Aihara}
  et~al.}{2018}]{Aihara2018PASJ}
{Aihara} H.,  et~al., 2018, \mn@doi [\pasj] {10.1093/pasj/psx081}, \href
  {http://adsabs.harvard.edu/abs/2018PASJ...70S...8A} {70, S8}

\bibitem[\protect\citeauthoryear{{Aihara} et~al.,}{{Aihara}
  et~al.}{2019}]{Aihara2019}
{Aihara} H.,  et~al., 2019, \mn@doi [\pasj] {10.1093/pasj/psz103}, \href
  {https://ui.adsabs.harvard.edu/abs/2019PASJ...71..114A} {71, 114}

\bibitem[\protect\citeauthoryear{{Alam} et~al.,}{{Alam}
  et~al.}{2015}]{Alam2015ApJS}
{Alam} S.,  et~al., 2015, \mn@doi [\apjs] {10.1088/0067-0049/219/1/12}, \href
  {https://ui.adsabs.harvard.edu/abs/2015ApJS..219...12A} {219, 12}

\bibitem[\protect\citeauthoryear{{Auger}, {Treu}, {Bolton}, {Gavazzi},
  {Koopmans}, {Marshall}, {Bundy}  \& {Moustakas}}{{Auger}
  et~al.}{2009}]{Auger2009ApJ}
{Auger} M.~W.,  {Treu} T.,  {Bolton} A.~S.,  {Gavazzi} R.,  {Koopmans}
  L.~V.~E.,  {Marshall} P.~J.,  {Bundy} K.,   {Moustakas} L.~A.,  2009, \mn@doi
  [\apj] {10.1088/0004-637X/705/2/1099}, \href
  {http://adsabs.harvard.edu/abs/2009ApJ...705.1099A} {705, 1099}

\bibitem[\protect\citeauthoryear{{Auger}, {Treu}, {Bolton}, {Gavazzi},
  {Koopmans}, {Marshall}, {Moustakas}  \& {Burles}}{{Auger}
  et~al.}{2010}]{Auger2010ApJ}
{Auger} M.~W.,  {Treu} T.,  {Bolton} A.~S.,  {Gavazzi} R.,  {Koopmans}
  L.~V.~E.,  {Marshall} P.~J.,  {Moustakas} L.~A.,   {Burles} S.,  2010,
  \mn@doi [\apj] {10.1088/0004-637X/724/1/511}, \href
  {https://ui.adsabs.harvard.edu/abs/2010ApJ...724..511A} {724, 511}

\bibitem[\protect\citeauthoryear{{Belli}, {Newman}  \& {Ellis}}{{Belli}
  et~al.}{2014a}]{Belli2014ApJ}
{Belli} S.,  {Newman} A.~B.,   {Ellis} R.~S.,  2014a, \mn@doi [\apj]
  {10.1088/0004-637X/783/2/117}, \href
  {http://adsabs.harvard.edu/abs/2014ApJ...783..117B} {783, 117}

\bibitem[\protect\citeauthoryear{{Belli}, {Newman}, {Ellis}  \&
  {Konidaris}}{{Belli} et~al.}{2014b}]{Belli2014bApJ}
{Belli} S.,  {Newman} A.~B.,  {Ellis} R.~S.,   {Konidaris} N.~P.,  2014b,
  \mn@doi [\apj] {10.1088/2041-8205/788/2/L29}, \href
  {https://ui.adsabs.harvard.edu/\#abs/2014ApJ...788L..29B} {788, L29}

\bibitem[\protect\citeauthoryear{{Belli}, {Newman}  \& {Ellis}}{{Belli}
  et~al.}{2017}]{Belli2017ApJ}
{Belli} S.,  {Newman} A.~B.,   {Ellis} R.~S.,  2017, \mn@doi [\apj]
  {10.3847/1538-4357/834/1/18}, \href
  {http://adsabs.harvard.edu/abs/2017ApJ...834...18B} {834, 18}

\bibitem[\protect\citeauthoryear{{Bernardi}, {Sheth}, {Dominguez-Sanchez},
  {Fischer}, {Chae}, {Huertas-Company}  \& {Shankar}}{{Bernardi}
  et~al.}{2018}]{Ber++18}
{Bernardi} M.,  {Sheth} R.~K.,  {Dominguez-Sanchez} H.,  {Fischer} J.~L.,
  {Chae} K.~H.,  {Huertas-Company} M.,   {Shankar} F.,  2018, \mn@doi [\mnras]
  {10.1093/mnras/sty781}, \href
  {https://ui.adsabs.harvard.edu/abs/2018MNRAS.477.2560B} {477, 2560}

\bibitem[\protect\citeauthoryear{{Bertin} \& {Arnouts}}{{Bertin} \&
  {Arnouts}}{1996}]{BertinArnouts1996A&AS}
{Bertin} E.,  {Arnouts} S.,  1996, \mn@doi [\aaps] {10.1051/aas:1996164}, \href
  {http://adsabs.harvard.edu/abs/1996A\%26AS..117..393B} {117, 393}

\bibitem[\protect\citeauthoryear{{Bezanson}, {van Dokkum}, {van de Sande},
  {Franx}  \& {Kriek}}{{Bezanson} et~al.}{2013a}]{Bezanson2013ApJ}
{Bezanson} R.,  {van Dokkum} P.,  {van de Sande} J.,  {Franx} M.,   {Kriek} M.,
   2013a, \mn@doi [\apjl] {10.1088/2041-8205/764/1/L8}, \href
  {http://adsabs.harvard.edu/abs/2013ApJ...764L...8B} {764, L8}

\bibitem[\protect\citeauthoryear{{Bezanson}, {van Dokkum}, {van de Sande},
  {Franx}, {Leja}  \& {Kriek}}{{Bezanson} et~al.}{2013b}]{Bezanson2013bApJl}
{Bezanson} R.,  {van Dokkum} P.~G.,  {van de Sande} J.,  {Franx} M.,  {Leja}
  J.,   {Kriek} M.,  2013b, \mn@doi [\apjl] {10.1088/2041-8205/779/2/L21},
  \href {http://adsabs.harvard.edu/abs/2013ApJ...779L..21B} {779, L21}

\bibitem[\protect\citeauthoryear{{Bezanson}, {Franx}  \& {van
  Dokkum}}{{Bezanson} et~al.}{2015}]{Bezanson2015ApJ}
{Bezanson} R.,  {Franx} M.,   {van Dokkum} P.~G.,  2015, \mn@doi [\apj]
  {10.1088/0004-637X/799/2/148}, \href
  {http://adsabs.harvard.edu/abs/2015ApJ...799..148B} {799, 148}

\bibitem[\protect\citeauthoryear{{Bolton}, {Treu}, {Koopmans}, {Gavazzi},
  {Moustakas}, {Burles}, {Schlegel}  \& {Wayth}}{{Bolton}
  et~al.}{2008}]{Bolton2008ApJ}
{Bolton} A.~S.,  {Treu} T.,  {Koopmans} L.~V.~E.,  {Gavazzi} R.,  {Moustakas}
  L.~A.,  {Burles} S.,  {Schlegel} D.~J.,   {Wayth} R.,  2008, \mn@doi [\apj]
  {10.1086/589989}, \href
  {https://ui.adsabs.harvard.edu/abs/2008ApJ...684..248B} {684, 248}

\bibitem[\protect\citeauthoryear{{Boylan-Kolchin}, {Ma}  \&
  {Quataert}}{{Boylan-Kolchin} et~al.}{2006}]{Boy06}
{Boylan-Kolchin} M.,  {Ma} C.-P.,   {Quataert} E.,  2006, \mn@doi [\mnras]
  {10.1111/j.1365-2966.2006.10379.x}, \href
  {https://ui.adsabs.harvard.edu/abs/2006MNRAS.369.1081B} {369, 1081}

\bibitem[\protect\citeauthoryear{{Bruzual} \& {Charlot}}{{Bruzual} \&
  {Charlot}}{2003}]{BruzualCharlot2003MNRAS}
{Bruzual} G.,  {Charlot} S.,  2003, \mn@doi [\mnras]
  {10.1046/j.1365-8711.2003.06897.x}, \href
  {http://adsabs.harvard.edu/abs/2003MNRAS.344.1000B} {344, 1000}

\bibitem[\protect\citeauthoryear{{Buchner} et~al.,}{{Buchner}
  et~al.}{2014}]{Buchner2014A&A}
{Buchner} J.,  et~al., 2014, \mn@doi [\aap] {10.1051/0004-6361/201322971},
  \href {https://ui.adsabs.harvard.edu/abs/2014A&A...564A.125B} {564, A125}

\bibitem[\protect\citeauthoryear{{Cannarozzo}, {Nipoti}, {Sonnenfeld},
  {Leauthaud}, {Huang}, {Diemer}  \& {Oyarz{\'u}n}}{{Cannarozzo}
  et~al.}{2020}]{Can20}
{Cannarozzo} C.,  {Nipoti} C.,  {Sonnenfeld} A.,  {Leauthaud} A.,  {Huang} S.,
  {Diemer} B.,   {Oyarz{\'u}n} G.,  2020, arXiv e-prints, \href
  {https://ui.adsabs.harvard.edu/abs/2020arXiv200605427C} {p. arXiv:2006.05427}

\bibitem[\protect\citeauthoryear{{Cappellari} et~al.,}{{Cappellari}
  et~al.}{2006}]{Cappellari2006MNRAS}
{Cappellari} M.,  et~al., 2006, \mn@doi [\mnras]
  {10.1111/j.1365-2966.2005.09981.x}, \href
  {http://adsabs.harvard.edu/abs/2006MNRAS.366.1126C} {366, 1126}

\bibitem[\protect\citeauthoryear{{Chabrier}}{{Chabrier}}{2003}]{Chabrier2003PASP}
{Chabrier} G.,  2003, \mn@doi [\pasp] {10.1086/376392}, \href
  {http://adsabs.harvard.edu/abs/2003PASP..115..763C} {115, 763}

\bibitem[\protect\citeauthoryear{{Cimatti}, {Nipoti}  \& {Cassata}}{{Cimatti}
  et~al.}{2012}]{Cimatti2012MNRAS}
{Cimatti} A.,  {Nipoti} C.,   {Cassata} P.,  2012, \mn@doi [\mnras]
  {10.1111/j.1745-3933.2012.01237.x}, \href
  {http://adsabs.harvard.edu/abs/2012MNRAS.422L..62C} {422, L62}

\bibitem[\protect\citeauthoryear{{Cimatti}, {Fraternali}  \&
  {Nipoti}}{{Cimatti} et~al.}{2019}]{CFN19}
{Cimatti} A.,  {Fraternali} F.,   {Nipoti} C.,  2019, {Introduction to galaxy
  formation and evolution. From primordial gas to present-day galaxies}.
Cambridge University Press

\bibitem[\protect\citeauthoryear{{Ciotti}, {Lanzoni}  \& {Volonteri}}{{Ciotti}
  et~al.}{2007}]{Ciotti2007ApJ}
{Ciotti} L.,  {Lanzoni} B.,   {Volonteri} M.,  2007, \mn@doi [\apj]
  {10.1086/510773}, \href {http://adsabs.harvard.edu/abs/2007ApJ...658...65C}
  {658, 65}

\bibitem[\protect\citeauthoryear{{Conroy}}{{Conroy}}{2013}]{Con++13}
{Conroy} C.,  2013, \mn@doi [\araa] {10.1146/annurev-astro-082812-141017},
  \href {https://ui.adsabs.harvard.edu/abs/2013ARA&A..51..393C} {51, 393}

\bibitem[\protect\citeauthoryear{{Damjanov}, {Zahid}, {Geller}, {Fabricant}  \&
  {Hwang}}{{Damjanov} et~al.}{2018}]{Damjanov2018ApJS}
{Damjanov} I.,  {Zahid} H.~J.,  {Geller} M.~J.,  {Fabricant} D.~G.,   {Hwang}
  H.~S.,  2018, \mn@doi [\apjs] {10.3847/1538-4365/aaa01c}, \href
  {https://ui.adsabs.harvard.edu/abs/2018ApJS..234...21D} {234, 21}

\bibitem[\protect\citeauthoryear{{Damjanov}, {Zahid}, {Geller}, {Utsumi},
  {Sohn}  \& {Souchereau}}{{Damjanov} et~al.}{2019}]{Dam19}
{Damjanov} I.,  {Zahid} H.~J.,  {Geller} M.~J.,  {Utsumi} Y.,  {Sohn} J.,
  {Souchereau} H.,  2019, \mn@doi [\apj] {10.3847/1538-4357/aaf97d}, \href
  {https://ui.adsabs.harvard.edu/abs/2019ApJ...872...91D} {872, 91}

\bibitem[\protect\citeauthoryear{{Djorgovski} \& {Davis}}{{Djorgovski} \&
  {Davis}}{1987}]{DjorgovskiDavis1987ApJ}
{Djorgovski} S.,  {Davis} M.,  1987, \mn@doi [\apj] {10.1086/164948}, \href
  {http://adsabs.harvard.edu/abs/1987ApJ...313...59D} {313, 59}

\bibitem[\protect\citeauthoryear{{Dom{\'\i}nguez S{\'a}nchez},
  {Huertas-Company}, {Bernardi}, {Tuccillo}  \& {Fischer}}{{Dom{\'\i}nguez
  S{\'a}nchez} et~al.}{2018}]{Dom18}
{Dom{\'\i}nguez S{\'a}nchez} H.,  {Huertas-Company} M.,  {Bernardi} M.,
  {Tuccillo} D.,   {Fischer} J.~L.,  2018, \mn@doi [\mnras]
  {10.1093/mnras/sty338}, \href
  {https://ui.adsabs.harvard.edu/abs/2018MNRAS.476.3661D} {476, 3661}

\bibitem[\protect\citeauthoryear{{Dom{\'\i}nguez S{\'a}nchez}, {Bernardi},
  {Brownstein}, {Drory}  \& {Sheth}}{{Dom{\'\i}nguez S{\'a}nchez}
  et~al.}{2019}]{Dom++19}
{Dom{\'\i}nguez S{\'a}nchez} H.,  {Bernardi} M.,  {Brownstein} J.~R.,  {Drory}
  N.,   {Sheth} R.~K.,  2019, \mn@doi [\mnras] {10.1093/mnras/stz2414}, \href
  {https://ui.adsabs.harvard.edu/abs/2019MNRAS.489.5612D} {489, 5612}

\bibitem[\protect\citeauthoryear{{Dressler}, {Lynden-Bell}, {Burstein},
  {Davies}, {Faber}, {Terlevich}  \& {Wegner}}{{Dressler}
  et~al.}{1987}]{Dressler1987ApJ}
{Dressler} A.,  {Lynden-Bell} D.,  {Burstein} D.,  {Davies} R.~L.,  {Faber}
  S.~M.,  {Terlevich} R.,   {Wegner} G.,  1987, \mn@doi [\apj]
  {10.1086/164947}, \href {http://adsabs.harvard.edu/abs/1987ApJ...313...42D}
  {313, 42}

\bibitem[\protect\citeauthoryear{{Eddington}}{{Eddington}}{1913}]{Eddington13}
{Eddington} A.~S.,  1913, \mn@doi [\mnras] {10.1093/mnras/73.5.359}, \href
  {http://adsabs.harvard.edu/abs/1913MNRAS..73..359E} {73, 359}

\bibitem[\protect\citeauthoryear{{Eisenstein} et~al.,}{{Eisenstein}
  et~al.}{2011}]{Eisenstein2011AJ}
{Eisenstein} D.~J.,  et~al., 2011, \mn@doi [\aj] {10.1088/0004-6256/142/3/72},
  \href {https://ui.adsabs.harvard.edu/abs/2011AJ....142...72E} {142, 72}

\bibitem[\protect\citeauthoryear{{Faber} \& {Jackson}}{{Faber} \&
  {Jackson}}{1976}]{FaberJackson1976ApJ}
{Faber} S.~M.,  {Jackson} R.~E.,  1976, \mn@doi [\apj] {10.1086/154215}, \href
  {http://adsabs.harvard.edu/abs/1976ApJ...204..668F} {204, 668}

\bibitem[\protect\citeauthoryear{{Fagotto}, {Bressan}, {Bertelli}  \&
  {Chiosi}}{{Fagotto} et~al.}{1994a}]{Fagotto1994aA&AS}
{Fagotto} F.,  {Bressan} A.,  {Bertelli} G.,   {Chiosi} C.,  1994a, \aaps,
  \href {https://ui.adsabs.harvard.edu/abs/1994A&AS..104..365F} {104, 365}

\bibitem[\protect\citeauthoryear{{Fagotto}, {Bressan}, {Bertelli}  \&
  {Chiosi}}{{Fagotto} et~al.}{1994b}]{Fagotto1994bA&AS}
{Fagotto} F.,  {Bressan} A.,  {Bertelli} G.,   {Chiosi} C.,  1994b, \aaps,
  \href {https://ui.adsabs.harvard.edu/abs/1994A&AS..105...29F} {105, 29}

\bibitem[\protect\citeauthoryear{{Fagotto}, {Bressan}, {Bertelli}  \&
  {Chiosi}}{{Fagotto} et~al.}{1994c}]{Fagotto1994cA&AS}
{Fagotto} F.,  {Bressan} A.,  {Bertelli} G.,   {Chiosi} C.,  1994c, \aaps,
  \href {https://ui.adsabs.harvard.edu/abs/1994A&AS..105...29F} {105, 29}

\bibitem[\protect\citeauthoryear{{Ferguson} et~al.,}{{Ferguson}
  et~al.}{2004}]{Fer04}
{Ferguson} H.~C.,  et~al., 2004, \mn@doi [\apj] {10.1086/378578}, \href
  {https://ui.adsabs.harvard.edu/abs/2004ApJ...600L.107F} {600, L107}

\bibitem[\protect\citeauthoryear{{Feroz} \& {Hobson}}{{Feroz} \&
  {Hobson}}{2008}]{FerozHobson2008MNRAS}
{Feroz} F.,  {Hobson} M.~P.,  2008, \mn@doi [\mnras]
  {10.1111/j.1365-2966.2007.12353.x}, \href
  {http://adsabs.harvard.edu/abs/2008MNRAS.384..449F} {384, 449}

\bibitem[\protect\citeauthoryear{{Feroz}, {Hobson}  \& {Bridges}}{{Feroz}
  et~al.}{2009}]{FerozHobsonBridges2009MNRAS}
{Feroz} F.,  {Hobson} M.~P.,   {Bridges} M.,  2009, \mn@doi [\mnras]
  {10.1111/j.1365-2966.2009.14548.x}, \href
  {http://adsabs.harvard.edu/abs/2009MNRAS.398.1601F} {398, 1601}

\bibitem[\protect\citeauthoryear{{Foreman-Mackey}, {Hogg}, {Lang}  \&
  {Goodman}}{{Foreman-Mackey} et~al.}{2013}]{Foreman-Mackey2013PASP}
{Foreman-Mackey} D.,  {Hogg} D.~W.,  {Lang} D.,   {Goodman} J.,  2013, \mn@doi
  [\pasp] {10.1086/670067}, \href
  {http://adsabs.harvard.edu/abs/2013PASP..125..306F} {125, 306}

\bibitem[\protect\citeauthoryear{{Frigo} \& {Balcells}}{{Frigo} \&
  {Balcells}}{2017}]{Fri17}
{Frigo} M.,  {Balcells} M.,  2017, \mn@doi [\mnras] {10.1093/mnras/stx875},
  \href {https://ui.adsabs.harvard.edu/abs/2017MNRAS.469.2184F} {469, 2184}

\bibitem[\protect\citeauthoryear{{Gallazzi}, {Charlot}, {Brinchmann}, {White}
  \& {Tremonti}}{{Gallazzi} et~al.}{2005}]{Gal++05}
{Gallazzi} A.,  {Charlot} S.,  {Brinchmann} J.,  {White} S.~D.~M.,   {Tremonti}
  C.~A.,  2005, \mn@doi [\mnras] {10.1111/j.1365-2966.2005.09321.x}, \href
  {http://adsabs.harvard.edu/abs/2005MNRAS.362...41G} {362, 41}

\bibitem[\protect\citeauthoryear{{Gargiulo}, {Saracco}, {Longhetti},
  {Tamburri}, {Lonoce}  \& {Ciocca}}{{Gargiulo} et~al.}{2015}]{Gargiulo2015A&A}
{Gargiulo} A.,  {Saracco} P.,  {Longhetti} M.,  {Tamburri} S.,  {Lonoce} I.,
  {Ciocca} F.,  2015, \mn@doi [\aap] {10.1051/0004-6361/201424235}, \href
  {http://adsabs.harvard.edu/abs/2015A%26A...573A.110G} {573, A110}

\bibitem[\protect\citeauthoryear{{Gargiulo}, {Saracco}, {Tamburri}, {Lonoce}
  \& {Ciocca}}{{Gargiulo} et~al.}{2016}]{Gargiulo2016AAP}
{Gargiulo} A.,  {Saracco} P.,  {Tamburri} S.,  {Lonoce} I.,   {Ciocca} F.,
  2016, \mn@doi [\aap] {10.1051/0004-6361/201526563}, \href
  {https://ui.adsabs.harvard.edu/abs/2016A&A...592A.132G} {592, A132}

\bibitem[\protect\citeauthoryear{{Geller}, {Dell'Antonio}, {Kurtz}, {Ramella},
  {Fabricant}, {Caldwell}, {Tyson}  \& {Wittman}}{{Geller}
  et~al.}{2005}]{Geller2005ApJ}
{Geller} M.~J.,  {Dell'Antonio} I.~P.,  {Kurtz} M.~J.,  {Ramella} M.,
  {Fabricant} D.~G.,  {Caldwell} N.,  {Tyson} J.~A.,   {Wittman} D.,  2005,
  \mn@doi [\apj] {10.1086/499399}, \href
  {https://ui.adsabs.harvard.edu/abs/2005ApJ...635L.125G} {635, L125}

\bibitem[\protect\citeauthoryear{{Goodman} \& {Weare}}{{Goodman} \&
  {Weare}}{2010}]{GW2010}
{Goodman} J.,  {Weare} J.,  2010, \mn@doi [Communications in Applied
  Mathematics and Computational Science] {10.2140/camcos.2010.5.65}, \href
  {https://ui.adsabs.harvard.edu/abs/2010CAMCS...5...65G} {5, 65}

\bibitem[\protect\citeauthoryear{{Hilz}, {Naab}  \& {Ostriker}}{{Hilz}
  et~al.}{2013}]{Hil13}
{Hilz} M.,  {Naab} T.,   {Ostriker} J.~P.,  2013, \mn@doi [\mnras]
  {10.1093/mnras/sts501}, \href
  {https://ui.adsabs.harvard.edu/abs/2013MNRAS.429.2924H} {429, 2924}

\bibitem[\protect\citeauthoryear{{Huang}, {Leauthaud}, {Greene}, {Bundy},
  {Lin}, {Tanaka}, {Miyazaki}  \& {Komiyama}}{{Huang} et~al.}{2018}]{Huang2018}
{Huang} S.,  {Leauthaud} A.,  {Greene} J.~E.,  {Bundy} K.,  {Lin} Y.-T.,
  {Tanaka} M.,  {Miyazaki} S.,   {Komiyama} Y.,  2018, \mn@doi [\mnras]
  {10.1093/mnras/stx3200}, \href
  {https://ui.adsabs.harvard.edu/abs/2018MNRAS.475.3348H} {475, 3348}

\bibitem[\protect\citeauthoryear{{Hyde} \& {Bernardi}}{{Hyde} \&
  {Bernardi}}{2009a}]{HydeBernardi2009MNRAS}
{Hyde} J.~B.,  {Bernardi} M.,  2009a, \mn@doi [\mnras]
  {10.1111/j.1365-2966.2009.14445.x}, \href
  {https://ui.adsabs.harvard.edu/abs/2009MNRAS.394.1978H} {394, 1978}

\bibitem[\protect\citeauthoryear{{Hyde} \& {Bernardi}}{{Hyde} \&
  {Bernardi}}{2009b}]{Hyd09b}
{Hyde} J.~B.,  {Bernardi} M.,  2009b, \mn@doi [\mnras]
  {10.1111/j.1365-2966.2009.14783.x}, \href
  {https://ui.adsabs.harvard.edu/abs/2009MNRAS.396.1171H} {396, 1171}

\bibitem[\protect\citeauthoryear{Jeffreys}{Jeffreys}{1961}]{Jeffreys1961ToP}
Jeffreys H.,  1961, Theory of Probability, third edn.
Oxford, Oxford, England

\bibitem[\protect\citeauthoryear{{Kormendy}}{{Kormendy}}{1977}]{Kormendy1977ApJ}
{Kormendy} J.,  1977, \mn@doi [\apj] {10.1086/155687}, \href
  {http://adsabs.harvard.edu/abs/1977ApJ...218..333K} {218, 333}

\bibitem[\protect\citeauthoryear{{Le F{\`e}vre} et~al.,}{{Le F{\`e}vre}
  et~al.}{2003}]{LeF++03}
{Le F{\`e}vre} O.,  et~al., 2003, in {Iye} M.,  {Moorwood} A. F.~M.,  eds,
  Society of Photo-Optical Instrumentation Engineers (SPIE) Conference Series
  Vol. 4841, Instrument Design and Performance for Optical/Infrared
  Ground-based Telescopes. pp 1670--1681, \mn@doi{10.1117/12.460959}

\bibitem[\protect\citeauthoryear{{Li} et~al.,}{{Li} et~al.}{2017}]{Li++17}
{Li} H.,  et~al., 2017, \mn@doi [\apj] {10.3847/1538-4357/aa662a}, \href
  {https://ui.adsabs.harvard.edu/abs/2017ApJ...838...77L} {838, 77}

\bibitem[\protect\citeauthoryear{{L{\'o}pez-Sanjuan}
  et~al.,}{{L{\'o}pez-Sanjuan} et~al.}{2012}]{Lop12}
{L{\'o}pez-Sanjuan} C.,  et~al., 2012, \mn@doi [\aap]
  {10.1051/0004-6361/201219085}, \href
  {https://ui.adsabs.harvard.edu/abs/2012A&A...548A...7L} {548, A7}

\bibitem[\protect\citeauthoryear{{Mason} et~al.,}{{Mason}
  et~al.}{2015}]{Mason2015ApJ}
{Mason} C.~A.,  et~al., 2015, \mn@doi [\apj] {10.1088/0004-637X/805/1/79},
  \href {https://ui.adsabs.harvard.edu/abs/2015ApJ...805...79M} {805, 79}

\bibitem[\protect\citeauthoryear{{Meert}, {Vikram}  \& {Bernardi}}{{Meert}
  et~al.}{2015}]{Meert2015MNRAS}
{Meert} A.,  {Vikram} V.,   {Bernardi} M.,  2015, \mn@doi [\mnras]
  {10.1093/mnras/stu2333}, \href
  {http://adsabs.harvard.edu/abs/2015MNRAS.446.3943M} {446, 3943}

\bibitem[\protect\citeauthoryear{{Mendel}, {Simard}, {Palmer}, {Ellison}  \&
  {Patton}}{{Mendel} et~al.}{2014}]{Mendel2014ApJS}
{Mendel} J.~T.,  {Simard} L.,  {Palmer} M.,  {Ellison} S.~L.,   {Patton} D.~R.,
   2014, \mn@doi [\apjs] {10.1088/0067-0049/210/1/3}, \href
  {http://adsabs.harvard.edu/abs/2014ApJS..210....3M} {210, 3}

\bibitem[\protect\citeauthoryear{{Miyazaki} et~al.,}{{Miyazaki}
  et~al.}{2018}]{Miy++18}
{Miyazaki} S.,  et~al., 2018, \mn@doi [\pasj] {10.1093/pasj/psx063}, \href
  {https://ui.adsabs.harvard.edu/abs/2018PASJ...70S...1M} {70, S1}

\bibitem[\protect\citeauthoryear{{Moresco} et~al.,}{{Moresco}
  et~al.}{2013}]{Mor++13}
{Moresco} M.,  et~al., 2013, \mn@doi [\aap] {10.1051/0004-6361/201321797},
  \href {https://ui.adsabs.harvard.edu/abs/2013A&A...558A..61M} {558, A61}

\bibitem[\protect\citeauthoryear{{Muzzin} et~al.,}{{Muzzin}
  et~al.}{2013}]{Muzzin2013ApJ}
{Muzzin} A.,  et~al., 2013, \mn@doi [\apj] {10.1088/0004-637X/777/1/18}, \href
  {http://adsabs.harvard.edu/abs/2013ApJ...777...18M} {777, 18}

\bibitem[\protect\citeauthoryear{{Naab} \& {Ostriker}}{{Naab} \&
  {Ostriker}}{2017}]{Naa17}
{Naab} T.,  {Ostriker} J.~P.,  2017, \mn@doi [\araa]
  {10.1146/annurev-astro-081913-040019}, \href
  {https://ui.adsabs.harvard.edu/abs/2017ARA&A..55...59N} {55, 59}

\bibitem[\protect\citeauthoryear{{Naab}, {Johansson}  \& {Ostriker}}{{Naab}
  et~al.}{2009}]{Naab2009ApJl}
{Naab} T.,  {Johansson} P.~H.,   {Ostriker} J.~P.,  2009, \mn@doi [\apjl]
  {10.1088/0004-637X/699/2/L178}, \href
  {http://adsabs.harvard.edu/abs/2009ApJ...699L.178N} {699, L178}

\bibitem[\protect\citeauthoryear{{Nipoti}, {Londrillo}  \& {Ciotti}}{{Nipoti}
  et~al.}{2003}]{Nipoti2003MNRAS}
{Nipoti} C.,  {Londrillo} P.,   {Ciotti} L.,  2003, \mn@doi [\mnras]
  {10.1046/j.1365-8711.2003.06554.x}, \href
  {http://adsabs.harvard.edu/abs/2003MNRAS.342..501N} {342, 501}

\bibitem[\protect\citeauthoryear{{Nipoti}, {Treu}  \& {Bolton}}{{Nipoti}
  et~al.}{2009a}]{Nipoti2009aApJ}
{Nipoti} C.,  {Treu} T.,   {Bolton} A.~S.,  2009a, \mn@doi [\apj]
  {10.1088/0004-637X/703/2/1531}, \href
  {http://adsabs.harvard.edu/abs/2009ApJ...703.1531N} {703, 1531}

\bibitem[\protect\citeauthoryear{{Nipoti}, {Treu}, {Auger}  \&
  {Bolton}}{{Nipoti} et~al.}{2009b}]{Nipoti2009bApJl}
{Nipoti} C.,  {Treu} T.,  {Auger} M.~W.,   {Bolton} A.~S.,  2009b, \mn@doi
  [\apjl] {10.1088/0004-637X/706/1/L86}, \href
  {http://adsabs.harvard.edu/abs/2009ApJ...706L..86N} {706, L86}

\bibitem[\protect\citeauthoryear{{Nipoti}, {Treu}, {Leauthaud}, {Bundy},
  {Newman}  \& {Auger}}{{Nipoti} et~al.}{2012}]{Nipoti2012MNRAS}
{Nipoti} C.,  {Treu} T.,  {Leauthaud} A.,  {Bundy} K.,  {Newman} A.~B.,
  {Auger} M.~W.,  2012, \mn@doi [\mnras] {10.1111/j.1365-2966.2012.20749.x},
  \href {http://adsabs.harvard.edu/abs/2012MNRAS.422.1714N} {422, 1714}

\bibitem[\protect\citeauthoryear{{Oogi} \& {Habe}}{{Oogi} \&
  {Habe}}{2013}]{Oog13}
{Oogi} T.,  {Habe} A.,  2013, \mn@doi [\mnras] {10.1093/mnras/sts047}, \href
  {https://ui.adsabs.harvard.edu/abs/2013MNRAS.428..641O} {428, 641}

\bibitem[\protect\citeauthoryear{{Posti}, {Nipoti}, {Stiavelli}  \&
  {Ciotti}}{{Posti} et~al.}{2014}]{Posti2014MNRAS}
{Posti} L.,  {Nipoti} C.,  {Stiavelli} M.,   {Ciotti} L.,  2014, \mn@doi
  [\mnras] {10.1093/mnras/stu301}, \href
  {http://adsabs.harvard.edu/abs/2014MNRAS.440..610P} {440, 610}

\bibitem[\protect\citeauthoryear{{Robertson}, {Cox}, {Hernquist}, {Franx},
  {Hopkins}, {Martini}  \& {Springel}}{{Robertson} et~al.}{2006}]{Rob06}
{Robertson} B.,  {Cox} T.~J.,  {Hernquist} L.,  {Franx} M.,  {Hopkins} P.~F.,
  {Martini} P.,   {Springel} V.,  2006, \mn@doi [\apj] {10.1086/500360}, \href
  {https://ui.adsabs.harvard.edu/abs/2006ApJ...641...21R} {641, 21}

\bibitem[\protect\citeauthoryear{{Scoville} et~al.,}{{Scoville}
  et~al.}{2007}]{Sco++07}
{Scoville} N.,  et~al., 2007, \mn@doi [\apjs] {10.1086/516585}, \href
  {https://ui.adsabs.harvard.edu/abs/2007ApJS..172....1S} {172, 1}

\bibitem[\protect\citeauthoryear{{S\'ersic}}{{S\'ersic}}{1968}]{Sersic68}
{S\'ersic} J.~L.,  1968, {Atlas de galaxias australes}

\bibitem[\protect\citeauthoryear{{Simard}, {Trevor Mendel}, {Patton}, {Ellison}
   \& {McConnachie}}{{Simard} et~al.}{2011}]{Simard2011Cat}
{Simard} L.,  {Trevor Mendel} J.,  {Patton} D.~R.,  {Ellison} S.~L.,
  {McConnachie} A.~W.,  2011, VizieR Online Data Catalog, \href
  {https://ui.adsabs.harvard.edu/abs/2011yCat..21960011S} {p. J/ApJS/196/11}

\bibitem[\protect\citeauthoryear{Skilling}{Skilling}{2004}]{Skilling2004AIP}
Skilling J.,  2004, \mn@doi [AIP Conference Proceedings] {10.1063/1.1835238},
  735, 395

\bibitem[\protect\citeauthoryear{{Somerville} \& {Dav{\'e}}}{{Somerville} \&
  {Dav{\'e}}}{2015}]{Som15}
{Somerville} R.~S.,  {Dav{\'e}} R.,  2015, \mn@doi [\araa]
  {10.1146/annurev-astro-082812-140951}, \href
  {https://ui.adsabs.harvard.edu/abs/2015ARA&A..53...51S} {53, 51}

\bibitem[\protect\citeauthoryear{{Sonnenfeld}, {Leauthaud}, {Auger}, {Gavazzi},
  {Treu}, {More}  \& {Komiyama}}{{Sonnenfeld} et~al.}{2018}]{Son++18b}
{Sonnenfeld} A.,  {Leauthaud} A.,  {Auger} M.~W.,  {Gavazzi} R.,  {Treu} T.,
  {More} S.,   {Komiyama} Y.,  2018, \mn@doi [\mnras] {10.1093/mnras/sty2262},
  \href {https://ui.adsabs.harvard.edu/abs/2018MNRAS.481..164S} {481, 164}

\bibitem[\protect\citeauthoryear{{Sonnenfeld}, {Wang}  \&
  {Bahcall}}{{Sonnenfeld} et~al.}{2019}]{Sonnenfeld2019A&A}
{Sonnenfeld} A.,  {Wang} W.,   {Bahcall} N.,  2019, \mn@doi [\aap]
  {10.1051/0004-6361/201834260}, \href
  {http://adsabs.harvard.edu/abs/2019A\%26A...622A..30S} {622, A30}

\bibitem[\protect\citeauthoryear{{Straatman} et~al.,}{{Straatman}
  et~al.}{2018}]{Straatman2018ApJS}
{Straatman} C.~M.~S.,  et~al., 2018, \mn@doi [\apjs]
  {10.3847/1538-4365/aae37a}, \href
  {http://adsabs.harvard.edu/abs/2018ApJS..239...27S} {239, 27}

\bibitem[\protect\citeauthoryear{{Strauss} et~al.,}{{Strauss}
  et~al.}{2002}]{Str++02}
{Strauss} M.~A.,  et~al., 2002, \mn@doi [\aj] {10.1086/342343}, \href
  {https://ui.adsabs.harvard.edu/abs/2002AJ....124.1810S} {124, 1810}

\bibitem[\protect\citeauthoryear{{Tanaka} et~al.,}{{Tanaka}
  et~al.}{2019}]{Tanaka2019arXiv}
{Tanaka} M.,  et~al., 2019, arXiv e-prints, \href
  {https://ui.adsabs.harvard.edu/abs/2019arXiv190910721T} {p. arXiv:1909.10721}

\bibitem[\protect\citeauthoryear{{Westera}, {Lejeune}, {Buser}, {Cuisinier}  \&
  {Bruzual}}{{Westera} et~al.}{2002}]{Wes++02}
{Westera} P.,  {Lejeune} T.,  {Buser} R.,  {Cuisinier} F.,   {Bruzual} G.,
  2002, \mn@doi [\aap] {10.1051/0004-6361:20011493}, \href
  {http://adsabs.harvard.edu/abs/2002A%26A...381..524W} {381, 524}

\bibitem[\protect\citeauthoryear{{Zahid} \& {Geller}}{{Zahid} \&
  {Geller}}{2017}]{Zah17}
{Zahid} H.~J.,  {Geller} M.~J.,  2017, \mn@doi [\apj]
  {10.3847/1538-4357/aa7056}, \href
  {https://ui.adsabs.harvard.edu/abs/2017ApJ...841...32Z} {841, 32}

\bibitem[\protect\citeauthoryear{{Zahid}, {Damjanov}, {Geller}, {Hwang}  \&
  {Fabricant}}{{Zahid} et~al.}{2016a}]{Zahid2016ApJ821}
{Zahid} H.~J.,  {Damjanov} I.,  {Geller} M.~J.,  {Hwang} H.~S.,   {Fabricant}
  D.~G.,  2016a, \mn@doi [\apj] {10.3847/0004-637X/821/2/101}, \href
  {https://ui.adsabs.harvard.edu/abs/2016ApJ...821..101Z} {821, 101}

\bibitem[\protect\citeauthoryear{{Zahid}, {Geller}, {Fabricant}  \&
  {Hwang}}{{Zahid} et~al.}{2016b}]{Zahid2016ApJ}
{Zahid} H.~J.,  {Geller} M.~J.,  {Fabricant} D.~G.,   {Hwang} H.~S.,  2016b,
  \mn@doi [\apj] {10.3847/0004-637X/832/2/203}, \href
  {http://adsabs.harvard.edu/abs/2016ApJ...832..203Z} {832, 203}

\bibitem[\protect\citeauthoryear{{van de Sande} et~al.,}{{van de Sande}
  et~al.}{2013}]{vandeSande2013ApJ}
{van de Sande} J.,  et~al., 2013, \mn@doi [\apj] {10.1088/0004-637X/771/2/85},
  \href {http://adsabs.harvard.edu/abs/2013ApJ...771...85V} {771, 85}

\bibitem[\protect\citeauthoryear{{van der Wel} et~al.,}{{van der Wel}
  et~al.}{2014}]{vanderWel14}
{van der Wel} A.,  et~al., 2014, \mn@doi [\apj] {10.1088/0004-637X/788/1/28},
  \href {https://ui.adsabs.harvard.edu/abs/2014ApJ...788...28V} {788, 28}

\bibitem[\protect\citeauthoryear{{van der Wel} et~al.,}{{van der Wel}
  et~al.}{2016}]{LEGA-C}
{van der Wel} A.,  et~al., 2016, \mn@doi [\apjs] {10.3847/0067-0049/223/2/29},
  \href {https://ui.adsabs.harvard.edu/abs/2016ApJS..223...29V} {223, 29}

\makeatother
\end{thebibliography}



\appendix
\onecolumn
\section{Comparison with independent estimates of the stellar mass of SDSS galaxies}
\label{app:comparison}
In this work we have estimated the stellar masses of the SDSS and LEGA-C galaxies anew in a self-consistent way. As a sanity check and for comparison with other works, it is useful to compare our estimates with others available in the literature.
For this purpose, we contrast here, for our SDSS sample of ETGs,  our values of $M_*$ with those obtained for the same galaxies by M14, who measured $M_*$ for $\approx$ 660,000 galaxies of the SDSS DR7 Legacy Survey, relying on the photometric analysis of \citet{Simard2011Cat} in  the $g$ and $r$ bands, extended by M14 also to the $u$, $i$ and $z$ bands (we took M14's stellar mass estimates from the \textsc{UPenn\_PhotDec\_MSTAR}\footnote{Available at \url{http://alan-meert-website-aws.s3-website-us-east-1.amazonaws.com/fit_catalog/download/index.html}.} catalogue of \citet{Meert2015MNRAS}).

\begin{figure}
        \includegraphics[width=0.5\columnwidth]{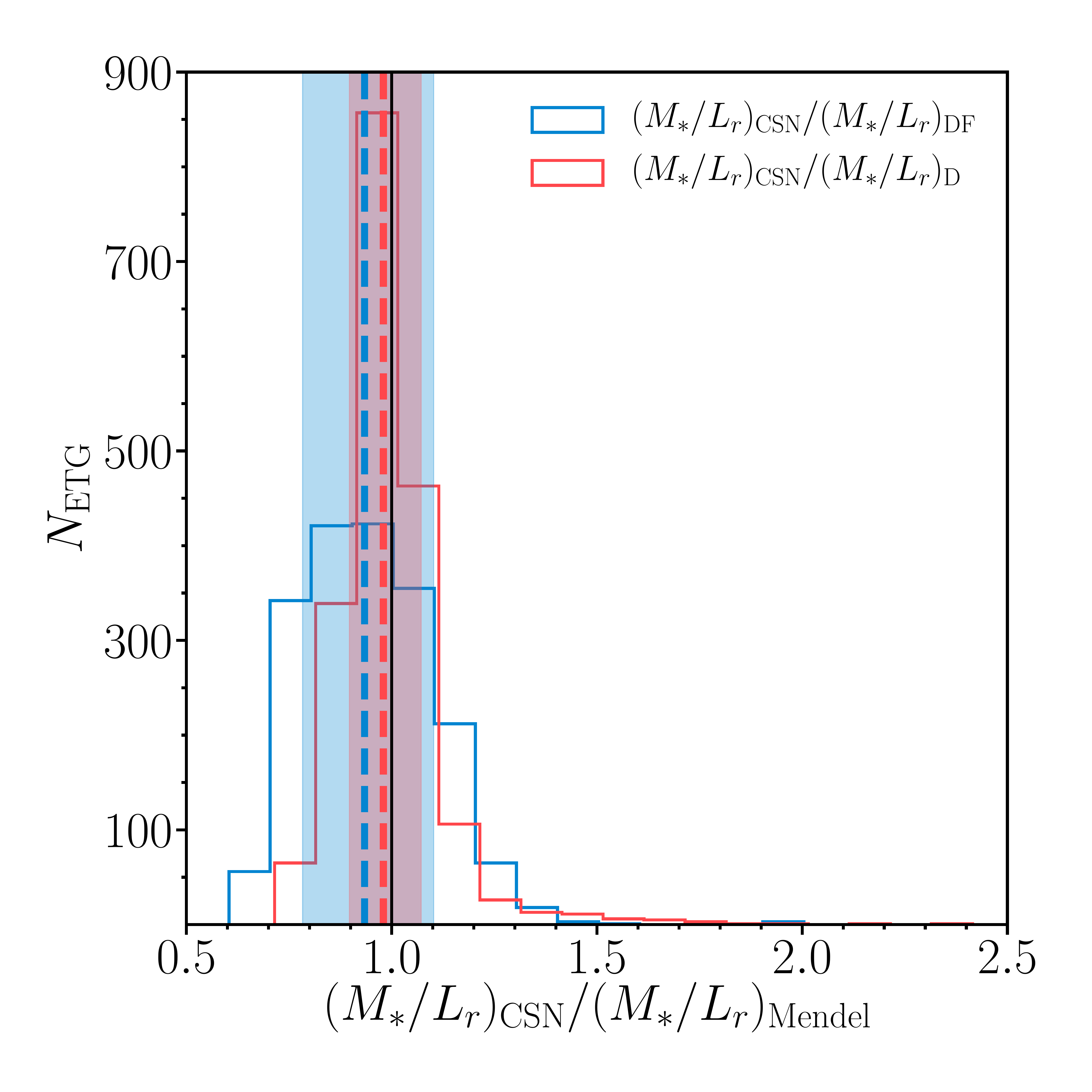}
        \includegraphics[width=0.5\columnwidth]{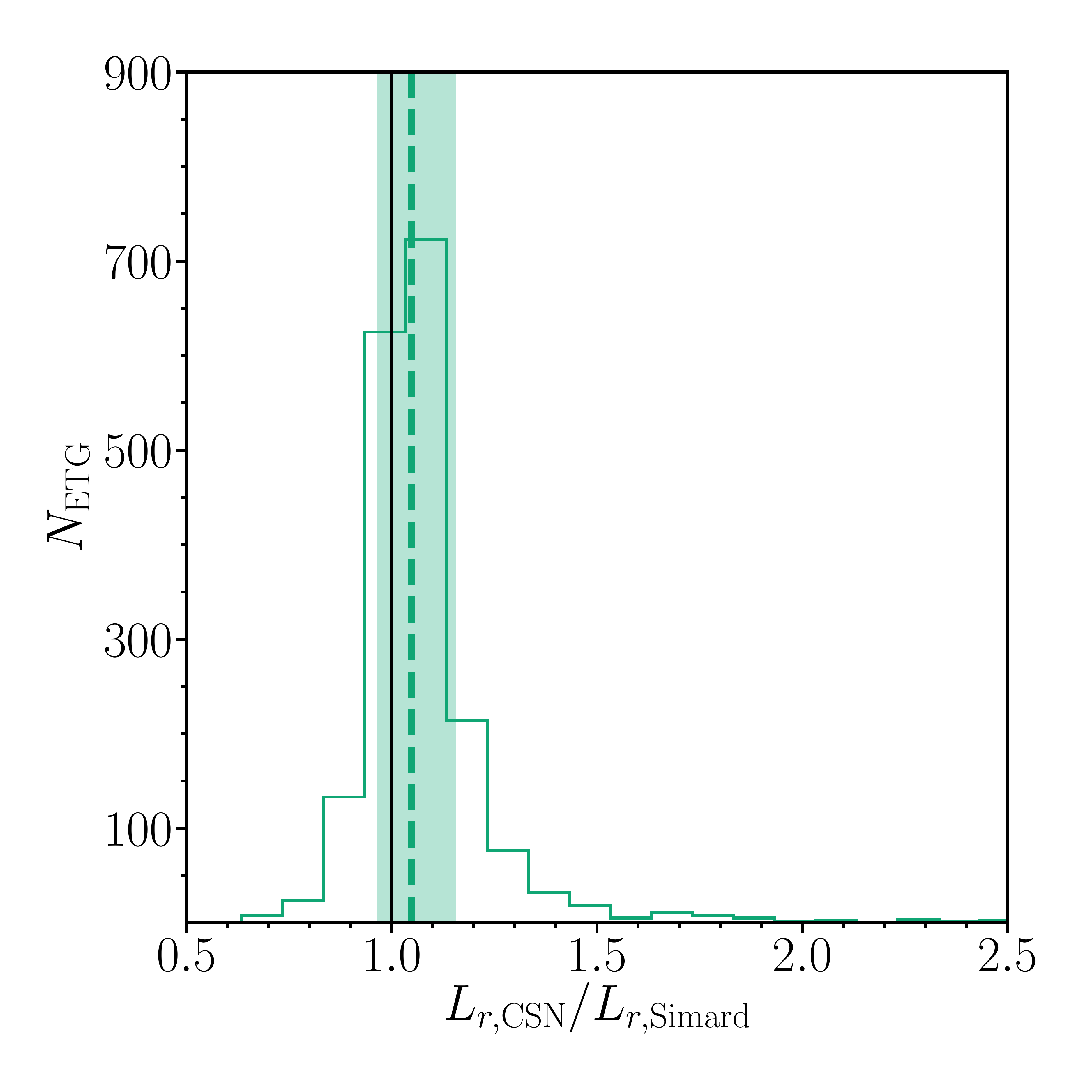} 
\caption{Left panel: distribution of the ratio between our $r$-band stellar mass-to-light ratio ($(M_*/L_r)_\mathrm{CSN}$) and those obtained by M14 assuming 
dust-free ($(M_*/L_r)_\mathrm{DF}$, blue histogram) and dusty ($(M_*/L_r)_\mathrm{D}$, red histogram) models for the SDSS galaxies of our sample. Right panel:
distribution of the ratio between our $r$-band luminosity ($L_{r,\mathrm{CSN}}$) and that of \citet{Simard2011Cat} ($L_{r,\mathrm{Simard}}$) for the same galaxies as in the left panel.
In both panels, the dashed lines represent the medians of the distributions, while the intervals between the 16-th and 84-th percentiles are indicated by the shaded areas.}
	\label{fig:comparison_ml_l}
\end{figure}

\begin{figure}
        \includegraphics[width=0.5\columnwidth]{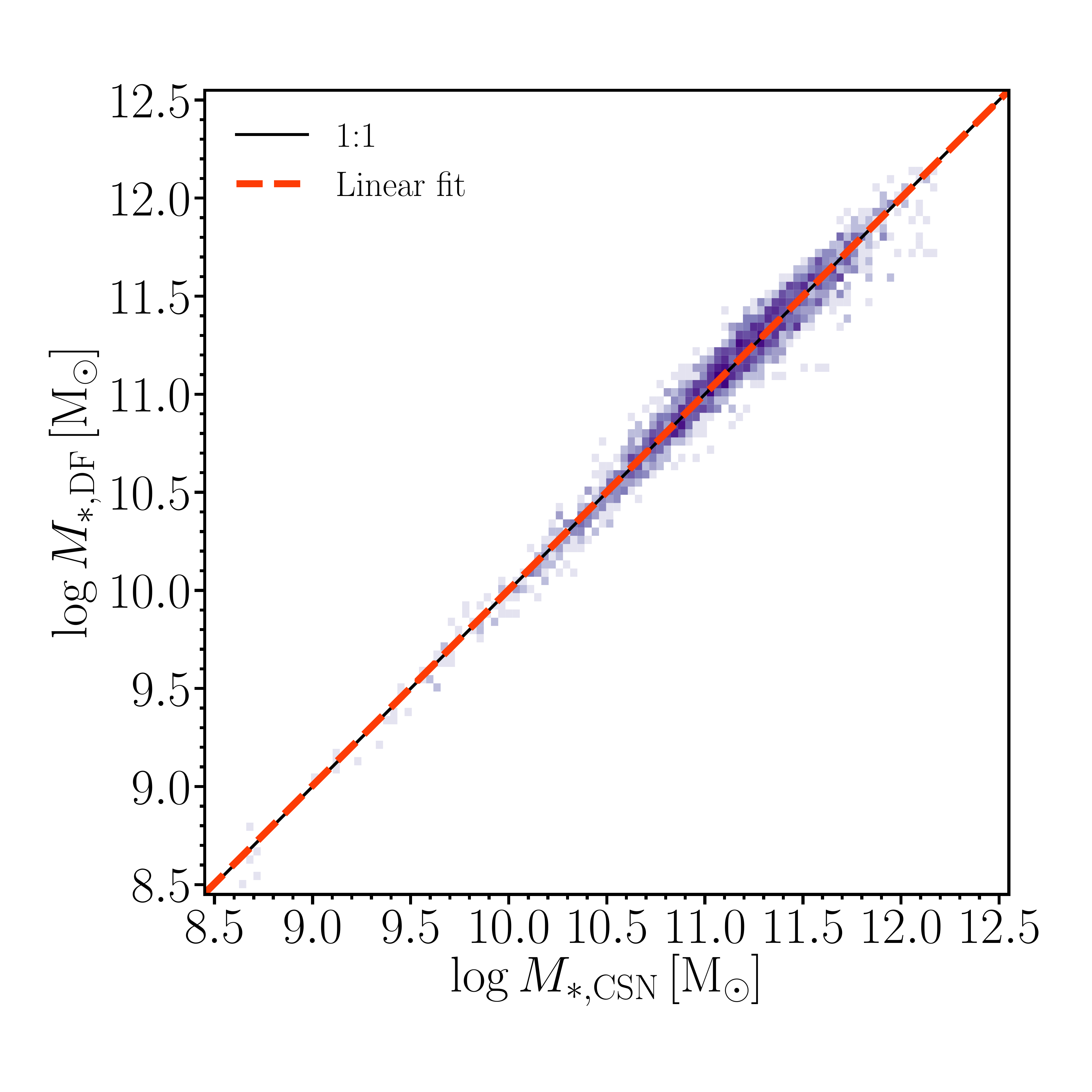}
        \includegraphics[width=0.5\columnwidth]{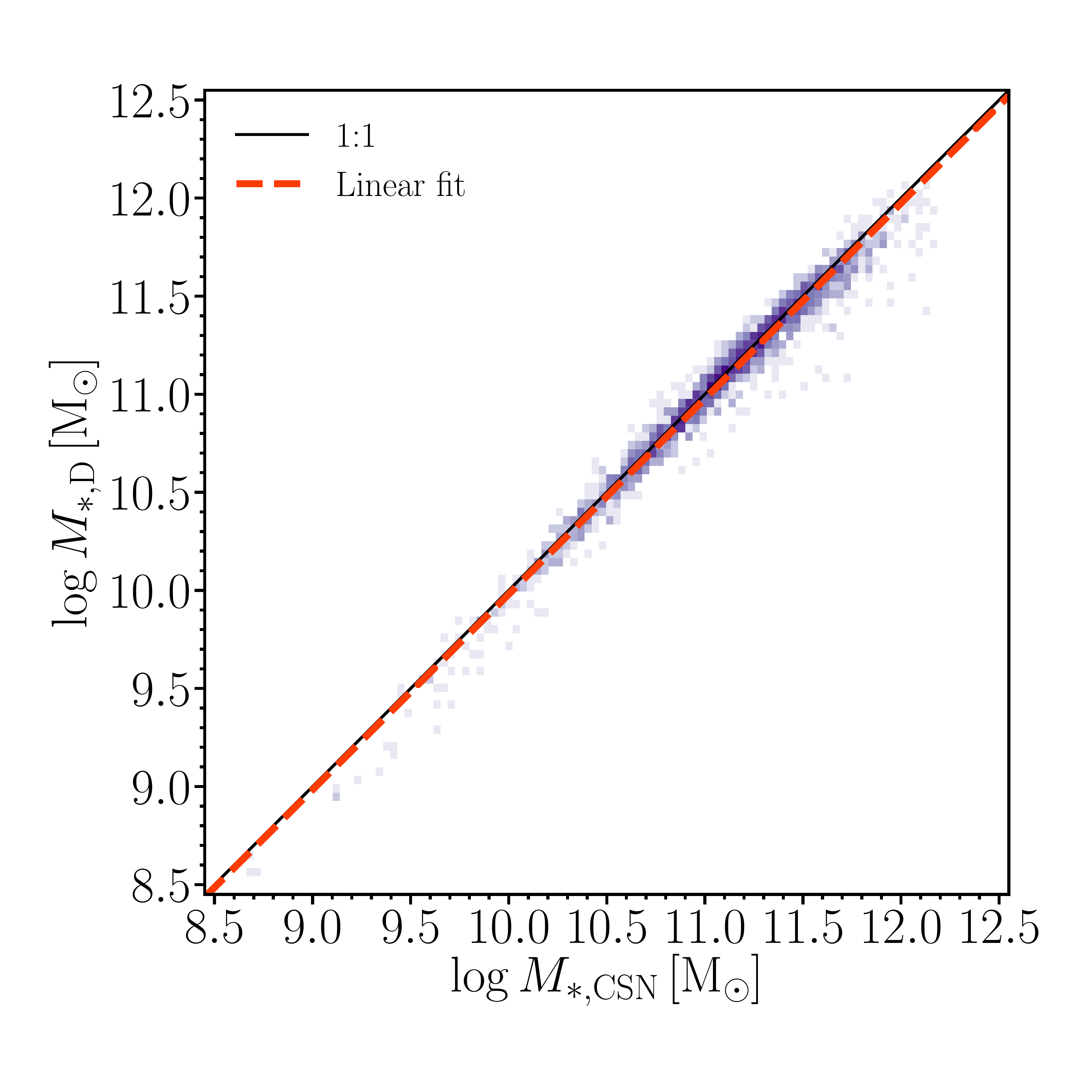} 
\caption{Pure Sérsic fit stellar masses based on the dust-free (left panel) and dusty (right panel)  models of M14 ($\log M_{*,\mathrm{DF}}$ and $\log M_{*,\mathrm{D}}$, respectively) as functions of our stellar masses ($\log M_{*,\mathrm{CSN}}$) for the SDSS galaxies of our sample.
The distributions are represented as two-dimensional histograms: the darkest the colour of the pixel the highest the number of galaxies. 
In each panel, the solid line represents the 1:1 relation, while the dashed line is the linear fit to the data.}
	\label{fig:comparison_masses}
\end{figure}

Both our and M14's stellar masses are obtained by multiplying the galaxy luminosity $L$ by the  stellar mass-to-light ratio $M_*/L$, so it is interesting to compare independently estimates of these two quantities.
Since our stellar masses are based on a Sérsic photometric fit, we limit our comparison to the stellar mass estimates of M14 based on the pure Sérsic fits of \citet{Simard2011Cat}.
We calculate M14's stellar mass-to-light ratio in the $r$ band $M_*/L_r$ related to two different models considered: one based on a "dust-free" model (assuming zero dust extinction) and the other on a "dusty" model (assuming non-zero dust extinction).
In \autoref{fig:comparison_ml_l}, we show, for the $\approx 2000$ galaxies of our SDSS sample (see \autoref{tab:sample_selection_steps}), the distributions of the ratios between our $M_*/L_r$ and those of M14 (left panel), and of the ratios between our $r$-band luminosities $L_r$ and those obtained by \citet{Simard2011Cat} for pure Sérsic fits. The overall agreement is good, though, on average, our $M_*/L_r$ tend to be slightly lower and our $L_r$ slightly higher than those of M14 and \citet{Simard2011Cat}, respectively.
In \autoref{fig:comparison_masses} we show, for the same galaxies as in \autoref{fig:comparison_ml_l}, the dust-free (left panel) and dusty  (right panel) stellar masses of M14 as functions of our stellar masses. In both cases, there is remarkably good agreement between our and M14's stellar masses: the linear fits are very close the 1:1 relation and the scatter is relatively small.

\section{Details of the calculation of the likelihood used in the model-data comparison}\label{app:likelihood}

Here we provide some steps of the calculation of the likelihood in \autoref{eq:Mobs_z_sigma_Phi}.
In this section all masses are in units of solar masses. 
By writing explicitly  each term in \autoref{eq:product} for our case, we obtain
\begin{equation}\label{eq:expliciting_Phi_Delta_appendix}
    \pdf(\data|\hyperp)=\prod_{i=1}^{n}\int\diff\logmi\,\diff\logsigmaei\,\diff z_i\,
    \pdf(\logmobsi,\logsigmaeobsi,z_i^\obs|\logmi,\logsigmaei,z_i)\,
    \pdf(\logmi,\logsigmaei,z_i|\hyperp).
\end{equation}
As explained in \autoref{subsec:pdf_and_hyperparameters}, we neglect the uncertainty on redshift, so that the first term on the right-side of \autoref{eq:expliciting_Phi_Delta_appendix} becomes
\begin{equation}
    \pdf(\logmobsi,\logsigmaeobsi,z_i^\obs|\logmi,\logsigmaei,z_i)=
    \pdf(\logmobsi|\logmi)\,\pdf(\logsigmaeobsi|\logsigmaei)\,\delta(z_i^\mathrm{obs}-z_i).
\end{equation}
Therefore, we can rewrite \autoref{eq:Mobs_z_sigma_Phi} for the $i$-th galaxy as follows:
\begin{equation}\label{eq:p_data_hyper}
\begin{aligned}
    \pp(\logmobsi,\logsigmaeobsi,z_i^\obs|\hyperp)=
&\bigintsss\diff\logmi\frac{\mathcal{A}(\logmi)}{\sqrt{2\pi{\errmass}}}\mathrm{exp}\left\{-\frac{(\logmi-\logmobsi)^2}{2{\errmass}}
\right\}\,
\frac{1}{\sqrt{2\pi{\sigma^2_*}}} \mathrm{exp}\left\{-\frac{(\logmi-\mu_*)^2}{2\sigma^2_*}\right\}\mathcal{E}(\logmi|\hyperp)\,\times\\
   \times&\bigintsss\diff\logsigmaei\frac{1}{\sqrt{2\pi{\errsigma}}}
   \mathrm{exp}\left\{-\frac{(\logsigmaei-\logsigmaeobsi)^2}{2{\errsigma}}\right\}
\frac{1}{\sqrt{2\pi{\errmu}}} \mathrm{exp}\left\{-\frac{(\logsigmaei-\mu_{\sigma,i})^2}{2{\errmu}}\right\},
\end{aligned}
\end{equation}
where
\begin{equation}
\mu_{\sigma,i}=\mu_0+\beta\log\left(\frac{M_{*,i}}{M_*^\mathrm{piv}}\right)+\zeta\log\left(\frac{1+z_i}{1+z^\mathrm{piv}}\right)
\end{equation}
and
\begin{equation}
\sigma_{\sigma_i}=\psi_0+\xi\log\left(\frac{1+z}{1+z^\mathrm{piv}}\right).  
\end{equation}
\autoref{eq:p_data_hyper}, the term $\mathcal{A}(\logm)$ allows to normalise the distribution over all values of the observed stellar mass. Specifically, $\mathcal{A}(\logm)$ ensures that the probability of having an ETG with $\logm^\obs$ between $\log M_{\rm *,min}$ (the lower bound of the considered observed stellar mass interval) and $+\infty$ is one:
\begin{equation}
\bigintsss_{\log M_{\rm *,min}}^\infty \diff\logmobsi\,\frac{\mathcal{A}(\logmi)}{\sqrt{2\pi{\errmass}}} 
\mathrm{exp}\left\{-\frac{(\logmi-\logmobsi)^2}{2{\errmass}}\right\}=1.
\end{equation}
Hence, $\mathcal{A}(\logmi)$ is given by
\begin{equation}
\mathcal{A}(\logmi)=\frac{1}{\displaystyle\bigintsss_{\log M_*^\mathit{min}}^\infty \diff\mathcal{M}'\,\frac{1}{\sqrt{2\pi{\errmass}}} 
\mathrm{exp}\left\{-\frac{(\logmi-\mathcal{M}')^2}{2{\errmass}}\right\}}=
\left[\frac{\sqrt{{\errmass}}}{\displaystyle\frac{1}{2}\,\errmass\,\mathrm{erf}\left(\frac{\sqrt{2}}{2}\frac{\logmi-\mathcal{M}}{\errmass}\right)}\right]_{\log M_*^\mathit{min}}^{+\infty}.
\end{equation}
In the previous two equations the term $\log M_{\rm *,min}$ is obtained from the mass-completeness limits at a given redshift for SDSS and LEGA-C ETGs (\autoref{ssec:complete}), while for the high-redshift sample galaxies we assume a constant value of $10.5$.

The integral term in $\diff\logsigmaei$ of \autoref{eq:p_data_hyper} can be written as
\begin{equation}
\frac{1}{\sqrt{2\pi({\errsigma}+{\errmu})}}\mathrm{exp}\left\{-\frac{(\logsigmaeobsi-\mu_{\sigma,i})^2}{2({\errsigma}+{\errmu})}
\right\}\bigintsss\diff\logsigmaei\,\frac{1}{\sqrt{2\pi{\tilde{\sigma}^2_i}}}\mathrm{exp}\left\{-\frac{(\logsigmaei-\tilde{\mu}_i)^2
}{2{\tilde{\sigma}^2_i}} \right\}=\frac{1}{\sqrt{2\pi({\errsigma}+{\errmu})}}\mathrm{exp}\left\{-\frac{(\logsigmaeobsi-\mu_{\sigma,i})^2}{2({\errsigma}+{\errmu})}
\right\},
\label{eq:dlogsigmaei}
\end{equation}
where
\begin{equation}
    \tilde{\mu}_i = \frac{\logsigmaeobsi\,\errmu+\mu_{\sigma,i}\,\errsigma}{\errmu+\errsigma}
    \qquad\qquad\textrm{and}\qquad\qquad
    \tilde{\sigma}_i=\sqrt{\frac{\errmu\errsigma}{\errmu+\errsigma}}.
\end{equation}
By writing $\mu_{\sigma,i}$ explicitly, \autoref{eq:p_data_hyper} becomes
\begin{equation}
\begin{aligned}    
\pp(\logmobsi,\logsigmaeobsi,z_i^\obs|\hyperp)=\bigintsss&\diff\logmi\frac{\mathcal{A}(\logmi)}{\sqrt{2\pi{\errmass}}}\mathrm{exp}\left\{-\frac{(\logmi-\logmobsi)^2}{2{\errmass}}
\right\}\,
\frac{1}{\sqrt{2\pi{\sigma^2_*}}} \mathrm{exp}\left\{-\frac{(\logmi-\mu_*)^2}{2\sigma^2_*}\right\}\mathcal{E}(\logmi|\hyperp)\,\times\\
\times\,&\frac{1}{\sqrt{2\pi}\sigma_{\mathrm{eff},i}|\beta|}\mathrm{exp}\left\{-\frac{(\logmi-\mu_{\mathrm{eff},i})^2}{2\sigma_{\mathrm{eff},i}^2}\right\}
\end{aligned}
\end{equation}
with
\begin{equation}
\mu_{\mathrm{eff},i}=\logm^\mathrm{piv}+\frac{\logsigmaeobsi-\mu_0-\zeta\left[\log(1+z_i)-\log(1+z^\mathrm{piv})\right]}{\beta}
\qquad\qquad \text{and} \qquad\qquad
\sigma_{\mathrm{eff},i}=\frac{({\errsigma}+{\errmu})}{\beta^2}.
\end{equation}
\begin{equation}\label{eq:final_likelihood}
\begin{aligned}
\pp(\logmobsi,\logsigmaeobsi,z_i^\obs|\hyperp)=\,\,&
\frac{1}{|\beta|}\frac{1}{\sqrt{2\pi({\errmass}+{\sigma^2_{\mathrm{eff},i}})}}\exp\left \{-\frac{(\logmobsi-\mu_{\mathrm{eff},i})^2}{2(\sigma_{M_{*,i
}}^2+\sigma_{\mathrm {eff},i}^2)}\right\}\times\\
\times&\bigintsss\diff\logmi\,\,
\frac{1}{\sqrt{2\pi{\sigma^{'}_i}^2}}\exp\left\{-\frac {(\logmi-\mu^{'}_{i})^2} {2{\sigma^{'}_i}^2} 
\right\}\,\,\mathcal{A}(\logmi)\,\,\mathcal{S}(\logmi),
\end{aligned}
\end{equation}
with
\begin{equation}
    \mu^{'}_{i}=\frac{\logmi\sigma^2_{\mathrm{eff},i}+\mu_{\mathrm{eff},i}\errmass}{{\sigma^2_{\mathrm{eff},i}\errmass}}
    \qquad\qquad \text{and} \qquad\qquad
    \sigma^{'}_{i}=\sqrt\frac{{\sigma^2_{\mathrm{eff},i}\errmass}}{{\sigma^2_{\mathrm{eff},i}+\errmass}}.
\end{equation}

We compute the integral term in \autoref{eq:final_likelihood} numerically, using the trapezoidal rule.

\section{Mock sample}
\label{sec:mock}

In order to check the reliability of our method, we performed some tests on  mock samples. 
In the following, we provide an example of our method applied to a mock sample of around 400 ETGs, with properties similar to our SDSS subsample, generated as follows (masses are in units of $\msun$ and velocity dispersions in units of $\kms$):
\begin{itemize}
\item stellar masses $M_*^\mathrm{t}$ are generated extracting $\log M_*^\mathrm{t}$ from a normal distribution with mean $11.321$ and standard deviation $0.358$;
\item the true velocity dispersions $\sigmae^\mathrm{t}$ are generated extracting $\logsigmae^\mathrm{t}\equiv\mu^\mathrm{t}$ from a normal distribution with mean
\begin{equation}
    \mu^\mathrm{t}=\mu_0^\mathrm{mock}+\beta_0^\mathrm{mock}\log\left(\frac{M_*^\mathrm{t}}{M_*^\mathrm{mock}}\right)
\end{equation}
and standard deviation $0.075$ dex, where $M_*^\mathrm{mock}=10^{11.321}$,  $\mu_0^\mathrm{mock}=2.287$, $\beta_0^\mathrm{mock}=0.176$;
\item the errors on the stellar masses $\sigma_{M_*}$ are extracted from a normal distribution with mean $0.760$ (the median stellar mass error in the SDSS sample) and standard deviation $\simeq0.009$ (the standard deviation of the stellar mass error distribution in the SDSS sample);
\item the errors on the velocity dispersions $\sigma_{\sigma_\mathrm{e}}$ are extracted from a normal distribution with mean $8.7$ (the median stellar velocity dispersion error in the SDSS sample) and standard deviation $2.995$ (the standard deviation of the stellar velocity dispersion error in the SDSS sample);
\item the values of $\logm^\obs$ and $\logsigmae^\obs$ are extracted from $\mathcal{N}(\mu=\logm^\mathrm{t},\,\sigma=\sigma_M)$ and $\mathcal{N}(\mu=\logsigmae^\mathrm{t},\,\sigma=\sigma_{\sigma_{\mathrm{e}}})$, respectively;
\item galaxies with $\logm^\obs<10.5$ are excluded, so for the mock $\log M_{\rm *,min}=10.5$.
\end{itemize}

For simplicity, we assume that all galaxies are at $z=z^\mathrm{piv}$, so that the mean and standard deviation of the skew prior in \autoref{eq:prior} used to model the stellar mass distribution reduces to $\mu_*=\mu_{*,0}$ and $\sigma_*=\sigma_{*,0}$.

\begin{figure*}
    \centering
    \includegraphics[width=0.9\columnwidth]{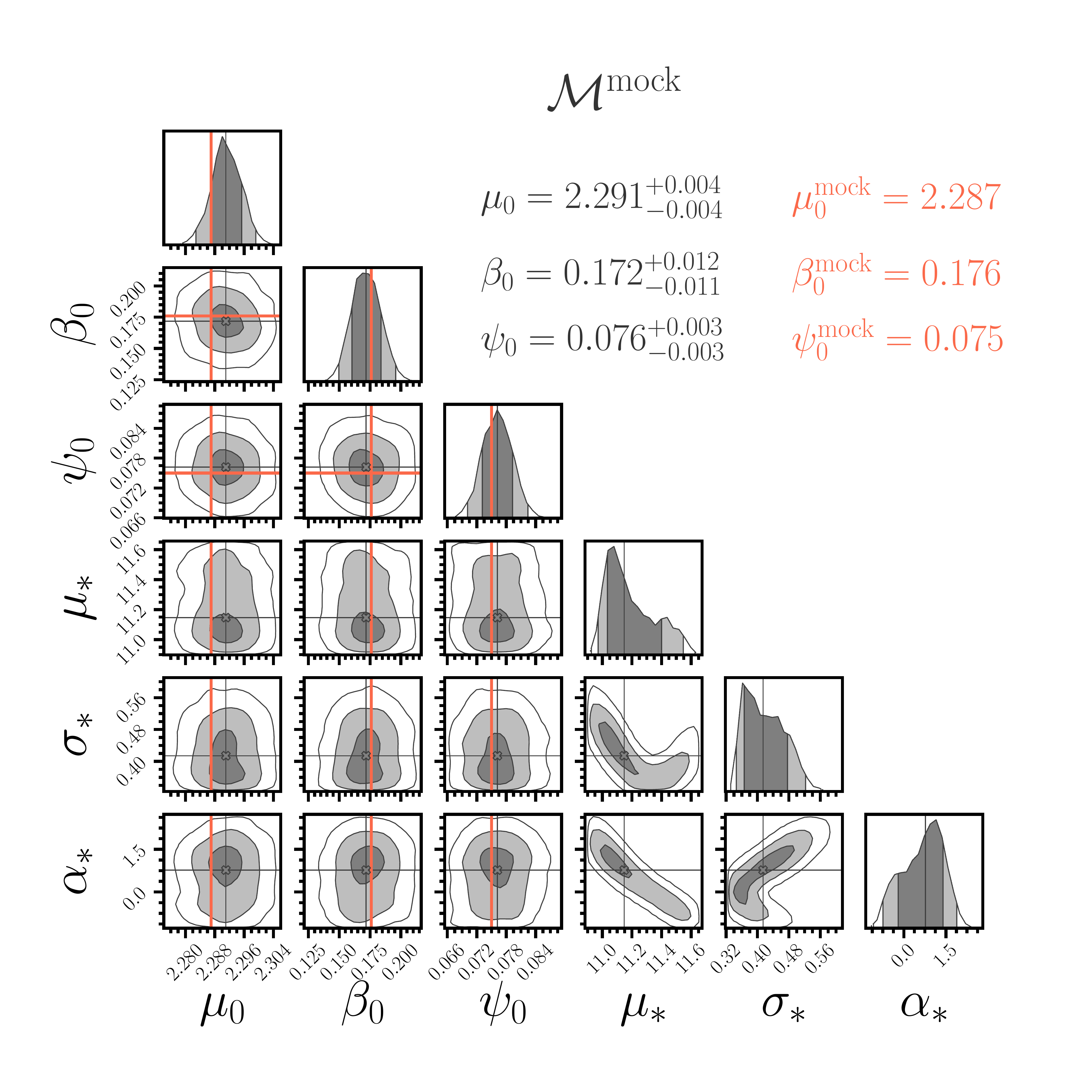}
    \caption{Same as \autoref{fig:mcmc_sdss}, but for model $\mathcal{M}^\mathrm{mock}$ (dark grey contours). The thick orange solid lines indicate the input values of the hyper-parameters $\mu_0^\mathrm{mock}$, $\beta_0^\mathrm{mock}$ and $\psi_0^\mathrm{mock}$.}
    \label{fig:mcmc_mock}
\end{figure*}

In order to sample the PDFs of the model applied to our mock catalogue (hereafter model $\mathcal{M}^\mathrm{mock}$), we perform a MCMC run (see \autoref{sec:sampling}), using $50$ random walkers running for $1000$ steps to reach the convergence of the hyper-parameter distribution. 
In \autoref{fig:mcmc_mock} we show the posterior PDFs of all hyper-parameters and report the median values of the hyper-parameters $\mu_0$, $\beta_0$ and $\psi_0$ with their $1\sigma$ uncertainties.  The input values of the hyper-parameters are all recovered within $1\sigma$.
\newpage
\bsp	
\label{lastpage}
\end{document}